\documentclass[aip, preprint, amsmath, onecolumn ]{revtex4-1}

\usepackage{graphicx}
\usepackage{float}
\usepackage{dcolumn}
\usepackage{bm}
\usepackage{units}
\usepackage{amsmath}
\usepackage{amssymb}
\usepackage{color}
\usepackage[version=4]{mhchem}
\usepackage{soul}
\usepackage{siunitx}
\usepackage[colorlinks = true, linkcolor = blue, urlcolor  = blue, citecolor = blue, anchorcolor = blue, unicode]{hyperref}
\usepackage{bookmark}

\let\tmpcmd\r
\def\ring{\tmpcmd}
\def\vr{\vec{r}}
\def\vk{\vec{k}}

\def\muB{\,\mu_B}
\def\Ry{\,\textrm{Ry}}
\def\cm{\,\textrm{cm}}
\def\nm{\,\textrm{nm}}
\def\eV{\,\textrm{eV}}
\def\meV{\,\textrm{meV}}
\def\GPa{\,\textrm{GPa}}

\def\VperAng{\,\textrm{V}/\textrm{\AA}}
\def\eVperAng{\,\textrm{eV}/\textrm{\AA}}

\def\etal{\textit{et$\,$al.}}

\def\cri{\ce{CrI_3}}
\def\bn{\ce{BN}}
\def\bncribn{$h$-\ce{BN}$\,|\,$\ce{2-CrI_3}$\,|\,h$-\ce{BN}}
\def\bncrigr{$h$-\ce{BN}$\,|\,$\ce{2-CrI_3}$\,|\,$\ce{graphene}}
\def\grcrigr{\ce{graphene}$\,|\,$\ce{2-CrI_3}$\,|\,$\ce{graphene}}
\def\mos{\ce{MoS_2}}

\def\BSTS{\ce{Bi Sb Te Se_2}}

\def\RPA{\textrm{RPA}}

\begin{document}

\author{Hai-Ping Cheng}
\email{hping@ufl.edu}
\affiliation{Department of Physics, University of Florida, Gainesville, FL 32611, USA}
\affiliation{Quantum Theory Project, University of Florida, Gainesville, FL 32611, USA}
\affiliation{Center for Molecular Magnetic Quantum Materials, University of Florida, Gainesville, FL 32611, USA}

\author{Shuanglong Liu}
\affiliation{Department of Physics, University of Florida, Gainesville, FL 32611, USA}
\affiliation{Quantum Theory Project, University of Florida, Gainesville, FL 32611, USA}

\author{Xiao Chen}
\affiliation{Department of Physics, University of Florida, Gainesville, FL 32611, USA}
\affiliation{Quantum Theory Project, University of Florida, Gainesville, FL 32611, USA} 

\author{Long Zhang}
\affiliation{Department of Physics, University of Florida, Gainesville, FL 32611, USA}
\affiliation{Quantum Theory Project, University of Florida, Gainesville, FL 32611, USA}

\author{James N Fry}
\affiliation{Department of Physics, University of Florida, Gainesville, FL 32611, USA}

\title{First-principles study of magnetism and electric field effects in 2D systems}

\begin{abstract}
This review article provides a bird's-eye view of what first-principles based methods can contribute to next-generation device design and simulation. After a brief overview of methods and capabilities in the area, we focus on published work by our group since 2015 and current work on \cri{}. We introduce both single- and dual-gate models in the framework of density functional theory and the constrained random phase approximation in estimating the Hubbard $U$ for 2D systems vs. their 3D counterparts. A wide range of systems, including graphene-based heterogeneous systems, transition metal dichalcogenides, and topological insulators, and a rich array of physical phenomena, including the macroscopic origin of polarization, field effects on magnetic order, interface state resonance induced peak in transmission coefficients, spin filtration, etc., are covered. For \cri{} we present our new results on bilayer systems such as the interplay between stacking and magnetic order, pressure dependence, and electric field induced magnetic phase transitions. We find that a bare bilayer \cri{},  graphene$\,|\,$bilayer \cri{}$\,|\,$graphene, $h$-BN$\,|\,$bilayer \cri{}$\,|\,h$-BN, and $h$-BN$\,|\,$bilayer \cri{}$\,|\,$graphene all have a different response at high field, while small field the difference is small except for graphene$\,|\,$bilayer \cri{}$\,|\,$graphene. We conclude 
%our results on bilayer \cri{} 
with discussion of some ongoing work and work planned in the near future, with the inclusion of further method development and applications.
\end{abstract}

\maketitle

% ==================================================
% Section 1: Introduction
% ==================================================

\section{Introduction}

Gating a junction with electric fields is a very common  experimental method to control functionality and properties of a system.
One example with the most significant societal impact is perhaps the field-effect transistor (FET) that, in conjunction with the development of metal-oxide-semiconductors (MOS), because of its high scalability led to a digital revolution in the 1950s.
First proposed in the mid-late 70s,\cite{RN3013} 
tunneling field effect transistors (TFETs) attracted much attention in mid 2000 because of nanostructures involving carbon nanotubes \cite{RN3015, RN3014} and the intense interest continues. In 2010 low-voltage tunneling FETs based on interband tunneling \cite{RN3012} were fully analyzed, followed by discovery of the single-layer \mos{} TFET. \cite{RN3016}
Shortly after, junctions with vertical geometry using layered $h$-BN 
and \mos{} 2D materials were built. \cite{RN1622, RN3052}
The performance characteristics, for example, the speed, of TFETs are not limited to the Maxwell-Boltzmann tail as in the conventional MOSFET.
To date, two-dimensional systems are regarded as promising materials for future low dissipation electronics that are not limited to transistor applications. \cite{RN3017,RN2459,RN3022, RN3020, RN3019, RN3018, RN3021,RN3025} Gating a system provide a easy method to applying an electric field and allows us to investigate the magnetic response of systems and thus magnetoelectric coupling, as well as a method to study charge-doping effects. 

\begin{figure}[htb!]
\centering
\includegraphics[width=0.45\textwidth]{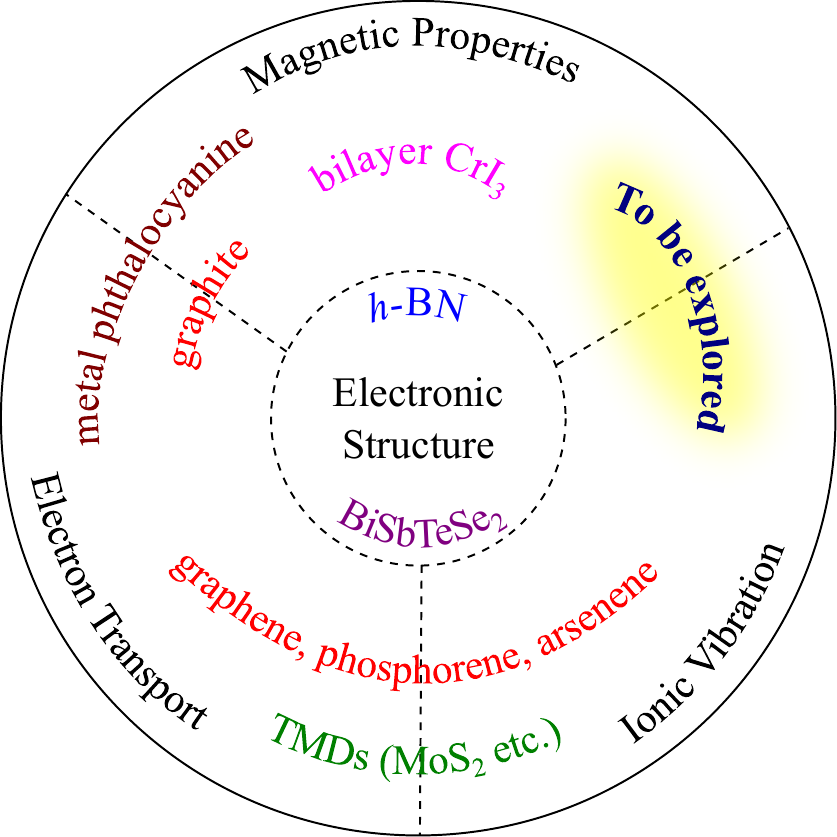}
\caption{\label{FIG:overview}
Physical properties and materials investigated by first-principles methods that deal with gate electric field. Selected references are as follows: graphene and graphite~\cite{RN20, RN265, RN14, RN103}, phosphorene and arsenene~\cite{RN154}, \textit{h}-\bn{}~\cite{RN20, RN233}, TMDs~\cite{RN157, RN154, RN160, RN100, RN155}, \ce{BiSbTeSe2}~\cite{RN3033}, and metal phthalocyanine~\cite{RN101, RN102}. 
} 
\end{figure}

Theoretically, modeling of gate field effects was motivated primarily by studies of MOSFETs. In early days, the circuit model or classical electromagnetic theory (for insulated-gate FETs) were used to model field-effect transistors. \cite{RN3023, RN3028}. Later, studies of the system at the electron level with inclusion of its quantum nature have been carried out within various approximations. The key is to solve the Poisson equation with appropriate boundary conditions. \cite{RN3026,RN3027} In this paper, our focus is on first-principles modeling and simulations of gate effects. Because of the limitations of computational power and algorithms, solution to the Poisson equation subject to prescribed boundary conditions as commonly used in classical E\&M have not been done until very recently, and to our best knowledge there are in practice only a few approaches that faithfully realize the E\&M principles. One approach has been developed by our group \cite{RN2199, RN2201} by employing the effective screening medium technique as the Poisson solver. \cite{RN92} In the last few years, we have applied this method to study both single- and dual-gate configurations, and a number of two-dimensional junctions and interfaces between bulk systems have been investigated. \cite{RN2193, RN3009, RN3010, RN2199, RN2201, RN2888, RN3060, RN3033} In the next section, we will highlight some important results from these studies. With this approach, boundary conditions in the $z$-direction are imposed according to the physical problem, which according to the uniqueness theorem guarantees the correct solution. The advantage of this approach is that it is combined with the non-equilibrium Green's function technique to study electron transport at finite bias, implemented in the TranSiesta package \cite{RN1155} in addition to the Quantum ESPRESSO package. \cite{RN1694} Besides TranSiesta, QuantumATK \cite{RN3047} can treat field effects by enabling Dirichlet (potential is held constant) and Neumann (electric field is held constant) boundary conditions, similar to our approach. For a generalized Poisson equation, Bani-Hashemian \textit{et$\,$al.} 
\cite{RN158} developed an algorithm to treat Dirichlet-type boundary conditions for first-principles device simulations. In addition to the boundary-condition-driven approach, several other methods should be mentioned because of their impact in current research. One is by Sohier \textit{et$\,$al.}, \cite{RN3032} in which the authors  truncate the Coulomb interaction in the direction perpendicular to the slab so that the charging of the slab can be simulated via field effects. An important development in this method is the treatment of flexural phonons in the presence of a field which breaks the mirror symmetry, thus allowing flexural-phonon-electron coupling. The method is also interfaced with the Quantum ESPRESSO package \cite{RN1694} for performing linear response calculations. Closely related to the truncated Coulomb approach is the dipole correction method by Brumme \textit{et$\,$al.}
\cite{RN162}, which introduces a way to include a charged plate within a system with periodic boundary conditions. In first-principles calculations, it is common to simply apply an electric field to understand field effects. Examples include but not limited to graphene-based 2D heterostructures. \cite{RN3051, RN3050, RN3049} It should be pointed out that applying an electric field to a system is very different from gating the system, since the former can be described as a closed system but the latter is certainly an open system. Experimentally, both gating and electric fields are used, and it is important to choose the appropriate tool to address the right problem. Figure~\ref{FIG:overview} is a sketch of materials and physical properties investigated using various method by the above-mentioned groups.
The remainder of this paper is organized as follows: In Section II we provide a brief description of our approach, models, and Poisson solver, and a review of prior applications; in Section III we present results from recent studies.
We conclude our effort of modeling field effects using first-principles method with an  outline of future developments and applications.   

% An overview of applications by us and others

% ==================================================
% Section 2: Review of our previous studies
% ==================================================

\section{Our Model and Previous Applications}

\begin{figure}[htb!]
\centering
\includegraphics[width=0.6\textwidth]{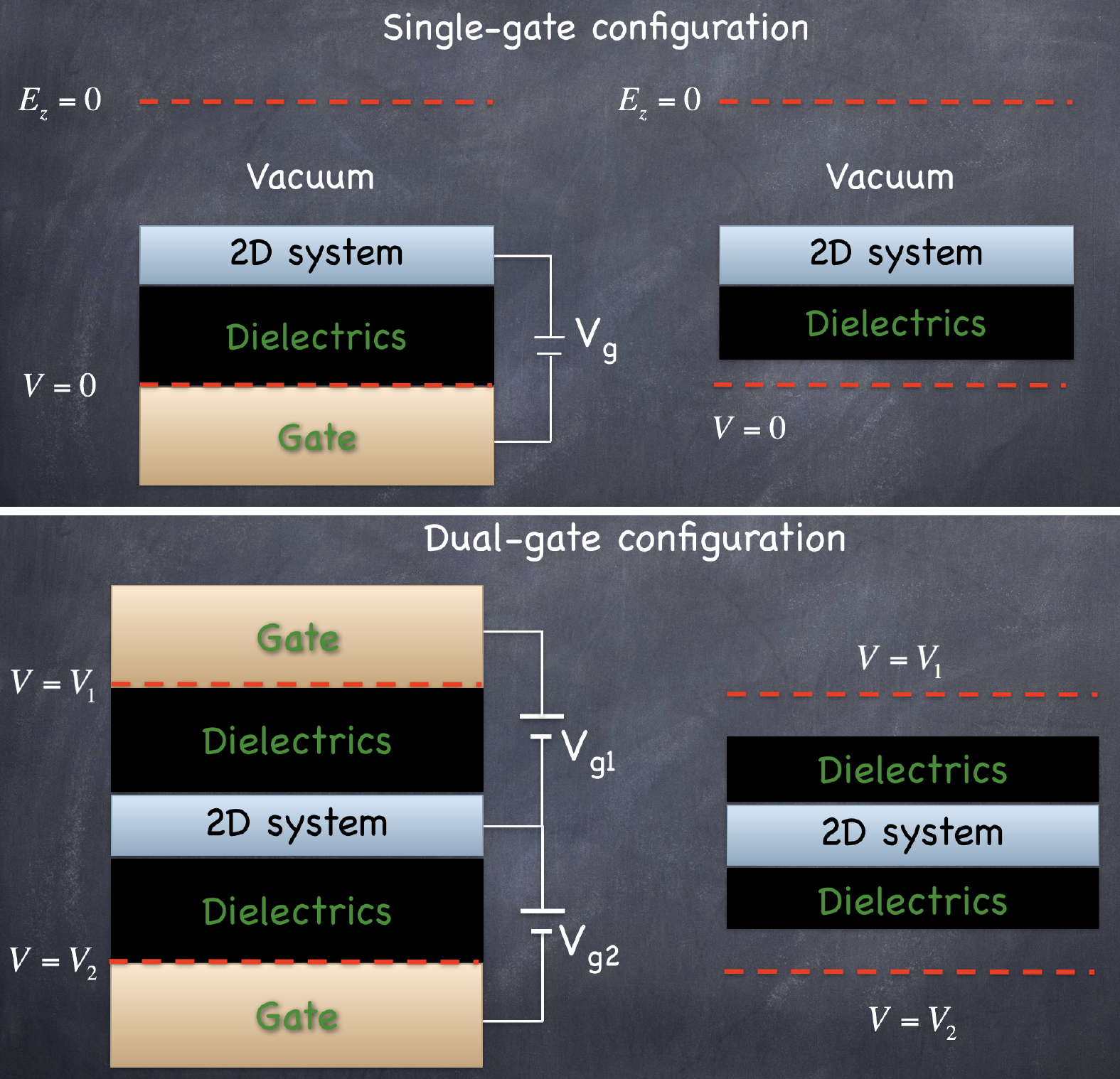}
\caption{
Device models: single-gate configuration (top) and dual gate configuration (bottom). 
\label{FIG:device}} 
\end{figure}

Similar to most other recent theoretical studies of field effects, we focus on two-dimensional systems, which comprise a very active research area. In our approach, experimental conditions are first identified, according to which we construct our simulation models. Figure~\ref{FIG:device} depicts single and dual gate configurations. For both the top and bottom panels, the left part is a sketch of the experiment, and the right is our model for simulation. Two key approximations in the simulation models are: 1) the dielectric layer is kept thin to reduce computational effort; and 2) a vacuum layer is inserted between electrodes (metal) and dielectrics to avoid complications caused by dielectric-metal interfaces, which have very little effect on the physical processes in the 2D systems.

The potential is the solution to the Poisson equation $ \nabla \cdot[\epsilon(\vr) \nabla]V(\vr) = -4\pi \rho_\textrm{tot}(\vr) $,
and the Green's function is defined as 
$ \nabla \cdot[\epsilon(\vr) \nabla]G(\vr,\vr^{\prime}) = -4\pi\delta(\vr-\vr^{\prime}) $.
With the given boundary condition, the Green's function is the potential of a point charge plus its image, 
\begin{equation}
G(\vec{g}_\parallel;z,z^{\prime})
= \frac{4\pi}{2 |g_\parallel|} \, \exp(-|g_\parallel| \, |z-z^{\prime}|)
 - \frac{4\pi}{2 |g_\parallel|} \, \exp[-|g_\parallel| \, (2z_1 - z-z^{\prime}) ]
\label{F2}
\end{equation}
In Equation \ref{F2}, $G$ is expanded in the momentum $\vec{g}$ within the 2D plane; $z$ is the  dimension in which the gate is applied, or the direction perpendicular to the plane, and $z_1$ is the position where the gate voltage is applied, or the position of the electrode.  Integrating $G$ over $z^{\prime}$ gives the electrostatic potential at any given point $z$. For the dual-gate configuration, there are two electrodes which lead to infinite number of image  charges; however, we know the analytical expression of the summation. The potential in the $x$-$y$ directions is easy to obtain in momentum space, since there are no image charges.  
This is the essence of the effective screening medium technique.~\cite{RN92} In the framework of density functional theory, \cite{RN925,RN8} the total energy of the system is  
\begin{eqnarray}
E[n] &=& K[n] + E_{xc}[n] + \frac12 \iint d\vr \, d\vr^{\prime} \,
 n(\vr) \, G(\vr,\vr^{\prime}) \, n(\vr^{\prime}) \nonumber \\
&& + \iint d\vr \, d\vr^{\prime} \, n(\vr) \, G(\vr,\vr^{\prime}) \, n_I(\vr^{\prime})
+ \frac12 \iint d\vr \, d\vr^{\prime} \, n_I(\vr) \, G(\vr,\vr^{\prime}) \, n_I(\vr^{\prime}) , 
\label{F1}
\end{eqnarray}
where $n$ is the electron density, $K$ is the single-particle,  non-interacting electron kinetic energy, 
$E_{xc}$ is the exchange-correlation functional, 
and $n_I$ is nuclear charge distribution. 
When analyzing interface properties, it is desirable to quantify the electric polarization as a function of the distance measured from the interface. For this purpose, we implemented the so-called hybrid Wannier function in our analysis code.
The conventional Wannier function is defined by the Fourier transform of the Bloch wave,
\begin{equation}
f(\vec{R},\vr) = \frac{\cal V}{(2\pi)^3} \int d\vk \,
 e^{i \vk \cdot \vec{R}} \, \psi_{n\vk}(\vr)
\end{equation}
The application of Wannier functions in solid state physics is now a common practice because of the availability of the Wannier90 package.\cite{RN278} 
However, the Wannier orbitals obtained via the 3D transformation are not adequate for interfaces between a metal and an insulator because of the delocalized wavefunction in the plane of the interface, and the procedure will not converge. The hybrid Wannier function technique was proposed to overcome this obstacle.\cite{RN3053} 
Along the $z$-axis, we calculate $ M_{mn}(\vk) = \mathrel{\bigl\langle} u_{m,\vk} \mathrel{\big|}
 u_{n,\vk+\vec{b}_\parallel} \mathrel{\bigr\rangle} $, where $u_{m,\vk}$ is the Bloch wave without the propagation exponential, $m,n$ are band indices, $\vk$ the crystal momentum, and  $\vec{b}_\parallel$ is the $\vk$-spacing in the $z$-direction. 
 A global matrix $ \Lambda(\vk_\perp)
= \prod_{j=0}^{N_\parallel-1} \tilde{M}(\vk_\perp + j \vec{b}_\parallel) $ is constructed whose eigenvalues $\lambda_m$ are related to the center of the Wannier function in the $z$-direction by 
$z_m = - (L/2 \pi) \, \mathop{\mathrm{Im}}(\ln \lambda_m) $, and $k_\perp$ is the crystal momentum $\vk$ in the $x$-$y$ plane. \cite{RN2887} The polarization $P$ in the $z$-direction for each $\vk_\perp$ is finally written as,
\begin{equation}
P(\vk_\perp) = \frac{1}{\cal V} \, 
\Bigl( -2e \sum_m \Delta z_{m, \vk_\perp}
 + \sum_\alpha Q_\alpha \Delta z_\alpha \Bigr) ,
\end{equation}
where the first term is the electron contribution to $P$ and the second term is the from ionic displacements, where $\Delta z_{m, \vk_\perp}$ is the difference between the ionic coordinate and the center of the Wannier orbital $m$, $Q_\alpha$ is the charge of ion $\alpha$, and $\Delta z_\alpha$ is the ionic displacement in the $z$-direction.

When DFT+$U$ is used in the calculations, the value of $U$ is often taken from the literature or sometimes used as adjustable parameters. If one would like to get a first principles estimate of $U$ for real materials, the constrained random phase approximation (cRPA) method can provide a fully quantum mechanical parameterization of $U$ based on the DFT ground state. The basic idea of cRPA \cite{PhysRevB_frequ_depend_U,U_from_cRPA} is to calculate a \textit{partial} RPA particle-hole polarization with the constrain of a physically motivated correlation window 
(\textit{e.g.} the $d$-like bands of transition metal atoms). One aims to estimate the screened Coulomb interaction for the correlation window. For this purpose, the particle-hole polarization between all possible pairs of occupied state and unoccupied state are taken into account. Within RPA, the \textit{full} particle-hole polarization can be written as: \cite{PhysRevLett.76.1212} 
\begin{equation}
P^{\RPA}(r,r^{\prime};\omega) = \sum_{i}^{\textrm{occ}} \sum_{j}^{\textrm{unocc}} \psi_{i}^{*}(r) \cdot \psi_{j}(r^\prime) \cdot \psi_{j}^{*}(r) \cdot \psi_{i}(r^\prime) \times \Bigl( \frac{1}{ \omega - \varepsilon_{j} + \varepsilon_{i} + i\delta} + \frac{1}{ \omega + \varepsilon_{j} - \varepsilon_{i} - i\delta} \Bigr) 
\end{equation}
where $\psi_{i}$ and $\varepsilon_{i}$ are the single particle eigenfunctions and eigenenergies of DFT. Summations on $i$ and $j$ are restricted such that $i$ must be an occupied state and $j$ must be an unoccupied state. 

The selected bands in the correlation window often have a strong orbital character, \textit{e.g.} $d$-like in our case. Following the convention in the literature, these bands are called $d$-space. If both the occupied and unoccupied states are within the $d$-space, then the polarization contributes to $P^{\RPA}_{d}(r,r^{\prime};\omega)$. All other pairs of occupied and unoccupied states contribute to $P^{\RPA}_{r}$, where $r$ stands for the rest of the polarization. Thus, the \textit{full} polarization is divided into two parts: $P^{\RPA} = P^{\RPA}_{d} + P^{\RPA}_{r}$. Here the \textit{partial} polarization $P^{\RPA}_{r}$ is the quantity related to the partially screened Coulomb interaction \cite{PhysRevB_frequ_depend_U}: 
\begin{equation}
W_{r}(\omega)=[1 - v \cdot P^{\RPA}_{r}(\omega)]^{-1} \cdot v 
\end{equation} 
In this expression, $v$ is the bare Coulomb interaction. According to the Hedin equations and the GW approximation, the full polarization, $P^{\RPA}$, screens the bare Coulomb interaction, $v$, to give the fully screened interaction $W$. 
With the same logic, $P^{\RPA}_{d}$ screens $W_{r}$ to give the fully screened interaction $W$. Thus, $W_{r}$ is identified as the screened on-site Coulomb interaction for the $d$-space, \textit{i.e.} $U(\omega) \equiv W_{r}(\omega)$, that includes the screening effect from the realistic environment of the material. In practise one calculates the \textit{partial} polarization $P^{\RPA}_{r}$ from the Kohn-Sham susceptibility, which is completely based on the DFT ground state, then derives  $U(\omega)$ from $P^{\RPA}_{r}$.

\subsection{Vertical Geometry}

\subsubsection{Graphene\texorpdfstring{$\,|\,h$}{Lg}-BN\texorpdfstring{$\,|\,$}{Lg}graphene: interface and transmission}

\begin{figure}[htb!]
\centering
\includegraphics[width=0.70\textwidth]{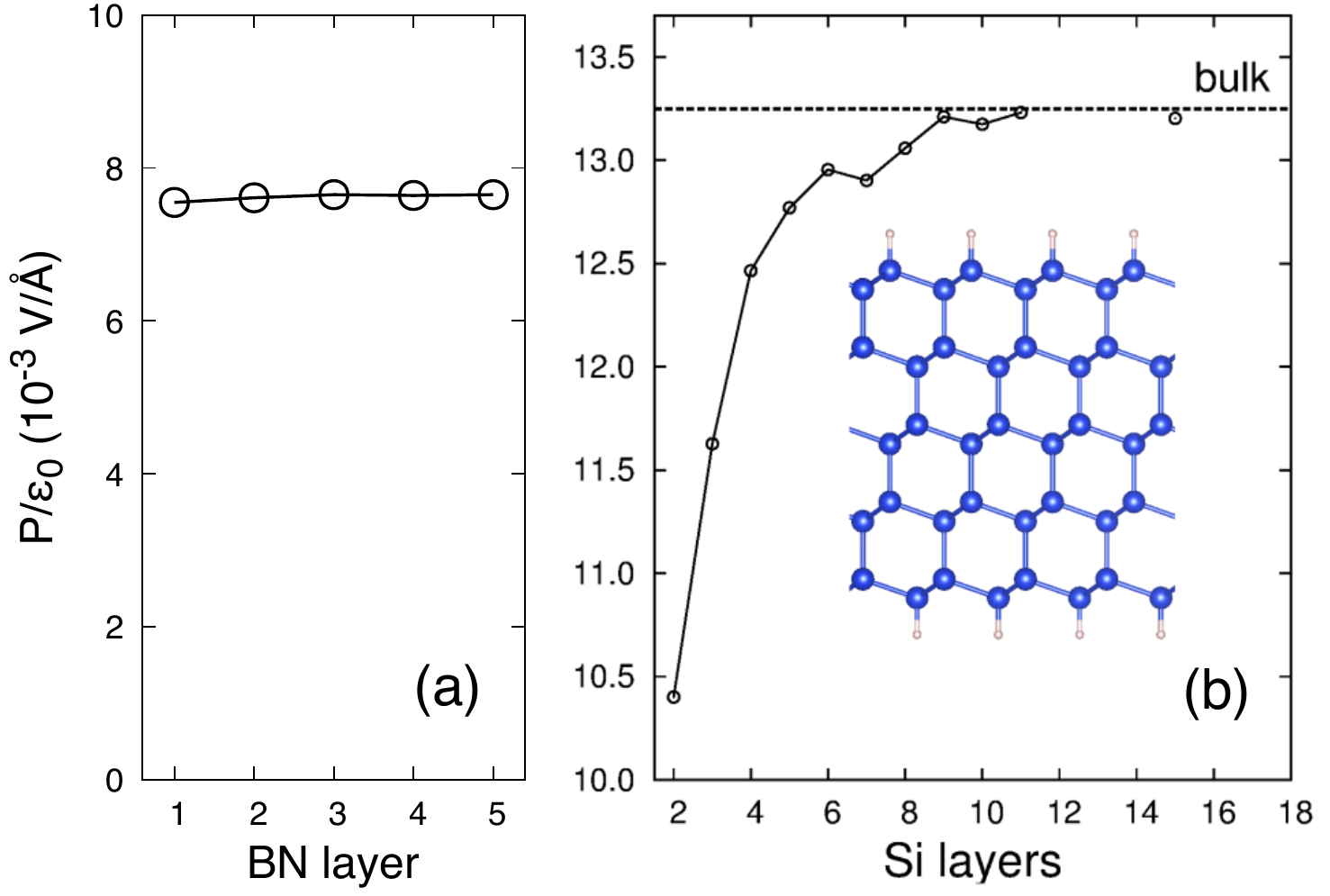}
\caption{
(a) Layer-by-layer polarization of five-layer $h$-BN between two graphene sheets and (b) layer-by-layer polarization of a H-terminated Si slab. 
Reprinted figure with permission from Y.-P. Wang and H.-P. Cheng, Physical Review B, 91, 245307 (2015), \href{https://journals.aps.org/prb/abstract/10.1103/PhysRevB.91.245307}{https://doi.org/10.1103/PhysRevB.91.245307}, Copyright (2015) by the American Physical Society. 
\label{FIG:Polar}} 
\end{figure}

\begin{figure}[htb!]
\centering
\includegraphics[width=0.85\textwidth]{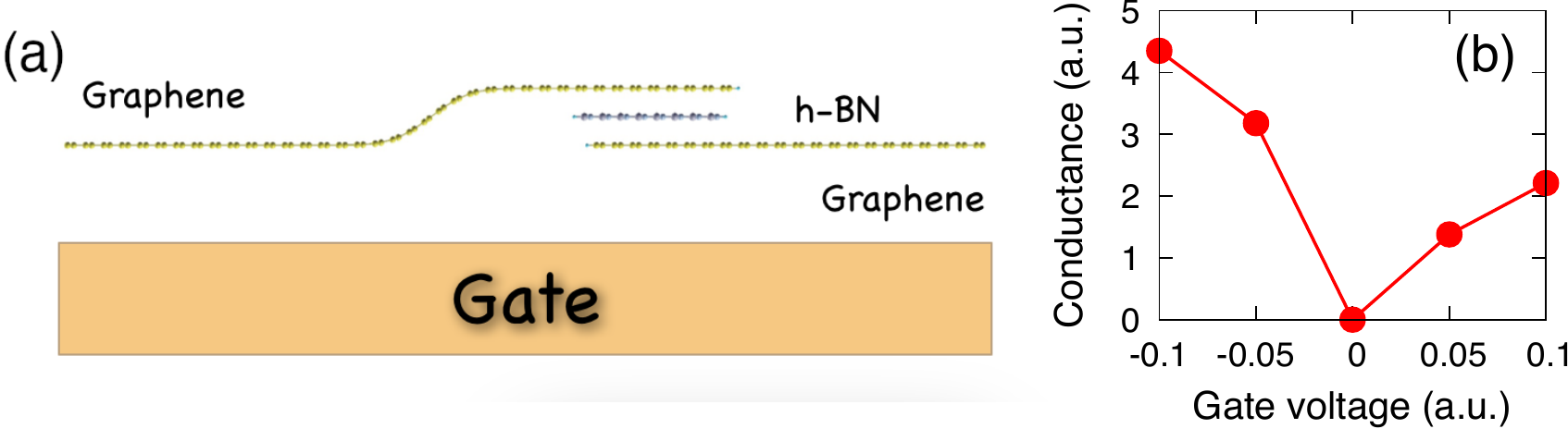}
\caption{
Monolayer $h$-BN between two graphene sheets: (a) model, and (b) Conductance as a function of applied gate voltage.
\label{FIG:G1}} 
\end{figure}

The graphene$\,|\,h$-BN$\,|\,$graphene heterostructure \cite{RN2199} was the first application of our approach, based on the single-gate system 
(as shown in Figure~\ref{FIG:device}, top panel) 
that was studied experimentally. \cite{RN1622} 
In this work, the gate effect at the $h$-BN and graphene interface was fully analyzed.
In order to see whether the immediate contact between graphene and $h$-BN makes a difference, layer-by-layer electric polarization analysis was performed for $h$-BN. 
The electric polarization for each $h$-BN layer is calculated by $P = - \sum_i e \delta z_i / Sd$. 
Here, $e$ is the unit charge, $\delta z_i$ is the change in the hybrid Wannier charge center upon applying a gate voltage in the direction of the gate field, $S$ is the area of the unit cell, and $d$ is the thickness of a $h$-BN layer which is set to $3.33\, \textrm{\AA}$. 
The summation is over all hybrid Wannier functions belonging to this $h$-BN layer within a unit cell. 
The layer-by-layer electric polarization analysis shows that the first $h$-BN layer in direct contact with the graphene sheet has a polarization similar to those further away from the interface (see Figure~\ref{FIG:Polar}), and this curve is also similar to that for the bare five-layer $h$-BN system (not shown). Its inert nature makes $h$-BN a perfect choice of a supporting material for graphene, let alone that the lattice mismatch is very small. It is clear that compared to $h$-BN, H-terminated Si shows much stronger interface effects (for simplicity we do not insert the Si slab between the graphene sheets). We also attempted to compute the transmission function as a function of gate voltage. Due to computational limitations at that time, only monolayer $h$-BN was considered to illustrate the point (see Figure \ref{FIG:G1}). However, the model we built is good for other investigations.

As expected from experimental measurements, first-principles calculations also show a Fermi Energy shift and gap opening. When there is hole doping, the graphene layer closer to the electrode is doped more than the one further away from the electrode, which is not surprising.

\subsubsection{Trilayer graphene}

\begin{figure}[htb!]
\centering
\includegraphics[width=0.8\textwidth]{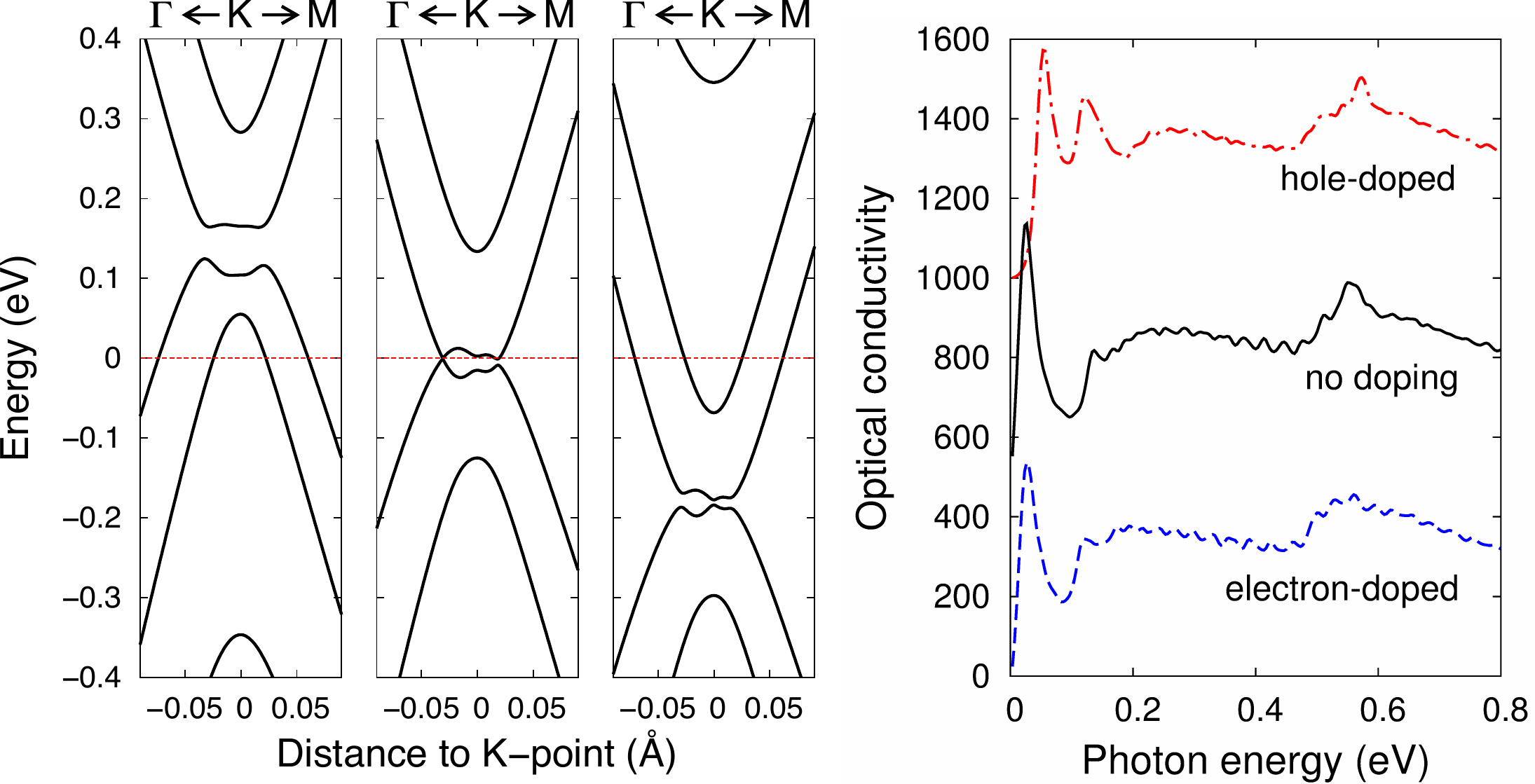}
\caption{
Trilayer graphene of ABA stacking order. Left: band structure as a function of the charge doping level. The electric field across the trilayer graphene is set to be $\epsilon_0 E = 0.015 \, \mathrm{C/m^2} $,and the net charge densities are respectively, from left to right, $ 9.5 $, $ 0$, and $ -9.5 $ $( \times 10^{12} \,\mathrm{cm^{-2}})$; right: The calculated real part of the optical conductivity for the same three net charge densities. 
As expected from experimental measurements, first-principles calculations also show Fermi Energy shift and gap opening. When there is hole doping, the graphene layer closer to the electrode is doped more than the one further away from the electrode, which is not surprising.
\label{FIG:Trilayer}} 
\end{figure}

Trilayer graphene was studied experimentally in the dual-gate configuration that inspired us to complete our implementation.\cite{RN2201} We examined field effects on both ABA stacking and ABC stacking orders, and our calculations reproduce the experimentally observed \cite{RN312,RN2610} gap opening in ABC stacking and band overlap in ABA stacking. The dual-gate configuration allows one to investigate effects of doping, and our calculations predict possible gap reopening upon doping in the ABA stacking as shown in Figure \ref{FIG:Trilayer}. We suggest that infrared optical conductivity measurements can confirm the calculated band gap reopening.

\subsubsection{Graphene \textbar azobenzene \textbar graphene: interface and multi-control}

\begin{figure}[htb!]
\centering
\includegraphics[width=0.9\textwidth]{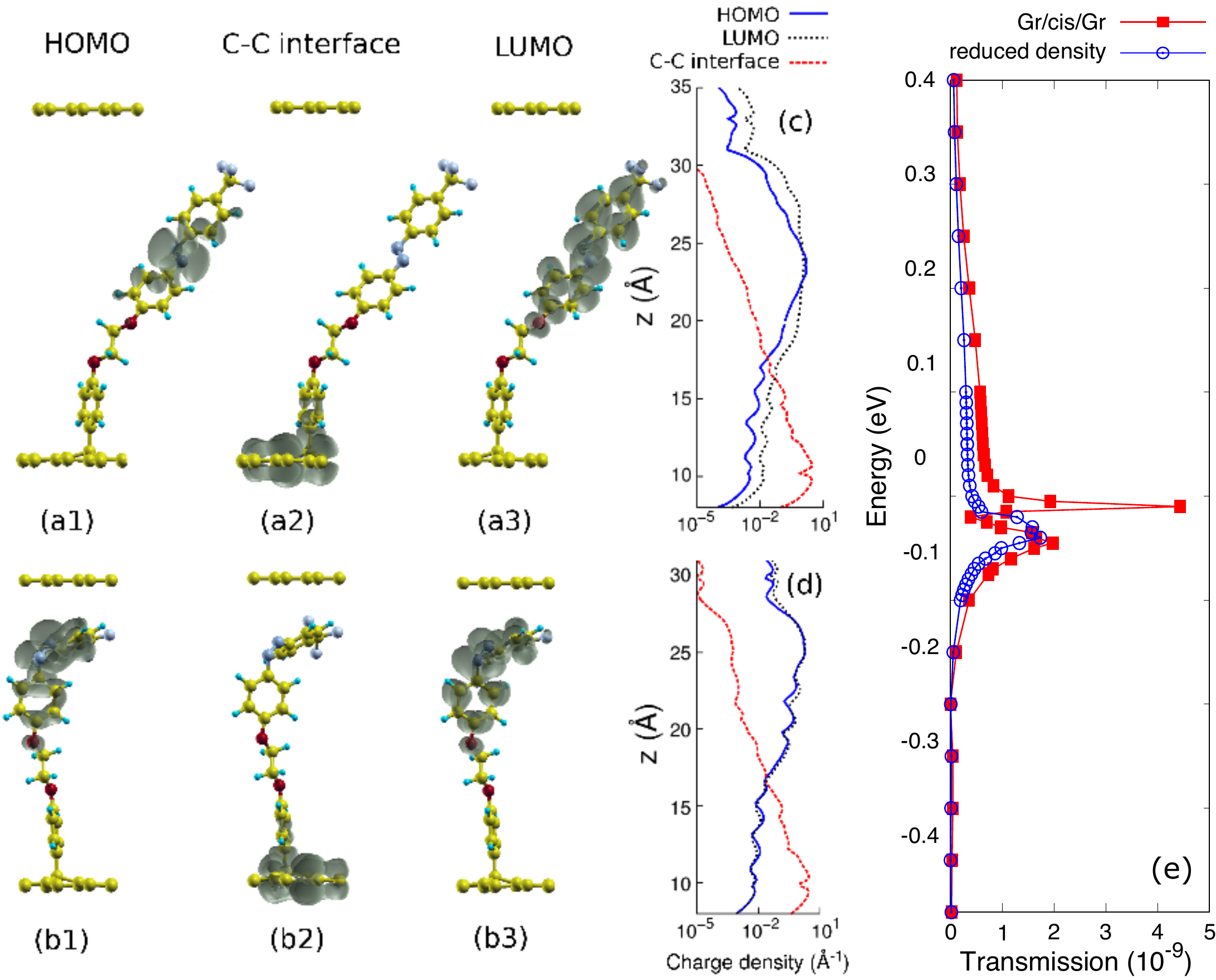}
\caption{
Charge density corresponding to the (a1,b1) HOMO, 
(a2,b2) C-C interface state, and (a3,b3) LUMO states 
of (a) {\it trans} and (b) {\it cis} junctions.
The charge density integrated over the $x$-$y$ plane (parallel to the graphene sheet) 
is shown along the $z$-direction (normal to graphene sheet)
for (blue solid lines) HOMO, (black dotted lines) LUMO and (red dashed lines) C-C interface states of (c) {\it trans} and (d) {\it cis} junctions.
(e) Transmission of the Gr$\,|\,\textit{cis}\,|\,$Gr junction 
and a junction with the density of \textit{cis} molecules reduced by half.
The gate voltage is $\epsilon_0 E_g = 12.9 \times 10^{-3} \,\textrm {C/m}^2$.
Reprinted with permission from Wang \etal{}, ACS Omega 2, 5824--5830 (2017), \href{https://pubs.acs.org/doi/full/10.1021/acsomega.7b00856}{https://doi.org/10.1021/acsomega.7b00856}, Copyright 2017 ACS Publications.
\label{FIG:AzoTransmission}} 
\end{figure}

This work was motivated by experimental studies of heterogeneous  junctions that consist of two graphene sheets bracketing a monolayer of azobeneze molecules. \cite{RN3056} The azobezenze molecule has two stable configurations, \textit{trans} and \textit{cis}, that can transform from one to another by optical excitation. \cite{RN2364, RN1616} We showed that these two forms of the molecule have different transport properties at zero bias and different $I$-$V$ characteristics in the one-dimensional configuration. \cite{RN484, RN1807}  Investigations of an azobenzene monolayer between two semi-infinite Au bulk leads indicate that the transport properties of the \textit{trans} and \textit{cis} molecules are more complicated. \cite{RN2606} Our analysis shows that chemisorption of molecules to the top Au lead to a different  \textit{trans} vs. \textit{cis} relation than a physisorption of molecules to the top Au lead. In addition, an ad-atom on an Au surface can change dramatically the $I$-$V$ characteristics, which explains the experimentally observed 
$I$-$V$ curve of an Au-azobenze-Au break junction. \cite{RN3057}  We further applied azobenzene molecules to modulate the interaction between the two nano-particles and studied the interaction with graphene surfaces in the two configurations. \cite{RN2194, RN1966} These studies demonstrated that one can use the configuration change to manipulate physical properties of a system. Simulations of the graphene$\,|\,$azobenzene$\,|\,$graphene vertical junction is therefore a natural extension of our long-standing interest in the added technique for gating the system. \cite{RN2888} We found a rich array of interesting phenomena.  The first noticeable finding is that, depending on the sign of the gate field,  the \textit{trans} and the \textit{cis} become more conducting only in one direction; and second, at some gate voltages, two peaks appear near the Fermi energy. Our analysis shows that gate voltage alters the energy levels in such a way that the interface state (the C-C bond between a molecule and graphene) moves closer to and the Dirac point of the top layer of graphene moves away from the graphene Fermi energy. Even more amazing, one interface state can interfere with another one, resulting in a second, stronger peak, which disappears when we reduce the coverage to 50\%. Figure~\ref{FIG:AzoTransmission} shows the interface state for the \textit{trans} and the \textit{cis} molecules in the vertical direction. The transmission functions in the panel (e) clearly show that at full coverage a very strong second peak appears as a result of interference of neighboring C-C interface states.

\subsubsection{Graphene \textbar TMD \textbar graphene junctions}

\begin{figure}[htb!]
\centering
\includegraphics[width=0.9\textwidth]{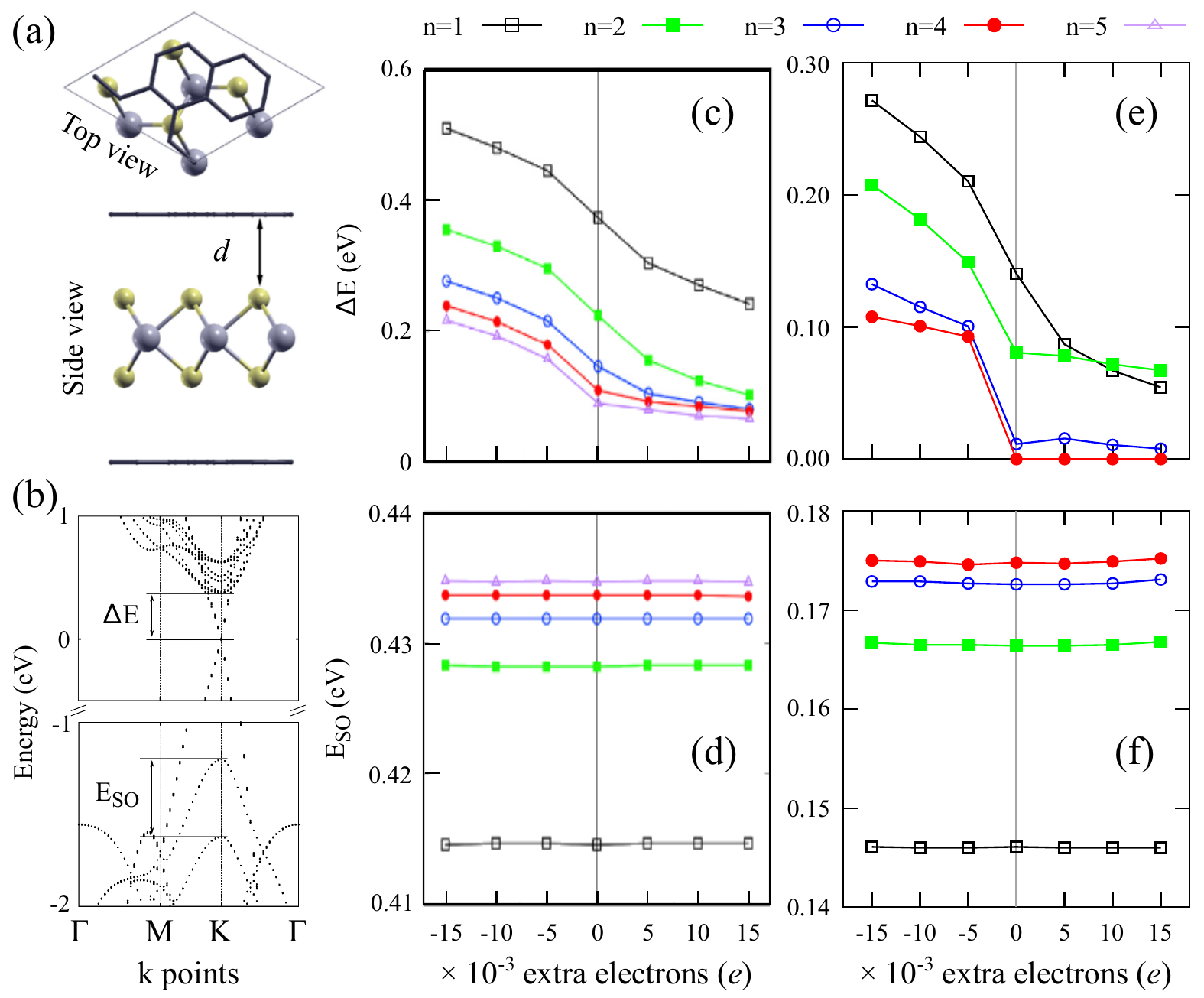}
\caption{\label{fig:CrTMD}
(a) Top view and side view of the atomic structure of the graphene$\,|\,$monolayer TMD$\,|\,$graphene hybrid system. 
$d = 3.50$ and $3.34 \, \textrm{\ring{A}} $ for \ce{WS_2} and \ce{MoS_2} respectively. 
(b) Band structure of graphene$\,|\,$monolayer \ce{WS_2}$\,|\,$graphene. 
$\Delta E$ is the distance between the conduction band of \ce{WS_2} and the Fermi level.  
$E_{\textrm{SO}}$ is the energy splitting of the valence band at K point due to the spin-orbit coupling. 
(c) $\Delta E$ and (d) $E_{\textrm{SO}}$ for graphene$\,|\, n$-layer \ce{WS_2}$\,|\,$graphene as a function of the number of doped electrons. 
(e) $\Delta E$ and (f) $E_{\textrm{SO}}$ for graphene$\,|\,n$-layer \ce{MoS_2}$\,|\,$graphene as a function of the number of doped electrons.
Reprinted from Journal of Physics and Chemistry of Solids, 128, Li \etal{}, Tunneling field-effect junctions with \ce{WS2} barrier, 343--350, \href{https://www.sciencedirect.com/science/article/pii/S0022369717315950}{https://doi.org/10.1016/j.jpcs.2017.12.005}, Copyright (2019), with permission from Elsevier. 
} 
\end{figure} 

Transition metal dichalcogenides (TMD) make up one group of 2D semiconductors that have  attracted much attention \cite{RN3063, RN3064, RN3065, RN3066} in the quest of TFETs. Compared to $h$-BN, the relatively smaller energy gap and strong spin-orbital coupling make them more interesting than $h$-BN, which has otherwise been a perfect choice for graphene supporting material. We studied field effects of graphene$\,|\,n$-layer TMD$\,|\,$graphene ($n$=1--5, TM=\ce{W}, \ce{Mo}) junctions. Figure~\ref{fig:CrTMD} compares WS$_2$ and MoS$_2$. Two critical quantities, the distance $\Delta E$ 
between the conduction band edge and the Dirac point (the Fermi level) and band splitting $E_\textrm{SO}$ due to the spin-orbital coupling, were computed as function of charge doping. The decrease of $\Delta E$ is responsible for the large on/off ratio observed reported by the experimental groups.~\cite{RN3016,RN2452}
%%%
For junctions with thinner \ce{WS_2} ($n = 1, 2$) or \ce{MoS_2} ($n=1$), $\Delta E$ is symmetric between hole doping and electron doping due to the symmetry in the Dirac cone to which electrons/holes are added. 
For junctions with thicker \ce{WS_2} or \ce{MoS_2}, holes are still added to the graphene only but electrons are added to both the graphene and the TMD, which causes the asymmetry in $\Delta E$. 
The gate field has little effect on $E_\textrm{so}$.
This is because both the two valence bands at K are mostly $d$ states from transition metal atoms~\cite{RN3067}. 
As such, the shift in the two bands are the same when a gate field is applied. 
In comparison, the junction with $n$-layer \ce{MoS_2} has a much smaller $\Delta E$ than the junction with $n$-layer \ce{WS_2}. 
Especially, $\Delta E$ is zero for the graphene$\,|\,4$-layer \ce{MoS_2}$\,|\,$graphene junction even under zero charge doping.

\subsubsection{Interface between Topological Insulators BSTS} 

\BSTS{} (BSTS) is a strong topological insulator with high bulk resistivity and robust surface states.  
It is promising for applications in high speed and low-energy-consumption spintronic devices. 
One of the device configurations is vertical tunneling junction, where two BSTS slabs are stacked together. 
A question arises for such a configuration that whether topological surface states survive at the interface between two BSTS slabs. 
First-principles calculations show that topological interface states are absent at the equilibrium inter-slab distance but can be preserved by inserting two or more layers of $h$-BN between the two BSTS slabs.~\cite{RN3033} 
Furthermore, experimental measurements revealed a weak dependence of the electron tunnelling current on the gate field at small bias voltages for a BSTS vertical tunneling junction.~\cite{RN3033} 
The dependence at high bias voltages is however stronger. 
Then, one may wonder how topological interface states (when present) respond to a gate electric field. 
In order to examine this problem, first-principles calculations were performed for the BSTS interface with bilayer $h$-BN.

\begin{figure}[htb!]
\centering
\includegraphics[width=0.85\textwidth]{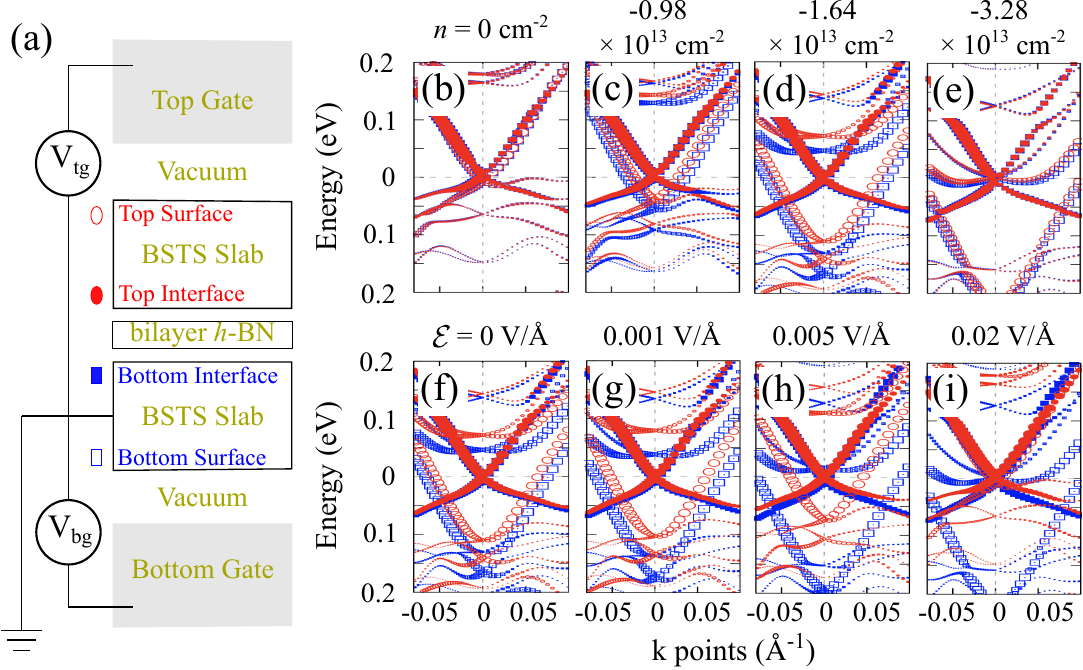}
\caption{\label{fig:BSTS} 
(a) Dual gate setup for a BSTS$\,|\,$bilayer $h$-BN$\,|\,$BSTS heterostructure. 
(b)--(e) The energy bands of the heterostructure under different charge doping concentration $n$. 
The top electrode and the bottom electrode potentials are the same. 
(f)--(i) The energy bands of the heterostructure under different average electric field $\mathcal{E}$.
The charge doping concentration is always $-1.64 \times 10^{13} \cm{}^{-2}$.
Reprinted with permission from Appl. Phys. Lett. 116, 031601 (2020); \href{https://aip.scitation.org/doi/10.1063/1.5127065}{https://doi.org/10.1063/1.5127065}. Copyright 2020 American Institute of Physics.
}
\end{figure}

Figure~\ref{fig:BSTS}a illustrates the BSTS$\,|\,$bilayer $h$-BN$\,|\,$BSTS interface under the influence of two gates. 
A dual gate setup permits independent control of the charge doping concentration $n$ and the average electric field between the two gate electrodes $\mathcal{E}$. 
Figures~\ref{fig:BSTS}b--\ref{fig:BSTS}e show the band structure of the interface under increasing electron doping levels but zero average electric field.
There are four species of Dirac states in the system, the top surface species, the bottom surface species, the top interface species, and the bottom interface species, which are represented by red empty circles, blue empty squares, red filled circles, and blue filled squares respectively. 
The larger a circle (a square) is, the more localized at the corresponding surface or interface the state is. 
At zero doping and zero average electric field, all four Dirac points are about the Fermi level, which is set to zero. 
When electrons are added to system, the top and the bottom surface bands move downward while the top and the bottom interface bands barely move, which remains true up to a doping concentration of 
$-3.28 \times 10^{13} \cm{}^{-2}$.  
This means that the added electrons mainly go to the top and the bottom surfaces, which helps to understand the weak dependence of the tunnelling current on the gate field at small bias voltages.
As the doping concentration increases, the conduction bands of bulk BSTS, which are above the Dirac cones in Figure~\ref{fig:BSTS}b, get closer to the Fermi level. 
Meanwhile, they extend more into the surface regions in real space. 
Eventually, the otherwise empty conduction bands become partially occupied at a doping concentration of about $-3.28 \times 10^{13} \cm{}^{-2}$. 
%%%
At a constant charge doping level of $-1.64 \times 10^{13} \cm{}^{-2}$, we applied different average electric fields and obtained the band structures shown in the Figures~\ref{fig:BSTS}f--\ref{fig:BSTS}i. 
The major effect of such an electric field is to separate the top surface bands from the bottom surface bands. 
Specifically, the top (bottom) surface bands are moved upward (downward), meaning that some electrons are transferred from the top  surface to the bottom surface. 
This is consistent with the intuition that a positive electric field along the $+z$ direction apples a force in the $-z$ direction on electrons. 
Again, the interface surface bands are not much affected. 
The conduction bands of bulk BSTS are also shifted downward by a positive electric field, and they cross the Fermi level at $\mathcal{E} = 0.02 \VperAng $ for $n = -1.64 \times 10^{13} \cm{}^{-2}$. 
Such a doping of the bulk states under a finite average electric field may correlate with the stronger dependence of the tunnelling current on the gate field at high bias voltages.

\subsection{Planar Geometry}
Planar geometry can be viewed as a special case of vertical geometry of minimal thickness, which simplifies the simulation model and reduces computational cost. Here we highlight a few systems that have been looked at in the last few years.

\subsubsection{Metal phthalocyanine 2D network and junctions}

\begin{figure}[htb!]
\centering
\includegraphics[width=1.0\textwidth]{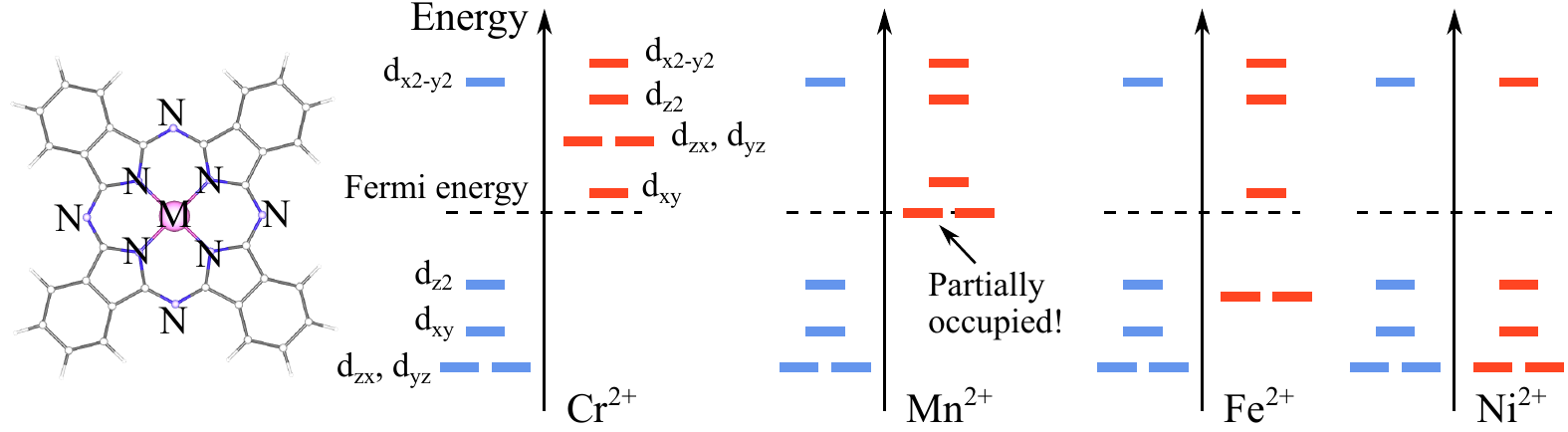}
\caption{Sketch of a MPc molecule (left) and our calculated Kohn-Sham orbitals near the HOMO-LUMO gap (right).  
\label{FIG:MetalPc}} 
\end{figure}

\begin{figure}[htb!]
\centering
\includegraphics[width=0.9\textwidth]{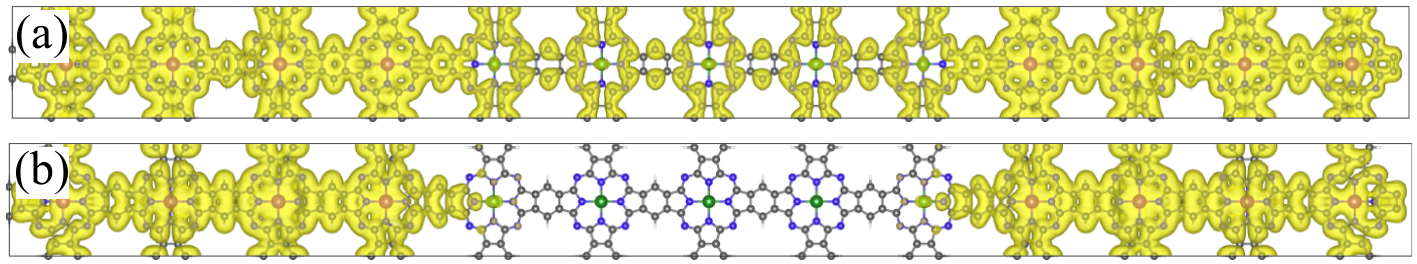}
\caption{The $k$-resolved charge density of a MnPc$\,|\,$NiPc$\,|\,$MnPc junction. Panel (a) indicates a conducting channel and (b) a non-conducting channel. 
Reprinted figure with permission from Liu \etal{}, Physical Review B, 97, 035409 (2018), \href{https://journals.aps.org/prb/abstract/10.1103/PhysRevB.97.035409}{https://doi.org/10.1103/PhysRevB.97.035409}, Copyright (2018) by the American Physical Society.
%
% \cite{RN3062}
\label{FIG:MetalPcTrans}} 
\end{figure}

Metal phthalocyanine (MPc) are planar molecules of nanometer size. 
Abel \textit{et al.} \cite{RN2124} attempted to synthesize a covalently bonded 2D framework using MPc molecules. Although it was later proven to be a 2D hydrogen-bonded network, covalently bonded 2D $10 \nm $ MnPc networks were  realized on Ag surfaces, \cite{RN2117} and 1D micrometer FePc wires have also been synthesized. \cite{RN3059} We reported field-effect studies of magnetic order in 2D MPc networks \cite{RN3060} and spin-dependent charge transport of MPc junctions. \cite{RN3062}
Figure~\ref{FIG:MetalPc} sketches a general metal phthalocyanine molecule, in which the transition metal ion transfer is at the center. We examined Cr, Mn, and Fe systems and find each of them loses two electrons to the organic framework. Interestingly, for \ce{Cr^{2+}}- and \ce{Fe^{2+}}-doped Pc, the $d$-orbital split is large enough that the 2D framework are semiconductors with a band gap, while 2D MnPc is a half metal because the MnPc has $d$ orbitals that are partially occupied (see the middle panel of Kohn-Sham levels in Figure~\ref{FIG:MetalPc}). More importantly, we found that the magnetic order of the FePc, MnPc and MnTPP [TPP: 5,10,15,20-tetra(phenyl)porphyrin] can be tuned by charge doping (via gating), and the interactions between two spins are mediated by itinerant electrons. Based on these findings, we designed planar MnPc$\,|\,$NiPc$\,|\,$MnPc 2D junctions and investigated 1) gate effects and 2) scattering region length dependence. It is found that this system can be used as a perfect spin filter when it is hole-doped. \cite{RN3062}. Further analysis shows that the role of gating is to align the energy level of the lead (MnPc) with the scattering region (NiPc), such that a conducting channel can appear (see Figure~\ref{FIG:MetalPcTrans}) 

\subsubsection{Graphene double-barrier junction and 1D interfaces}

The idea of patterning graphene or a graphitic material into functioning circuitry has been a scientific and engineering focus since the time of discovery of carbon nanotubes \cite{RN1232} and single-layer or few-layer graphene. \cite{RN1336}

\begin{figure}[htb!]
\centering
\includegraphics[width=0.75\textwidth]{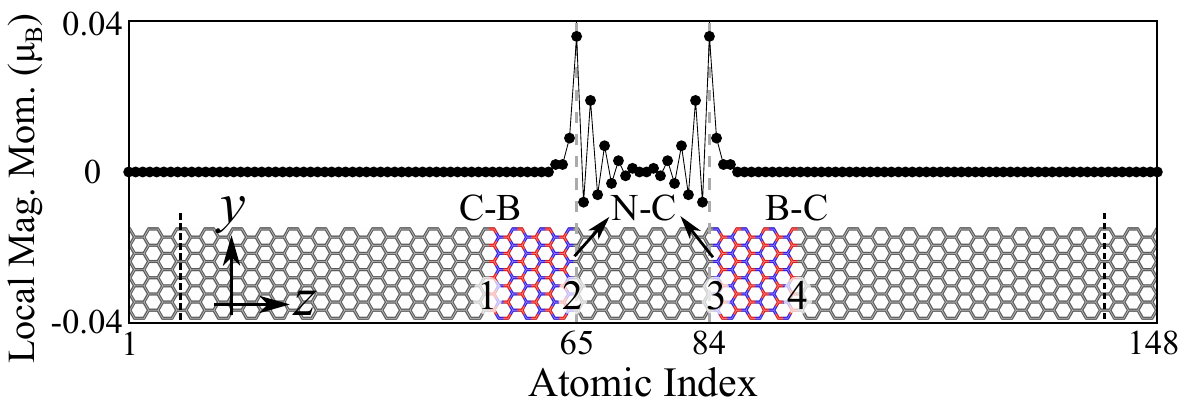}
\caption{$h$-BN double-barrier in patterned graphene. The local magnetic moment is plotted as a function of $z$. The dashed vertical lines mark the scattering region. 
Reprinted from Carbon, 144, Liu \etal{}, Spin dependent resonant electron tunneling through planar graphene barriers, 362--369, \href{https://www.sciencedirect.com/science/article/abs/pii/S0008622318311667}{https://doi.org/10.1016/j.carbon.2018.12.035}, Copyright (2019), with permission from Elsevier. 
\label{FIG:GraphBN}} 
\end{figure}

\begin{figure}[htb!]
\centering
\includegraphics[width=0.8\textwidth]{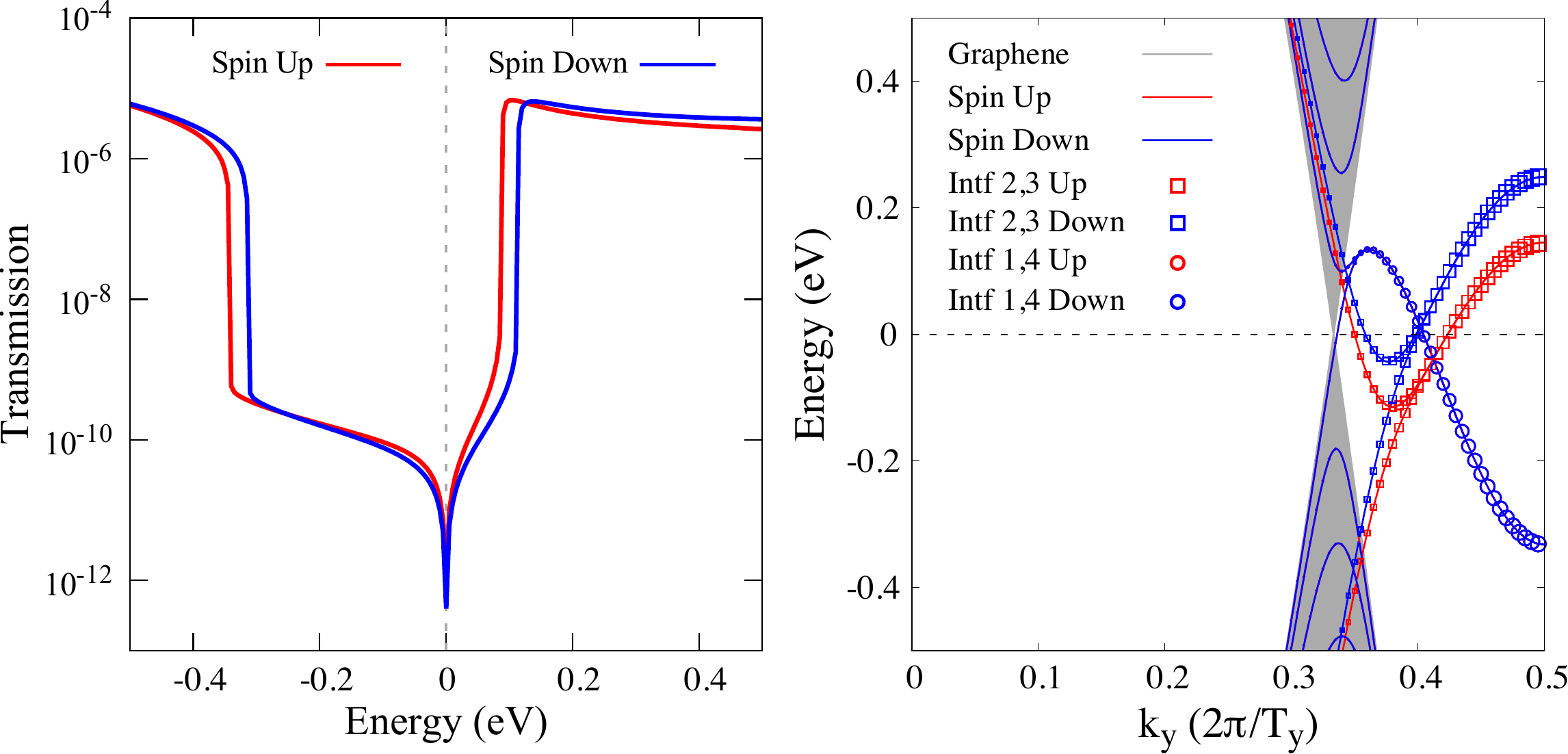}
\caption{The left panel shows the $h$-BN double-barrier transmission function as a function of energy and the right panel the band structure along the $y$-direction.
The size of the red and blue squares represents the degree of localization to the graphene and $h$-BN interface.
Reprinted from Carbon, 144, Liu \etal{}, Spin dependent resonant electron tunneling through planar graphene barriers, 362--369, \href{https://www.sciencedirect.com/science/article/abs/pii/S0008622318311667}{https://doi.org/10.1016/j.carbon.2018.12.035}, Copyright (2019), with permission from Elsevier.  
\label{FIG:GraphBNTrans}} 
\end{figure}

We carried out first-principles investigations of a double barrier in graphene framework. Figure~\ref{FIG:GraphBN} depicts the model for a double barrier that consists of $h$-BN. The zigzag edge was chosen and the interface composition was discussed: on one side we chose the C-B bond and on the other the N-C bond. Vacuum provides another system, in which the zigzag graphene edge was terminated by H atoms. With DFT calculations, details such as inter-edge spin couplings can be considered.
The transmission coefficient function (see left panel of Figure~\ref{FIG:GraphBNTrans}) shows a shift in rising/falling transmission coefficient between the two spin channels, 
indicating spin filtering can be achieved by a bias voltage. 
Band analysis (see right panel of Figure~\ref{FIG:GraphBNTrans}) indicates that this originates from the difference between the bands (of the middle graphene ribbon) of the two spins. Only when the graphene ribbon bands intersect with the Dirac cone can the transmission coefficient rise to a non-negligible value. We call this a resonance, which is even more pronounced if the double barrier is made of vacuum. It is interesting to see that the center graphene ribbon is conducting, with sizable contribution from interface state as the bands cross the Fermi level. Such interface enhancement or induced conducting behavior was observed and characterized in another study from our group where we found that a quasi-1D conducting wire can form at the interface of two semiconductors. \cite{RN2193}

\section{Current Work: Bilayer \texorpdfstring{CrI\textsubscript{3}}{CrI3}} 

\begin{figure}[htb!]
\centering
\includegraphics[width=0.8\textwidth]{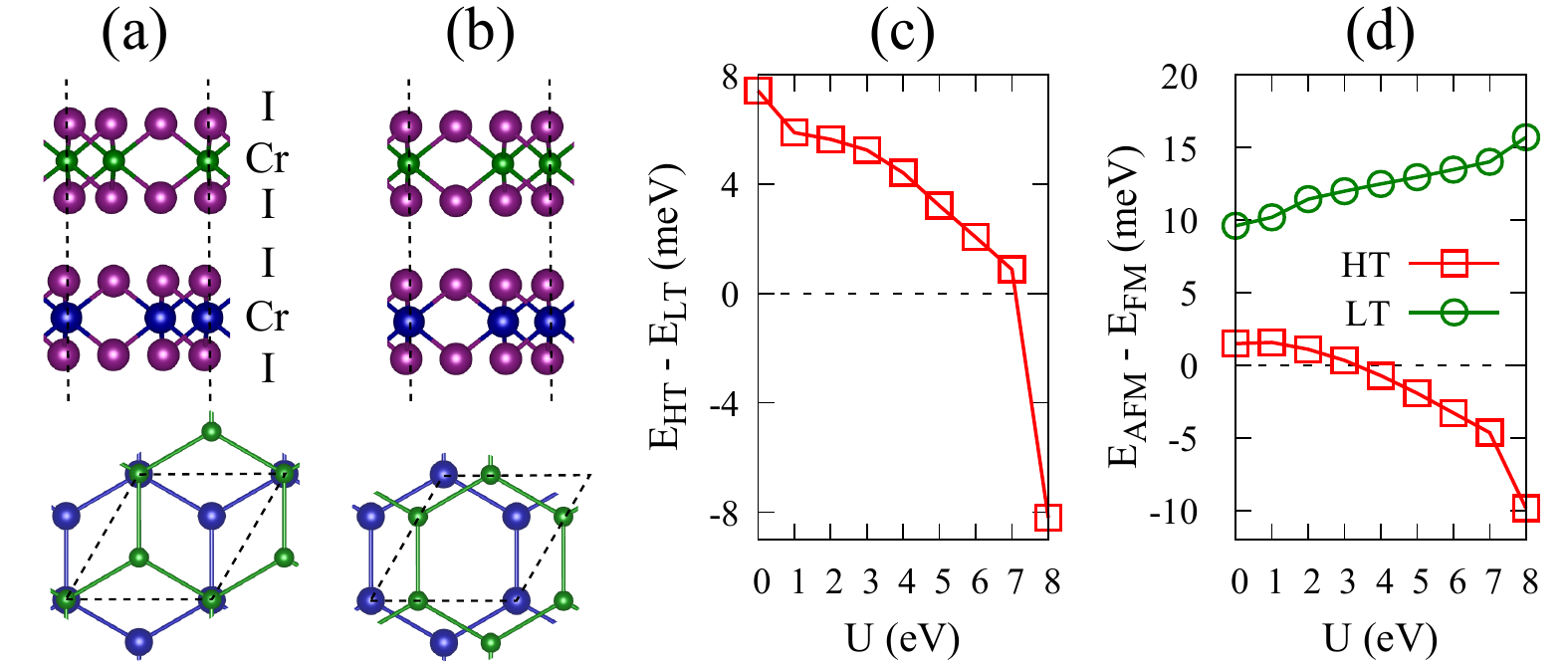}
\caption{\label{fig:CrI3_U}
(a)/(b) Atomic configuration of bilayer \cri{} showing high/low temperature (HT/LT) stacking. 
The dashed lines mark the boundary of a unit cell. 
(c) Energy difference between the HT stacking \cri{} and the LT stacking \cri{} versus the Hubbard $U$ parameter in the PBE$+U$ method.
(d) Energy difference between the interlayer antiferromagnetic (AFM) configuration and the interlayer ferromagnetic configuration of both the HT and the LT stacking \cri{} versus the Hubbard $U$ parameter. 
} 
\end{figure}

Other than previously studied materials, many more two dimensional magnetic materials can be employed in spintronic devices, such as high capacity information storage.
Recent discoveries of 2D magnetic material include \ce{FePS_3}~\cite{RN250}, \ce{Cr_2Ge_2Te_6}~\cite{RN246}, \ce{MnSe_2}~\cite{RN248}, and \ce{CrI_3}~\cite{RN247}.
Bulk \cri{} is a layered van der Waals material with ferromagnetic order at low temperatures. 
It has a high temperature (HT) monoclinic phase with space group $C2/m$ and a low temperature (LT) rhombohedral phase with space group $R\bar{3}$.~\cite{RN255} 
The two phases differ in the interlayer stacking, as shown in Figures~\ref{fig:CrI3_U}a and \ref{fig:CrI3_U}b. 
Interestingly, bilayer \cri{} exhibits antiferromagnetic (AFM) interlayer coupling~\cite{RN247}, which can be tuned by a magnetic field~\cite{RN247,RN251,RN254}, gate electric field~\cite{RN251,RN254}, and pressure~\cite{RN252, ISI:000497968400013}. 
In efforts to explain the experimentally observed AFM magnetic order, Sivadas \etal{} reported stacking-dependent magnetism~\cite{RN257} and Jang \etal{} analyzed the interaction between localized $e_g$ and $t_{2g}$ orbitals~\cite{RN258}. 
Among these and other efforts, we will examine the role of the local Coulomb interaction in Subsection \ref{section:CrI3_U}. 
In some experimental setups to apply a gate field\cite{RN251,RN254,RN253}, bilayer \cri{} is in contact with graphene or hexagonal boron nitride ($h$-BN), which motivated us to model gate field effects on heterostructures such as graphene$\,|\,$bilayer \cri{}$\,|\,$graphene, BN$\,|\,$bilayer \cri{}$\,|\,$BN, and BN$\,|\,$bilayer \cri{}$\,|\,$graphene. 
These results will be presented in Subsection \ref{section:CrI3_Gate}, emphasizing the interface effects on the magnetic phase transition. 
For these heterostructures, we will denote bilayer \cri{} by \ce{2-CrI_3} for brevity. 
We will also discuss how pressure affects structural and magnetic properties of bilayer \cri{} in Subsection \ref{section:CrI3_Pressure}.

\subsection{Enhancement of local Coulomb interaction}

\label{section:CrI3_U}

We relaxed the atomic structure of bilayer \cri{} based on density functional theory~\cite{RN925, RN8} as implemented in the Vienna Ab initio Simulation Package (VASP)~\cite{RN2539}. 
The computational details can be found in the reference~\footnote{
We set an energy cutoff of $450 \eV$ for plane waves and adopted the Perdew-Burke-Ernzerhof exchange correlation energy functional \cite{RN8} together with PAW pseudopotentials. \cite{RN2495}
A $9\times 9 \times 1$ Monkhorst-Pack mesh for sampling the first Brillouin zone was applied. 
The van der Waals interaction was taken into account via the PBE-D3 method. 
An energy tolerance of $1\times 10^{-6} \eV$ and a force tolerance of $0.001 \, \textrm{eV}/\textrm{\AA}$ were used for self-consistent and ionic relaxations, respectively. 
A vacuum region separates periodic images of the 2D system along the out-of-plane direction by at least $12\,\textrm{\AA}$ to eliminate any interaction.
}. 
Then, we applied the DFT$+U$ method~\cite{RN262} (the details of estimating the relative $U$ values will be discussed shortly) in VASP to obtain the total energies of bilayer \cri{} in both the HT and LT stacking. 
Figure~\ref{fig:CrI3a}c shows the energy difference between the HT and LT stackings versus the Hubbard $U$ parameter. 
The energy difference changes sign as the strength of the local Coulomb interaction increases. 
If $U$ is small, the LT stacking is energetically preferred, which is the case for bulk \cri{}. 
If $U$ is larger than $7 \eV$, the HT stacking has lower energy than the LT stacking. 
With such a large value of the $U$ parameter, the HT stacked bilayer \cri{} energetically prefers the AFM configuration to the FM configuration, as shown in Figure~\ref{fig:CrI3a}d. 
In contrast, the LT stacked bilayer \cri{} always prefers the FM configuration, no matter how large or small the $U$ parameter is. 
Therefore, the HT stacking together with a pronounced local Coulomb interaction seem to be responsible for the experimentally observed interlayer AFM magnetism. 

It is generally accepted that magnetic moments are associated with localized electrons whose behavior is determined by the competition between the kinetic energy of electrons and the strength of the local Coulomb interaction. Dimensional confinement can dramatically affect the competition, yielding unusual properties. For example, bulk $\ce{SrVO}_{3}$ is known to be a strongly correlated metal. In standard DFT calculations, the isolated three $t_{2g}$ bands of vanadium determine the low-energy properties of this material. In DFT+DMFT calculations, the Coulomb interactions are explicitly taken into account and the ground state of bulk $ \ce{SrVO}_{3}$ is still found to be a metal (with smaller band width). When the system is under dimensional confinement, orbital re-occupation can happen along with an enhanced Coulomb interaction. In a charge density self-consistent (CSC) DFT+DMFT \cite{Amadon_2012,PhysRevB.90.235103,PhysRevB.94.155131} study, 2D mono-layer $ \ce{SrVO}_{3}$ was found to be insulating \cite{PhysRevB.94.155131}, with a band gap of about $2 \eV$. Actually, already at the DFT level, the in-plane $d_{xy}$ band is no longer degenerate with the other two $t_{2g}$ bands. After DMFT and charge density consistency, the $d_{xy}$ orbital is found to be half-filled, and the other two out-of-plane $t_{2g}$ bands are almost empty. The change from a bulk correlated metal state to a mono-layer Mott insulating state is accompanied with an increase of Coulomb interaction from about $ 4 \eV$ to $ 5.5 \eV$. 

In order to confirm that the local Coulomb interaction increases as the dimension of \cri{} reduces from 3D to 2D, we calculated the $U$-matrix within the Kanamori parameterization from first principles using the constrained random phase approximation (cRPA) method as implemented in the FP-LAPW DFT code, a modified version of ELK code \cite{ELK_code,Anton_IEEE_paper_2010}. The code has been benchmarked \cite{PhysRevB.100.035104,U_from_MLWF_cRPA_NiOCoOFeOMnO_2013} with other implementations using late transition metal monoxides, and we obtain  consistent results. In our calculation, both bulk and mono-layer \cri{} have isolated Cr $d$-like bands around the Fermi level. We choose the five $d$-like bands as our correlation window. The ground state includes 100 empty bands to make a reasonable estimation of the partial particle-hole polarization, $P^{\RPA}_{r}$. The resulting averaged intra-orbital $U$ increases from about $1.9 \eV$ for bulk \cri{} to about $2.8 \eV$ for the mono-layer \cri{}. The calculated $U$ parameters are not large enough to bear the AFM ground state of bilayer \cri{}. However one should keep in mind the cRPA calculation is based on paramagnetic ground states of the two structures, and the Pauli exclusion principle (which is a different mechanism to give rise to on-site repulsion between electrons, especially for magnetic systems) is not taken into account. It is still  a nontrivial job to incorporate the Pauli principle with the current cRPA method in one calculation scheme. Here our observations confirmed an enhancement of about $1 \eV$ in the Coulomb interaction when the structure of \cri{} reduces from 3D to 2D.

\subsection{Field induced magnetic phase transition}

\label{section:CrI3_Gate}

\begin{figure}[htb!]
\centering
\includegraphics[width=0.8\textwidth]{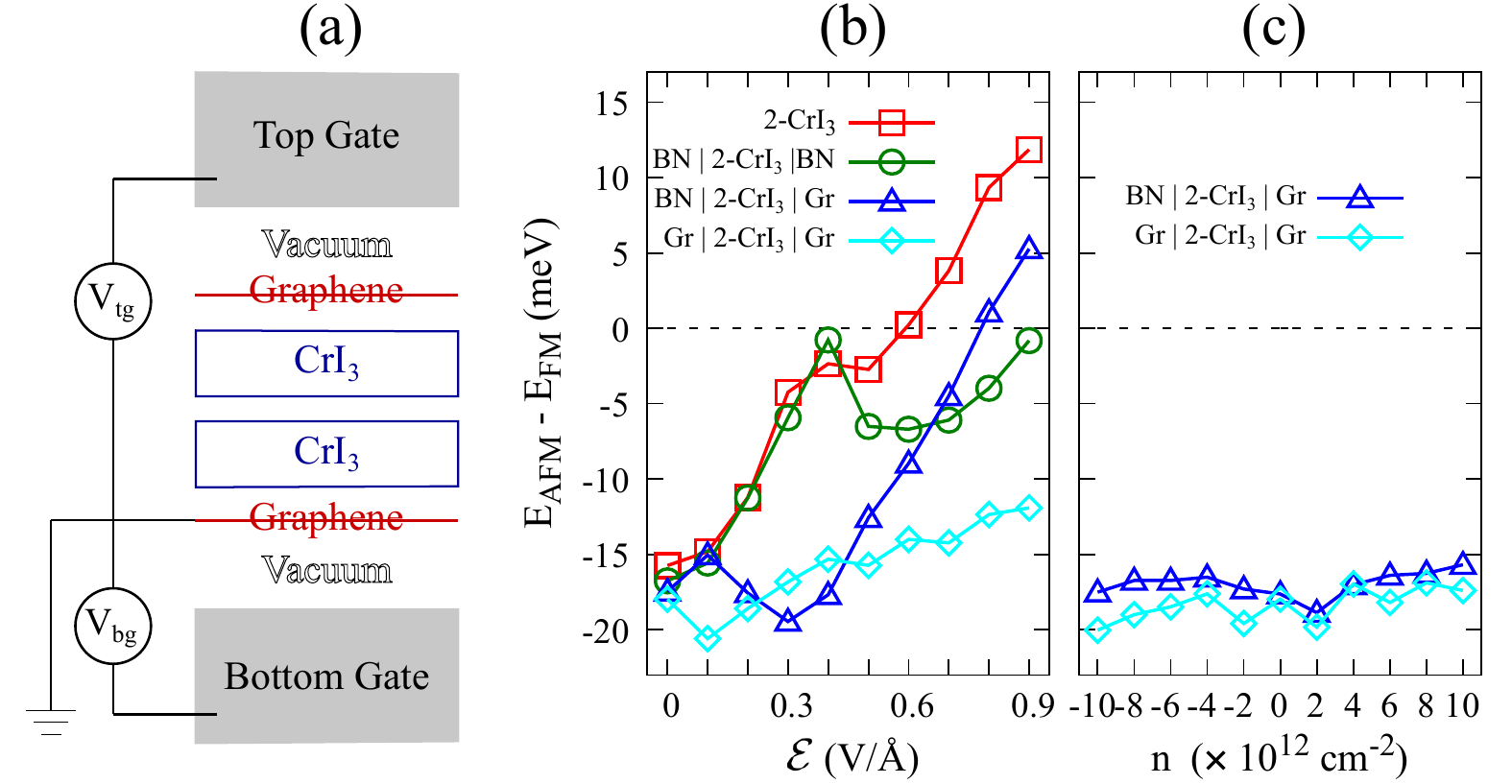}
\caption{\label{fig:CrI3a}
(a) Illustration of \grcrigr{} under a dual gate configuration in our simulation. 
(b) Energy difference between the AFM and the FM magnetic configurations of bilayer \cri{}, \bncribn{}, \bncrigr{}, and \grcrigr{} versus electric field. 
(c) Energy difference between the AFM and the FM magnetic configurations of \bncrigr{} and \grcrigr{} versus extra charge density $n$. 
} 
\end{figure}

Figure~\ref{fig:CrI3a}a illustrates a \grcrigr{} heterostructure subject to a dual gate setup, which permits a vertical electric field and charge doping. 
For such a dual gate setup, the average electric field between the gate electrodes will be what we call the electric field $\mathcal{E}$,  
\begin{equation}
    \mathcal{E}=(V_{\textrm{TG}} - V_{\textrm{BG}})/L, 
\end{equation}
where $V_{\textrm{TG}}/V_{\textrm{BG}}$ is the electrostatic potential of the top/bottom gate, 
and $L$ is the distance between the two gate electrodes. 
A metallic part of the heterostructure, which is graphene in the case of Figure~\ref{fig:CrI3a}a, is considered to be grounded so that extra charge can be introduced from the environment to the heterostructure. 
The electric field $\mathcal{E}$ and the extra charge density $n$ can be viewed as two independent variables for a dual gate setup. 
If the heterostructure is insulating, extra charges can hardly be added to the system due to the lack of states around the Fermi energy. 
Therefore, we will fix the extra charge density $n$ to be zero and tune only the electric field $\mathcal{E}$ in our simulations.

We study gate field effects on bare bilayer \cri{}, \bncribn{}, \bncrigr{}, and \grcrigr{}. 
Our calculations are based on density functional theory in conjunction with the effective screening medium method as implemented in the SIESTA package~\cite{RN304}.
The computational details of gate calculations with SIESTA can be found in the reference.~\footnote{
We used a double-$\zeta$ basis set for \ce{Cr} $3d$ orbitals, a single-$\zeta$ polarized basis set for \ce{Cr} $4s$ orbitals, and a single-$\zeta$ basis set for \ce{I} $5s$ and $5p$ orbitals. 
We applied the Perdew-Burke-Ernzerhof exchange correlation energy functional and norm-conserving pseudo-potentials. 
A $51 \times 51 \times 1$ Monkhorst-Pack $k$-mesh was used to sample the reciprocal space. 
Such a $k$-mesh was tested to be dense enough to capture the interlayer charge transfer between graphene and \cri{}. 
A MeshCutoff of $150 \Ry{}$ was applied for the real space sampling. 
The Hubbard $U$ parameter in the DFT$+U$ method was set to $4\eV$. 
For insulating or semiconducting systems, we adopted a Fermi-Dirac function with $T=10\, \textrm{K}$ to determine the occupation of Kohn-Sham orbitals. 
For metallic systems, we adopted the $4$th order Methfessel-Paxton smearing method with $T=200 \, \textrm{K}$ to calculate the electron distribution accurately. 
}
The unit cell of the heterostructures is as large as a $\sqrt{3} \times \sqrt{3}$ supercell of the bare bilayer \cri{}, or a $5 \times 5$ supercell of graphene. 
Graphene (\bn{}) is compressed by $0.7\%$ ($1.1\%$) to fit with the lattice constant of bilayer \cri{}. 
The atomic structures of the heterostructures were relaxed by VASP using the same parameters for the bare bilayer \cri{} except that the force tolerance is set to $0.02 \eVperAng{}$ and the $k$-point mesh is $5 \times 5 \times 1$. 
Here, we consider only the HT stacking bilayer \cri{}, since it is likely what is seen in experiments. 
First, we consider the effects of electric field under the condition of zero charge doping. 
Figure~\ref{fig:CrI3a}b shows the energy difference between the AFM and the FM magnetic configurations for these systems.
At zero electric field, the energy of the AFM state is lower than that of the FM state by $15$--$18 \meV{}$ per unit cell (of bilayer \cri{}).~\footnote{The energy difference $E_{\textrm{AFM}} - E_{\textrm{FM}}$ in Figure~\ref{fig:CrI3a}b differs from that in \ref{fig:CrI3a}d because the former is calculated using a localized basis set (SIESTA package) but the latter using plane waves (VASP package). Plane wave results are considered to be more accurate.} 
As electric field increases, a magnetic phase transition from the AFM state to the FM state occurs at $\sim0.6 \VperAng{}$ for bare bilayer \cri{}. 
Such an AFM-to-FM magnetic phase transition was also reported in previous experimental~\cite{RN251} and theoretical~\cite{RN256} studies.
If bilayer \cri{} is covered by graphene on both the bottom and the top sides, the AFM state is always energetically preferred. 
Actually, the energy difference $E_\textrm{AFM} - E_\textrm{FM}$ is less than $-10 \meV$ up to an electric field of $0.9 \VperAng{}$. 
The magnetic phase transition is also absent for the heterostructure \bncribn{}. 
However, the AFM and the FM state are quite close in energy at an electric field of $0.4 \VperAng{}$. 
Immediately after $\mathcal{E} = 0.4 \VperAng{}$, $E_\textrm{AFM} - E_\textrm{FM}$ decreases and reaches about $-6 \meV$ at $0.5 \VperAng{}$. 
If bilayer \cri{} is covered by \bn{} on the bottom side and graphene on the top side, an AFM-to-FM magnetic phase transition was calculated to occur at around $0.8 \VperAng{}$. 
Since the heterostructure \bncrigr{} is asymmetric in the out-of-plane direction, it is sensitive to the direction of electric field. 
Based on our calculations, a negative electric field also tends to stabilize the FM state; however it doesn't induce any magnetic phase transition down to $-0.9 \VperAng{}$. 
Second, we consider the effects of charge doping under the condition of zero electric field ($V_\textrm{TG} = V_\textrm{BG}$). 
Figure~\ref{fig:CrI3a}c shows the energy difference $E_\textrm{AFM} - E_\textrm{FM}$ versus the extra charge density for \bncrigr{} and \grcrigr{}. 
These two system are always in the AFM state under both electron doping and hole doping conditions. 
The energy difference $E_\textrm{AFM} - E_\textrm{FM}$ varies between $-20 \meV{}$ and $-15 \meV{}$ for the doping level range of $[-10^{13}:+10^{13}] \cm{}^{-2}$.

\begin{figure}[H]
\centering
\includegraphics[width=0.8\textwidth]{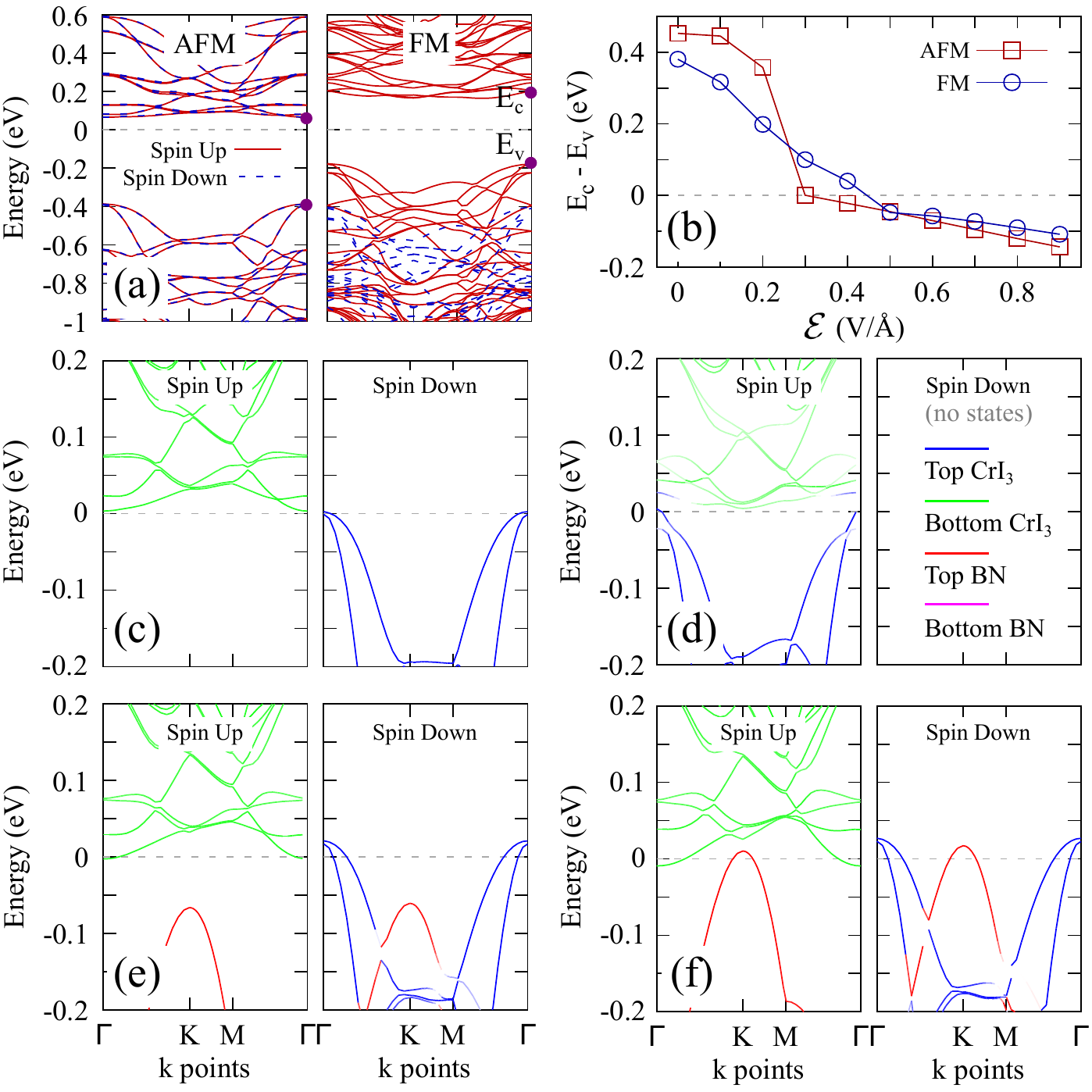}
\caption{\label{fig:CrI3b}
(a) Band structure of bilayer \cri{} in both AFM and FM states. 
(b) $E_c - E_v$ versus electric field across bilayer \cri{}. 
(c) Energy bands for AFM bilayer \cri{} for $\mathcal{E} = 0.3 \VperAng$. 
(d) Energy bands for FM bilayer \cri{} for $\mathcal{E} = 0.5 \VperAng$. 
(e) Energy bands for AFM \bncribn{} for $ \mathcal{E} = 0.4 \VperAng$. 
(f) Energy bands for AFM \bncribn{} for $ \mathcal{E} = 0.5 \VperAng$. 
} 
\end{figure}

So far, we have examined the magnetic phase transition in pure and hybrid bilayer \cri{} systems via total energy calculations. 
Next, we will explain the interfacial effects on the energy diagrams of Figures~\ref{fig:CrI3a}b and \ref{fig:CrI3a}c by detailed electronic structure. 
Figure~\ref{fig:CrI3b}a shows the energy bands of the HT stacking bilayer \cri{} in both AFM and FM states. 
The purple circles highlight the conduction band $E_c$ 
and the valence band $E_v$ at the $\Gamma$-point. 
Figure~\ref{fig:CrI3b} shows the energy difference between $E_c$ and $E_v$ versus electric field. 
Under zero electric field, $E_c - E_v$ is $0.45 \eV$ for the AFM state and $0.38 \eV$ for the FM state.
The difference decreases as the electric field increases for both states and reaches zero at an electric field of $\sim 0.3 \VperAng{}$ for the AFM state. 
The slope of the $E_c(\mathcal{E}) - E_v(\mathcal{E})$ curve for the AFM state changes significantly at $\mathcal{E} \sim 0.3 \VperAng{}$.
This corresponds to the direct band gap closing for the AFM bilayer \cri{}, which is shown in Figure~\ref{fig:CrI3b}c.
In Figure~\ref{fig:CrI3b}c, we see that both the conduction band (the spin-up electrons of the bottom \cri{} layer) and the valence band (the spin-down electrons of the top \cri{} layer) touch the Fermi level.
Past $\mathcal{E}=0.3 \VperAng$, $E_c - E_v$ decreases almost linearly with the electric field. 
The FM bilayer \cri{} experiences an indirect band gap closing at  $\mathcal{E} \sim 0.5 \VperAng{}$, where both the conduction and the valence electrons at the Fermi level are spin-up electrons, see Figure~\ref{fig:CrI3b}d. %
The band gap closing seems to be correlated with the plateau of the $E_\textrm{AFM}(\mathcal{E}) - E_\textrm{FM}(\mathcal{E})$ curve between $0.3 \VperAng{}$ and $0.5 \VperAng{}$. 
Figure~\ref{fig:CrI3b}e (Figure~\ref{fig:CrI3b}f) shows the band structure of \bncribn{} under an electric field of $0.4 \VperAng$ ($0.5 \VperAng$). 
The valence band of the top \bn{} layer crosses the Fermi level at $\mathcal{E} = 0.5 \VperAng$ but not at $\mathcal{E} = 0.4 \VperAng$. 
Such a band crossing is likely the reason for the significant reduction of the energy difference between the AFM and the FM states, as depicted in Figure~\ref{fig:CrI3a}b. 
Similarly for the heterostructure \bncrigr{}, the bottom \bn{} layer starts to lose electrons to the top graphene layer at an electric field around $-0.4 \VperAng{}$. 
The electron transfer results in an induced electric field which is opposite to the direction of the gate electric field, and thus it weakens the net electric field across the bilayer \cri{}. 
As a result, the AFM-to-FM magnetic phase transition is hindered.

\begin{figure}[htb!]
\centering
\includegraphics[width=0.9\textwidth]{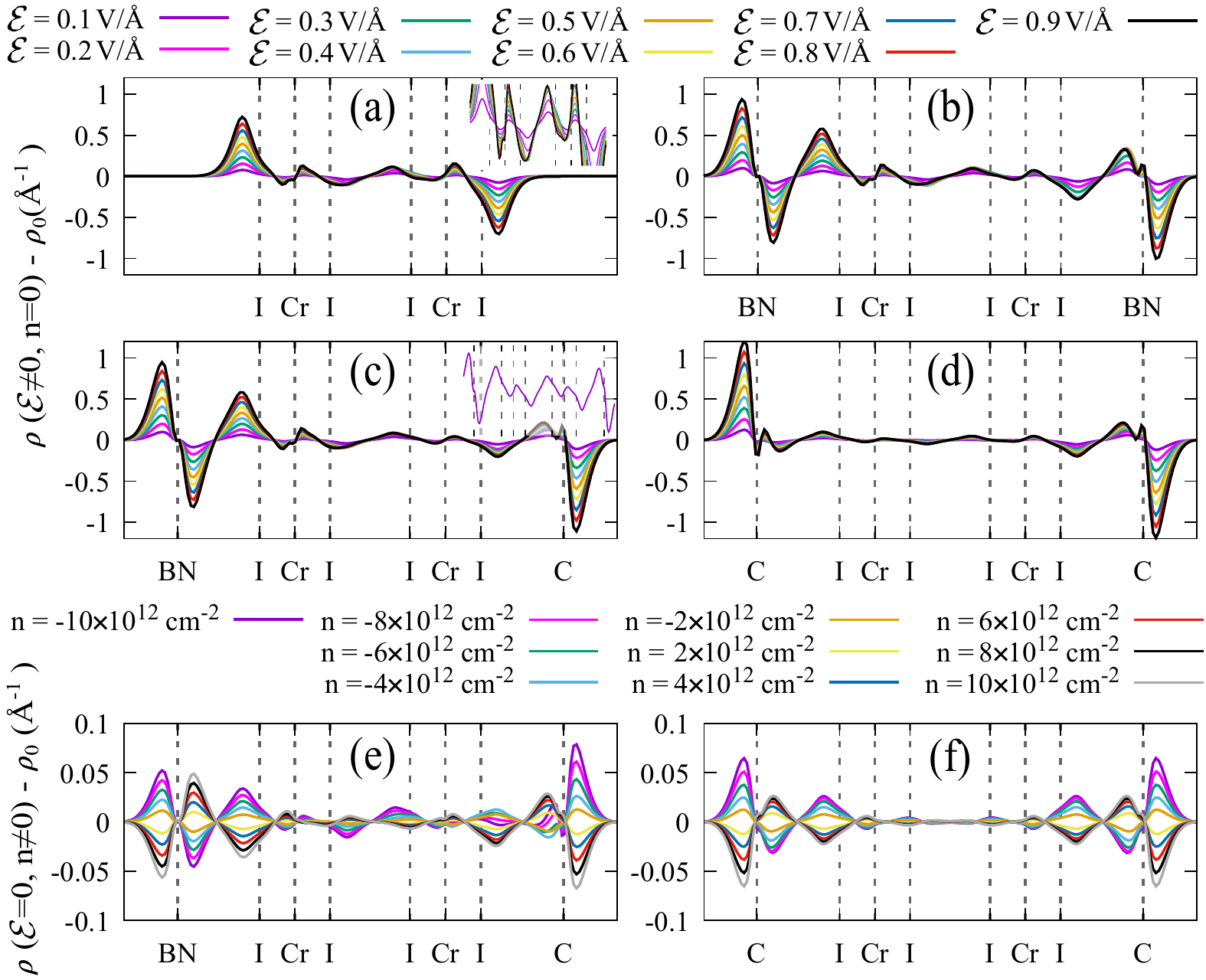}
\caption{\label{fig:CrI3c}
(a)--(d) Plane-averaged electron density difference of bilayer \cri{}, \bncribn{}, \bncrigr{}, and \grcrigr{}, respectively, under different electric fields. 
$\rho(\mathcal{E}, n)$ is the plane-averaged electron density under an electric field $\mathcal{E}$ and with some extra charge density $n$. 
Negative $n$ means electron doping.
$\rho_0 \equiv \rho(\mathcal{E}=0, n=0)$. 
The inset in the panel (a) or the panel (c) shows the same curves with a smaller $y$ range of $[-0.12:0.12] \, \textrm{\AA}^{-1}$. 
(e)--(f) Plane averaged electron density difference of \bncrigr{}, and \grcrigr{}, respectively, under various doping levels. 
} 
\end{figure}

In order to examine the electron redistribution of the systems in response to a gate field, we plot the plane-averaged electron density difference $\rho(\mathcal{E}, n) - \rho_0$ in Figure~\ref{fig:CrI3c}, where $\rho_0$ is the electron density without any gate field.  
Since the averaged electron density difference for the AFM and the FM states are quite similar, we will present only that for the AFM state. 
Figure~\ref{fig:CrI3c}a--\ref{fig:CrI3c}d shows the effects of electric field for bilayer \cri{}, \bncribn{}, \bncrigr{}, and \grcrigr{}, respectively, under zero charge doping.
For bilayer \cri{}, the major electron transfer is from the topmost iodine atomic layer to the bottommost iodine atomic layer. 
This electron transfer increases gradually from $\mathcal{E}=0.1 \VperAng{}$ to $\mathcal{E}=0.9 \VperAng{}$, which does not signal the band gap closing at around $\mathcal{E}=0.3 \VperAng{}$. 
In comparison, the inner chromium and iodine atomic layers experience a relatively small change in electron density. 
Especially, the band gap closing is signaled by the change in the electron density between the inner two \cri{} layers. 
This electron density change increases gradually at small electric field but saturates after the band gap closing, as shown in the inset of Figure~\ref{fig:CrI3c}a. 
%%%
For \bncribn{}, the electron transfer is similar to that of bare bilayer \cri{} when the electric field is smaller than $0.4 \VperAng{}$. 
Within this range of electric field, a local electronic dipole forms for each \bn{} atomic layer without significant electron transfer between \bn{} and \cri{}. 
Consequently, the energy difference between the AFM and the FM states for \bncribn{} is quite close to that for bare bilayer \cri{} when $\mathcal{E} < 0.4 \VperAng{}$ (see Figure~\ref{fig:CrI3a}b). 
However, the major electron transfer is from the top-most \bn{} atomic layer and the bottom-most iodine atomic layer after $\mathcal{E}=0.4 \VperAng{}$, which is consistent with the band structure in Figure~\ref{fig:CrI3b}. 
%%%
The inset of Figure~\ref{fig:CrI3c}c shows the electron density variation for \bncrigr{} under $0.1 \VperAng{}$ electric field, a local electronic dipole also forms around the top graphene layer rendering a small amount of electron transfer between graphene and the remaining insulating part of the heterostructure. 
Again, the energy difference between the AFM and the FM states of \bncrigr{} is close to that of bare bilayer \cri{} at $\mathcal{E} = 0.1 \VperAng{}$. 
As the electric field further increases, the electron transfer between the top graphene layer and the bottommost iodine atomic layer gradually becomes dominant.
This behavior is correlated with the observation that the magnetic phase transition occurs at a larger electric field for \bncrigr{} compared with bare bilayer \cri{}. 
In contrast to the local dipole formation around graphene in case of $0.1 \VperAng{}$ electric field, a $-0.1 \VperAng{}$ electric field results in a significant amount of electron transfer between the bottom \cri{} layer and the top graphene layer. 
%%%
When bilayer \cri{} is covered by graphene on both the top and the bottom sides, the inter-\cri{}-layer electron transfer is greatly reduced due to the electrostatic shielding of the graphene layers. 
As a result, there is no magnetic phase transition up to an electric field of $0.9 \VperAng{}$. 
Figures~\ref{fig:CrI3c}e and \ref{fig:CrI3c}f show the effects of charge doping for \bncrigr{} and \grcrigr{}, respectively, under the condition of zero electric field ($V_\textrm{TG} = V_\textrm{BG}$).
Upon the addition of electrons or holes, the extra charges go to both the top and the bottom graphene layers resulting in a small or even negligible inter-\cri{}-layer electron transfer. 
This is the major difference from the case of applying electric field, where there is significant inter-\cri{}-layer electron transfer. 
The lack of significant inter-\cri{}-layer electron transfer seems the reason why $E_\textrm{AFM}(n) - E_\textrm{FM}(n)$ does not change much with extra charge density $n$. 
Furthermore, there is still some inter-\cri{}-layer electron transfer of \bncrigr{} due to the asymmetry between the \bn{} and the graphene layers. 
This inter-\cri{}-layer electron transfer, although small by itself, is larger than that of \grcrigr{}.
Such a comparison could explain why $E_\textrm{AFM}(n) - E_\textrm{FM}(n)$ of \bncrigr{} is slightly larger than $E_\textrm{AFM}(n) - E_\textrm{FM}(n)$ of \grcrigr{} as shown in Figure~\ref{fig:CrI3a}c.

\begin{figure}[htb!]
\centering
\includegraphics[width=0.9\textwidth]{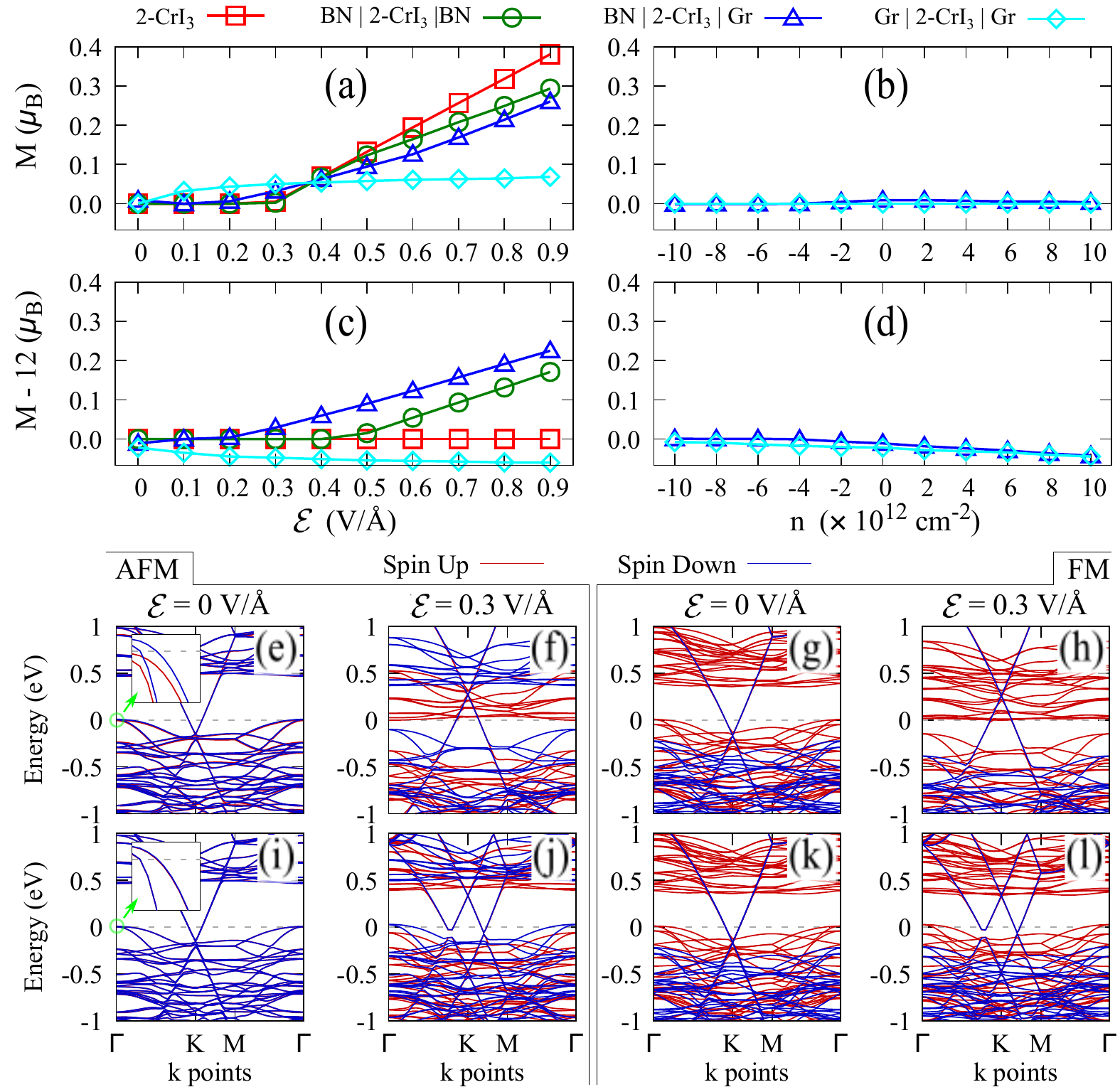}
\caption{\label{fig:CrI3d}
(a)/(c) Magnetic moment of bilayer \cri{}, \bncribn{}, \bncrigr{}, and \grcrigr{} in AFM/FM magnetic configuration versus electric field.
(b)/(d) Magnetic moment of \bncrigr{} and \grcrigr{} in AFM/FM magnetic configuration versus extra charge density. 
Panel (b)/(d) shares the same $y$ axis with panel (a)/(c). 
Gr in the legend stands for graphene. 
(e)-(f) [(g)-(h)] Energy bands for the AFM [the FM] \bncrigr{} under zero and $0.3 \VperAng{}$ electric fields.
(i)-(j) [(k)-(l)] Energy bands for the AFM [the FM] \grcrigr{} under zero and $0.3 \VperAng{}$ electric fields.
%
% The Dirac point for graphene is at the K point. 
%
The inset in (e) and (i): zoomed-in view of the energy bands with an energy range of $[-30:10] \meV{}$ at the $\Gamma$ point. 
} 
\end{figure}

Figure~\ref{fig:CrI3d}a (\ref{fig:CrI3d}c) shows the magnetic moment $M$ of bare bilayer \cri{}, \bncribn{}, \bncrigr{}, and \grcrigr{} systems in the AFM (FM) state versus electric field. 
$M$ is measured per unit cell of bilayer \cri{} with four chromium atoms. 
For AFM bare bilayer \cri{}, the magnetic moment remains zero until $\mathcal{E} > 0.3 \VperAng{}$, where the band gap closes and inter-spin electron transfer transpires, as shown in Figure~\ref{fig:CrI3b}c. 
The magnetic moment $M$ increases linearly with the electric field after $0.3 \VperAng{}$ and reaches $\sim 0.4 \muB$ per unit cell at $\mathcal{E} = 0.9 \VperAng{}$. 
An AFM-to-FM magnetic phase transition occurs at around $0.6 \VperAng{}$, where the magnetic moment changes from $\sim 0.2 \muB$ to $12 \muB$. 
$M$ is insensitive to the electric field for FM bilayer \cri{} because there is no inter-spin electron transfer even if the band gap closes, as shown in Figure~\ref{fig:CrI3b}d. 
%%%
For AFM \bncribn{}, the magnetic moment also becomes finite after $0.3 \VperAng{}$ for the same reason as with the case of bare bilayer \cri{}, which is evidenced by the band structure of AFM \bncribn{} under $\mathcal{E} = 0.4 \VperAng{}$ shown in Figure~\ref{fig:CrI3b}e. 
After $\mathcal{E} = 0.5 \VperAng{}$, the magnetic moment of AFM \bncribn{} is smaller than that of AFM bilayer \cri{}. 
This can be understood since the total inter-layer electron transfer of \bncribn{} consists of both intra-spin and inter-spin electron transfer, while that of bilayer \cri{} consists merely of inter-spin electron transfer. 
The former shows a smaller change in the magnetic moment than the latter under the assumption that the total inter-layer electron transfer is the same, which should be a good approximation for these two systems under the same electric field. 
The magnetic moment of FM \bncribn{} does not increase until $\mathcal{E} \geqslant 0.5 \VperAng{}$, which is due to the electron transfer from the spin down channel of the top \bn{} layer to the spin up channel of the bottom \cri{} layer, as shown in Figure~\ref{fig:CrI3b}f. 
%%%
For \bncrigr{} under zero electric field, graphene is slightly doped with electrons regardless of the magnetic state, which can be seen from the band structures in Figures~\ref{fig:CrI3d}e and \ref{fig:CrI3d}g. 
Beginning at $\mathcal{E} = 0.3 \VperAng{}$, the magnetic moment increases significantly for both the AFM and the FM states.
This is because graphene loses electrons from both the spin up and the spin down channels to the spin up channel of \cri{} as shown in Figures~\ref{fig:CrI3d}f and \ref{fig:CrI3d}h. 
%%%
For \grcrigr{} under zero electric field, both the top and bottom graphene layers gain the same amount of electrons from \cri{} as evidenced by the same shift of the Dirac cones in Figures~\ref{fig:CrI3d}i and \ref{fig:CrI3d}k for the AFM and the FM states respectively. 
The magnetic moment of AFM \grcrigr{} under zero electric field is exactly zero $\muB$ due to the spin degeneracy. 
In contrast, the magnetic moment of FM \grcrigr{} under zero electric field is not exact $12 \muB$, which is the value for bare bilayer \cri{}. 
This is because the valence bands of the FM bilayer \cri{} are spin split by $\sim 0.2 \eV$ and only the spin-up energy bands are hole doped. 
Figure~\ref{fig:CrI3d}j (\ref{fig:CrI3d}l) shows the band structure of the AFM (the FM) \grcrigr{} under an electric field of $0.3 \VperAng{}$. 
The major change in the band structure is the shift of graphene bands that corresponds to the process of the top graphene layer losing electrons to the bottom graphene layer. 
A finite electric field breaks the spin degeneracy of the AFM state leaving the spin-up valence band fully occupied and the spin-down valence band slightly doped with holes. 
As a result, the magnetic moment of the AFM \grcrigr{} state under finite electric field is slightly above zero $\muB$. 
On the contrary, a finite electric field slightly enhances the hole doping of the spin-up valence band of FM \grcrigr{} and, in consequence, the magnetic moment reduces a bit. 
%%%
The inset of Figure~\ref{fig:CrI3d}e (\ref{fig:CrI3d}i) shows a zoomed-in plot of the band structure of \bncrigr{} (\grcrigr{}) in the energy range of $[-30:10] \meV$ at the $\Gamma$ point. 
Note that the energy bands are spin degenerate for \grcrigr{} but not for \bncrigr{}. 
Such a comparison indicates that the spin degeneracy of bare AFM bilayer \cri{} is preserved (broken) by the symmetrical (asymmetrical) surrounding chemical environment.

\subsection{Bilayer \texorpdfstring{CrI\textsubscript{3}}{CrI3} under pressure}

\label{section:CrI3_Pressure}

Recently it has been experimentally observed~\cite{RN252} that the state of bilayer \cri{} at zero external magnetic field switches from an interlayer AFM state to an interlayer FM state on increasing the pressure. 
A corresponding structural transition from HT stacking to LT stacking has also been confirmed by Raman spectroscopy. %
Interestingly, having experienced an increase in pressure up to $2.70 \GPa$ and then taken out of the pressure cell, the bilayer \cri{} sample remains in the LT structure and the FM state. 
This implies that the structural transition that occurs during the pressurizing stage is irreversible.

\begin{figure}[htb!]
\centering
\includegraphics[width=0.9\textwidth]{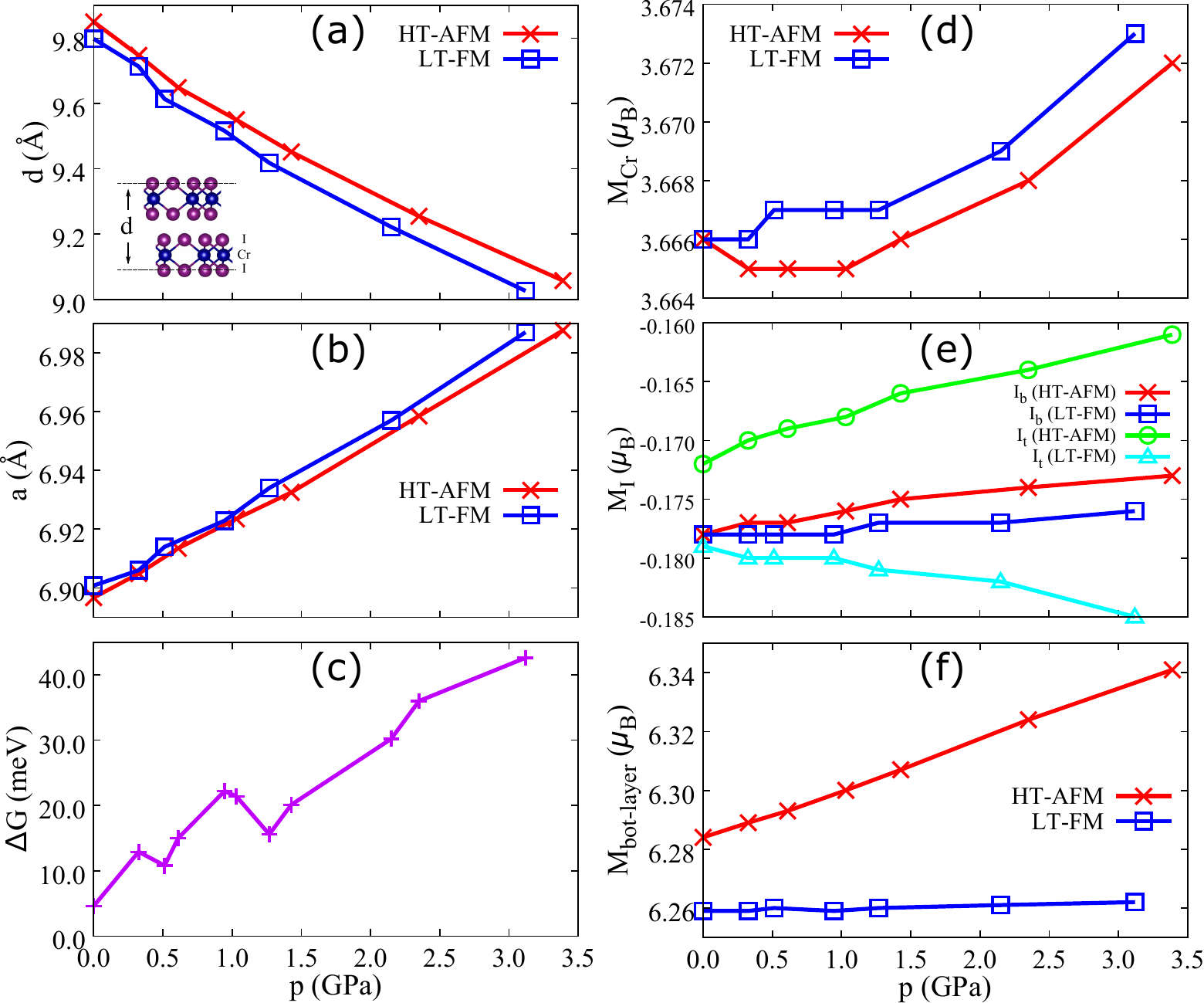}
\caption{\label{FIG:CrI3pressure}
(a) Distance $d$ between the topmost layer of I atoms and the bottommost layer of I atoms, as displayed in inset, as a function of pressure for each of the HT-AFM and LT-FM states. (b) Hexagonal lattice parameter $a$ as a function of pressure for each of the HT-AFM and LT-FM states. (c) Gibbs free energy difference $\Delta G = G_\textrm{HT-AFM} - G_\textrm{LT-FM}$. This curve shows that the HT-AFM state is less stable than the LT-FM state at all experimental pressures and becomes increasingly unstable compared to LT-FM state with increasing pressure. (d) Magnetization of a single Cr atom on the bottom \cri{} layer. (e) Magnetization of a single I atom in each of the two mono-atomic I layers of the bottom \cri{} layer. (f) Magnitude of total magnetization in a single layer of \cri{} per unit cell for each of the HT-AFM and LT-FM states.
}
\end{figure}

In order to study a possible structural transition and to investigate the variation of magnetism in each state, we have calculated the Gibbs free energies of the FM state in the LT structure (LT-FM) and the AFM state in the HT structure (HT-AFM) as functions of pressure. 
Our calculations are based on density functional theory as implemented in the VASP package with the same computational details as in the reference$^{92}$. 
To simulate the effect of pressure, the distance $d$ between the topmost monoatomic and bottommost monoatomic layers of I atoms (see inset of Figure~\ref{FIG:CrI3pressure}(a)) has been tuned, while the in-plane coordinates of these I atoms, all coordinates of the other atoms and the lattice constant $a$ of the hexagonal unit cell are fully relaxed. 
In the end, at each $d$ the pressure is calculated by summing up the atomic forces on the topmost or bottommost monoatomic layer of I atoms and then dividing the result by the in-plane area of the unit cell. In Figure~\ref{FIG:CrI3pressure}(a) and \ref{FIG:CrI3pressure}(b), lattice-related parameters $d$ and $a$ are plotted respectively. As expected, $d$, as a measure of interlayer distance, decreases with increasing pressure, while the lattice constant $a$ increases. 
The Gibbs free energy difference between the LT-FM and HT-AFM states, $\Delta G = G_\textrm{HT-AFM} - G_\textrm{LT-FM}$, obtained from the linearly-interpolated Gibbs free energy $G = E + pV$ as a function of $p$ in each state, is shown in Figure~\ref{FIG:CrI3pressure}(c). 
Here $E$ is the total energy of the state, $p$ is the pressure, and the volume is $V=Ad$, where $A=\sqrt{3}\,a^{2}/2$ is the in-plane area of the unit cell. 
The free energy shows that the HT-AFM state is less stable than the LT-FM state at all experimental pressures and becomes increasingly unstable compared to the LT-FM state with increasing pressure. 
It worth mentioning that the Gibbs free energy difference $\Delta G$ depends on the choice of the Hubbard $U$ parameter of the DFT$+U$ method, and $U$ is $5 \eV$ here. 
However, We have checked that the trend of $\Delta G$ as a function of pressure $p$ doesn't change for $U = 8 \eV$. 
We conjecture that this increasing relative instability leads to a transition from an initial metastable HT-AFM state to the more stable LT-AFM state at some intermediate pressure, so that after the sample is taken out of the pressure cell, the LT-FM state persists. Spin-orbit coupling has been taken into account in the calculation, and the result shows that the magnetizations of all atoms are in the out-of-plane direction. The magnetizations of several single atoms in the bottom \cri{} layer, which also range over all the different atoms in the sense of symmetry, are shown in Figure~\ref{FIG:CrI3pressure}(d) and (e). Notice that in both the HT-AFM and LT-FM states, the magnetizations of I atoms are in the opposite direction from that of the Cr atoms in the same \cri{} layer. Finally the magnitude of total magnetization per unit cell on each \cri{} layer as a sum of the magnetizations of individual atoms is plotted in Figure~\ref{FIG:CrI3pressure}(f), from which we see that only the strength of the AFM order is enhanced by pressure, while there is essentially no change in that of the FM order.

% ==================================================
% Section 4: Conclusion
% ==================================================

\section{conclusions, discussions, and outlook} 

So far we have demonstrated that a first-principles description of the gate field effect provides physical insight that can not be revealed by empirical methods. As an experimental tool, gating is a convenient knob for controlling the electronic, magnetic, and electron transport properties of two-dimensional materials. As we reduce the size of systems or devices to nanometer scale and if  quantum mechanical laws are governing the physical processes, it is inevitable to deal with system performance at the electron level and its response to gate fields in various configurations.  Much of the scientific outcomes were discussed previous publications; here we add a few lines to conclude the study of bilayer \cri{} and the heterogeneous systems between bilayer \cri{} and graphene/\bn{}. 
We showed that the local Coulomb interaction (the Hubbard $U$ parameter in specific) increases by about $1 \eV$ as the dimension of \cri{} reduces from 3D to 2D, which stabilizes the AFM state of the high temperature stacking.
Our calculations also show that the FM state of the low temperature stacking can be stabilized by increasing pressure, which agrees with experiments. 
Inserting bilayer \cri{} between two graphene sheets prohibits a magnetic phase transition in electric fields up to at least $0.9 \VperAng$ due to electrostatic shielding. Note that the current experimental limit is only about $0.1 \VperAng$. 
If bilayer \cri{} is placed between a $h$-BN sheet (the bottom) and a graphene sheet (the top), a magnetic phase transition can be driven by an electric field but the field strength required for the phase transition is larger than that for bare bilayer \cri{}. 
According to our simulations, charge doping does not induce a magnetic phase transition, at least for doping concentrations below $1 \times 10^{13} \cm{}^{-2}$. 
For the \bncribn{} system, the top \bn{} layer becomes hole doped at $\sim 0.4 \VperAng{}$, which inhibits the magnetic phase transition. 
The magnetic moment of bare bilayer in the AFM state does not increase until the band gap closes at $\sim 0.3 \VperAng{}$. 
In contrast, the magnetic moment of \bncrigr{} in the AFM state increases at a smaller electric field due to charge transfer between graphene and bilayer \cri{}.

% Future work

Before the end of the paper, we present the outlook for ongoing and near future projects. The tunability of the interlayer magnetic order by a magnetic field permits giant tunneling magnetoresistance.~\cite{RN253} One of our ongoing projects is exploration of gate field effects on spin-dependent electron tunneling properties through bilayer \cri{} in a graphene$\,|\,$bilayer \cri{}$\,|\,$graphene vertical tunneling junction. 
%%%
Inspired by experimental activities in the Center for Molecular Magnetic Quantum Materials (M$^2$QM), we plan to study adsorption of \ce{Mn_12} molecules and other single-molecule magnets onto monolayer \ce{MoS_2} which has not yet been reported. It was shown that charging is an effective way to tune the magnetic anisotropy of a magnetic molecule;~\cite{RN104} and we have previously simulated the electron transfer between graphene and \ce{Mn_12} molecules without a gate field.~\cite{RN259} Following this thread, we started investigations to address how charge doping affects the magnetic anisotropy of single-molecule magnets such as \ce{Mn_12}. 
%%%
A glance in the field of first-principles transport studies, there is very little theoretical work ~\cite{RN263, RN264} addressing the role of spin-orbit coupling of molecules or 2D junction under finite gate or bias voltages. 
One of the reasons for such studies being scarce is the lack of computational tools. Another one is the computational cost. 
At the frontier of methodology development in our group, a few things are on the horizon: 1) First-principles modeling and algorithm development for computing Schottky barrier, implementing spin-orbital coupling in ESM and transport calculations, and adding phonon-electron coupling in non-equilibrium Green's function and the ESM framework. In particular, the capability of including the spin-orbital coupling allows us to look at Janus monolayer \ce{MoSSe}, which was recently synthesized in experiments.~\cite{RN260, RN261} An intrinsic out-of-plane potential buildup exists in  \ce{MoSSe} since \ce{S} and \ce{Se} atoms have different electron affinity. Our plan of making shared tools for the community will be based on the packages TranSIESTA and QuantumEspresso. Looking forward, we will then test and apply these tools to study in depth the roles of spin-orbit coupling, gate effects on magnetoelectric coupling and magnetorestriction, spin-phonon coupling, and manifestations of these couplings and effects in transport measurements.

\begin{acknowledgements}

This work was supported by the US Department of Energy (DOE), Office of Basic Energy Sciences (BES), under Contract No. DE-FG02-02ER45995. 
Computations were done using the utilities of the National Energy Research Scientific Computing Center and University of Florida Research Computing.

\end{acknowledgements}

\section*{Data Availability}

The data that supports the findings of this study are available within the article. 

\bibliographystyle{apsrev4-2}

\bibliography{refs}

%apsrev4-2.bst 2019-01-14 (MD) hand-edited version of apsrev4-1.bst
%Control: key (0)
%Control: author (72) initials jnrlst
%Control: editor formatted (1) identically to author
%Control: production of article title (-1) disabled
%Control: page (0) single
%Control: year (1) truncated
%Control: production of eprint (0) enabled
\begin{thebibliography}{113}%
\makeatletter
\providecommand \@ifxundefined [1]{%
 \@ifx{#1\undefined}
}%
\providecommand \@ifnum [1]{%
 \ifnum #1\expandafter \@firstoftwo
 \else \expandafter \@secondoftwo
 \fi
}%
\providecommand \@ifx [1]{%
 \ifx #1\expandafter \@firstoftwo
 \else \expandafter \@secondoftwo
 \fi
}%
\providecommand \natexlab [1]{#1}%
\providecommand \enquote  [1]{``#1''}%
\providecommand \bibnamefont  [1]{#1}%
\providecommand \bibfnamefont [1]{#1}%
\providecommand \citenamefont [1]{#1}%
\providecommand \href@noop [0]{\@secondoftwo}%
\providecommand \href [0]{\begingroup \@sanitize@url \@href}%
\providecommand \@href[1]{\@@startlink{#1}\@@href}%
\providecommand \@@href[1]{\endgroup#1\@@endlink}%
\providecommand \@sanitize@url [0]{\catcode `\\12\catcode `\$12\catcode
  `\&12\catcode `\#12\catcode `\^12\catcode `\_12\catcode `\%12\relax}%
\providecommand \@@startlink[1]{}%
\providecommand \@@endlink[0]{}%
\providecommand \url  [0]{\begingroup\@sanitize@url \@url }%
\providecommand \@url [1]{\endgroup\@href {#1}{\urlprefix }}%
\providecommand \urlprefix  [0]{URL }%
\providecommand \Eprint [0]{\href }%
\providecommand \doibase [0]{https://doi.org/}%
\providecommand \selectlanguage [0]{\@gobble}%
\providecommand \bibinfo  [0]{\@secondoftwo}%
\providecommand \bibfield  [0]{\@secondoftwo}%
\providecommand \translation [1]{[#1]}%
\providecommand \BibitemOpen [0]{}%
\providecommand \bibitemStop [0]{}%
\providecommand \bibitemNoStop [0]{.\EOS\space}%
\providecommand \EOS [0]{\spacefactor3000\relax}%
\providecommand \BibitemShut  [1]{\csname bibitem#1\endcsname}%
\let\auto@bib@innerbib\@empty
%</preamble>
\bibitem [{\citenamefont {Chang}\ and\ \citenamefont {Esaki}(1977)}]{RN3013}%
  \BibitemOpen
  \bibfield  {author} {\bibinfo {author} {\bibfnamefont {L.~L.}\ \bibnamefont
  {Chang}}\ and\ \bibinfo {author} {\bibfnamefont {L.}~\bibnamefont {Esaki}},\
  }\href {https://doi.org/10.1063/1.89505} {\bibfield  {journal} {\bibinfo
  {journal} {Applied Physics Letters}\ }\textbf {\bibinfo {volume} {31}},\
  \bibinfo {pages} {687} (\bibinfo {year} {1977})}\BibitemShut {NoStop}%
\bibitem [{\citenamefont {Appenzeller}\ \emph {et~al.}(2004)\citenamefont
  {Appenzeller}, \citenamefont {Lin}, \citenamefont {Knoch},\ and\
  \citenamefont {Avouris}}]{RN3015}%
  \BibitemOpen
  \bibfield  {author} {\bibinfo {author} {\bibfnamefont {J.}~\bibnamefont
  {Appenzeller}}, \bibinfo {author} {\bibfnamefont {Y.~M.}\ \bibnamefont
  {Lin}}, \bibinfo {author} {\bibfnamefont {J.}~\bibnamefont {Knoch}},\ and\
  \bibinfo {author} {\bibfnamefont {P.}~\bibnamefont {Avouris}},\ }\href
  {https://doi.org/10.1103/PhysRevLett.93.196805} {\bibfield  {journal}
  {\bibinfo  {journal} {Physical Review Letters}\ }\textbf {\bibinfo {volume}
  {93}},\ \bibinfo {pages} {4} (\bibinfo {year} {2004})}\BibitemShut {NoStop}%
\bibitem [{\citenamefont {Koswatta}\ \emph {et~al.}(2007)\citenamefont
  {Koswatta}, \citenamefont {Lundstrom},\ and\ \citenamefont
  {Nikonov}}]{RN3014}%
  \BibitemOpen
  \bibfield  {author} {\bibinfo {author} {\bibfnamefont {S.~O.}\ \bibnamefont
  {Koswatta}}, \bibinfo {author} {\bibfnamefont {M.~S.}\ \bibnamefont
  {Lundstrom}},\ and\ \bibinfo {author} {\bibfnamefont {D.~E.}\ \bibnamefont
  {Nikonov}},\ }\href {https://doi.org/10.1021/nl062843f} {\bibfield  {journal}
  {\bibinfo  {journal} {Nano Letters}\ }\textbf {\bibinfo {volume} {7}},\
  \bibinfo {pages} {1160} (\bibinfo {year} {2007})}\BibitemShut {NoStop}%
\bibitem [{\citenamefont {Seabaugh}\ and\ \citenamefont
  {Zhang}(2010)}]{RN3012}%
  \BibitemOpen
  \bibfield  {author} {\bibinfo {author} {\bibfnamefont {A.~C.}\ \bibnamefont
  {Seabaugh}}\ and\ \bibinfo {author} {\bibfnamefont {Q.}~\bibnamefont
  {Zhang}},\ }\href {https://doi.org/10.1109/JPROC.2010.2070470} {\bibfield
  {journal} {\bibinfo  {journal} {Proceedings of the IEEE}\ }\textbf {\bibinfo
  {volume} {98}},\ \bibinfo {pages} {2095} (\bibinfo {year}
  {2010})}\BibitemShut {NoStop}%
\bibitem [{\citenamefont {Radisavljevic}\ \emph
  {et~al.}(2011{\natexlab{a}})\citenamefont {Radisavljevic}, \citenamefont
  {Radenovic}, \citenamefont {Brivio}, \citenamefont {Giacometti},\ and\
  \citenamefont {Kis}}]{RN3016}%
  \BibitemOpen
  \bibfield  {author} {\bibinfo {author} {\bibfnamefont {B.}~\bibnamefont
  {Radisavljevic}}, \bibinfo {author} {\bibfnamefont {A.}~\bibnamefont
  {Radenovic}}, \bibinfo {author} {\bibfnamefont {J.}~\bibnamefont {Brivio}},
  \bibinfo {author} {\bibfnamefont {V.}~\bibnamefont {Giacometti}},\ and\
  \bibinfo {author} {\bibfnamefont {A.}~\bibnamefont {Kis}},\ }\href
  {https://doi.org/10.1038/nnano.2010.279} {\bibfield  {journal} {\bibinfo
  {journal} {Nature Nanotechnology}\ }\textbf {\bibinfo {volume} {6}},\
  \bibinfo {pages} {147} (\bibinfo {year} {2011}{\natexlab{a}})}\BibitemShut
  {NoStop}%
\bibitem [{\citenamefont {Britnell}\ \emph {et~al.}(2012)\citenamefont
  {Britnell}, \citenamefont {Gorbachev}, \citenamefont {Jalil}, \citenamefont
  {Belle}, \citenamefont {Schedin}, \citenamefont {Mishchenko}, \citenamefont
  {Georgiou}, \citenamefont {Katsnelson}, \citenamefont {Eaves}, \citenamefont
  {Morozov}, \citenamefont {Peres}, \citenamefont {Leist}, \citenamefont
  {Geim}, \citenamefont {Novoselov},\ and\ \citenamefont
  {Ponomarenko}}]{RN1622}%
  \BibitemOpen
  \bibfield  {author} {\bibinfo {author} {\bibfnamefont {L.}~\bibnamefont
  {Britnell}}, \bibinfo {author} {\bibfnamefont {R.~V.}\ \bibnamefont
  {Gorbachev}}, \bibinfo {author} {\bibfnamefont {R.}~\bibnamefont {Jalil}},
  \bibinfo {author} {\bibfnamefont {B.~D.}\ \bibnamefont {Belle}}, \bibinfo
  {author} {\bibfnamefont {F.}~\bibnamefont {Schedin}}, \bibinfo {author}
  {\bibfnamefont {A.}~\bibnamefont {Mishchenko}}, \bibinfo {author}
  {\bibfnamefont {T.}~\bibnamefont {Georgiou}}, \bibinfo {author}
  {\bibfnamefont {M.~I.}\ \bibnamefont {Katsnelson}}, \bibinfo {author}
  {\bibfnamefont {L.}~\bibnamefont {Eaves}}, \bibinfo {author} {\bibfnamefont
  {S.~V.}\ \bibnamefont {Morozov}}, \bibinfo {author} {\bibfnamefont
  {N.~M.~R.}\ \bibnamefont {Peres}}, \bibinfo {author} {\bibfnamefont
  {J.}~\bibnamefont {Leist}}, \bibinfo {author} {\bibfnamefont {A.~K.}\
  \bibnamefont {Geim}}, \bibinfo {author} {\bibfnamefont {K.~S.}\ \bibnamefont
  {Novoselov}},\ and\ \bibinfo {author} {\bibfnamefont {L.~A.}\ \bibnamefont
  {Ponomarenko}},\ }\href {https://doi.org/10.1126/science.1218461} {\bibfield
  {journal} {\bibinfo  {journal} {Science}\ }\textbf {\bibinfo {volume}
  {335}},\ \bibinfo {pages} {947} (\bibinfo {year} {2012})}\BibitemShut
  {NoStop}%
\bibitem [{\citenamefont {Roy}\ \emph {et~al.}(2015)\citenamefont {Roy},
  \citenamefont {Tosun}, \citenamefont {Cao}, \citenamefont {Fang},
  \citenamefont {Lien}, \citenamefont {Zhao}, \citenamefont {Chen},
  \citenamefont {Chueh}, \citenamefont {Guo},\ and\ \citenamefont
  {Javey}}]{RN3052}%
  \BibitemOpen
  \bibfield  {author} {\bibinfo {author} {\bibfnamefont {T.}~\bibnamefont
  {Roy}}, \bibinfo {author} {\bibfnamefont {M.}~\bibnamefont {Tosun}}, \bibinfo
  {author} {\bibfnamefont {X.}~\bibnamefont {Cao}}, \bibinfo {author}
  {\bibfnamefont {H.}~\bibnamefont {Fang}}, \bibinfo {author} {\bibfnamefont
  {D.~H.}\ \bibnamefont {Lien}}, \bibinfo {author} {\bibfnamefont {P.~D.}\
  \bibnamefont {Zhao}}, \bibinfo {author} {\bibfnamefont {Y.~Z.}\ \bibnamefont
  {Chen}}, \bibinfo {author} {\bibfnamefont {Y.~L.}\ \bibnamefont {Chueh}},
  \bibinfo {author} {\bibfnamefont {J.}~\bibnamefont {Guo}},\ and\ \bibinfo
  {author} {\bibfnamefont {A.}~\bibnamefont {Javey}},\ }\href
  {https://doi.org/10.1021/nn507278b} {\bibfield  {journal} {\bibinfo
  {journal} {Acs Nano}\ }\textbf {\bibinfo {volume} {9}},\ \bibinfo {pages}
  {2071} (\bibinfo {year} {2015})}\BibitemShut {NoStop}%
\bibitem [{\citenamefont {Fiori}\ \emph {et~al.}(2014)\citenamefont {Fiori},
  \citenamefont {Bonaccorso}, \citenamefont {Iannaccone}, \citenamefont
  {Palacios}, \citenamefont {Neumaier}, \citenamefont {Seabaugh}, \citenamefont
  {Banerjee},\ and\ \citenamefont {Colombo}}]{RN3017}%
  \BibitemOpen
  \bibfield  {author} {\bibinfo {author} {\bibfnamefont {G.}~\bibnamefont
  {Fiori}}, \bibinfo {author} {\bibfnamefont {F.}~\bibnamefont {Bonaccorso}},
  \bibinfo {author} {\bibfnamefont {G.}~\bibnamefont {Iannaccone}}, \bibinfo
  {author} {\bibfnamefont {T.}~\bibnamefont {Palacios}}, \bibinfo {author}
  {\bibfnamefont {D.}~\bibnamefont {Neumaier}}, \bibinfo {author}
  {\bibfnamefont {A.}~\bibnamefont {Seabaugh}}, \bibinfo {author}
  {\bibfnamefont {S.~K.}\ \bibnamefont {Banerjee}},\ and\ \bibinfo {author}
  {\bibfnamefont {L.}~\bibnamefont {Colombo}},\ }\href
  {https://doi.org/10.1038/nnano.2014.207} {\bibfield  {journal} {\bibinfo
  {journal} {Nature Nanotechnology}\ }\textbf {\bibinfo {volume} {9}},\
  \bibinfo {pages} {768} (\bibinfo {year} {2014})}\BibitemShut {NoStop}%
\bibitem [{\citenamefont {Wang}\ \emph {et~al.}(2012)\citenamefont {Wang},
  \citenamefont {Kalantar-Zadeh}, \citenamefont {Kis}, \citenamefont
  {Coleman},\ and\ \citenamefont {Strano}}]{RN2459}%
  \BibitemOpen
  \bibfield  {author} {\bibinfo {author} {\bibfnamefont {Q.~H.}\ \bibnamefont
  {Wang}}, \bibinfo {author} {\bibfnamefont {K.}~\bibnamefont
  {Kalantar-Zadeh}}, \bibinfo {author} {\bibfnamefont {A.}~\bibnamefont {Kis}},
  \bibinfo {author} {\bibfnamefont {J.~N.}\ \bibnamefont {Coleman}},\ and\
  \bibinfo {author} {\bibfnamefont {M.~S.}\ \bibnamefont {Strano}},\ }\href
  {https://doi.org/10.1038/nnano.2012.193} {\bibfield  {journal} {\bibinfo
  {journal} {Nature Nanotechnology}\ }\textbf {\bibinfo {volume} {7}},\
  \bibinfo {pages} {699} (\bibinfo {year} {2012})}\BibitemShut {NoStop}%
\bibitem [{\citenamefont {Choi}\ \emph {et~al.}(2010)\citenamefont {Choi},
  \citenamefont {Lahiri}, \citenamefont {Seelaboyina},\ and\ \citenamefont
  {Kang}}]{RN3022}%
  \BibitemOpen
  \bibfield  {author} {\bibinfo {author} {\bibfnamefont {W.}~\bibnamefont
  {Choi}}, \bibinfo {author} {\bibfnamefont {I.}~\bibnamefont {Lahiri}},
  \bibinfo {author} {\bibfnamefont {R.}~\bibnamefont {Seelaboyina}},\ and\
  \bibinfo {author} {\bibfnamefont {Y.~S.}\ \bibnamefont {Kang}},\ }\href
  {https://doi.org/10.1080/10408430903505036} {\bibfield  {journal} {\bibinfo
  {journal} {Critical Reviews in Solid State and Materials Sciences}\ }\textbf
  {\bibinfo {volume} {35}},\ \bibinfo {pages} {52} (\bibinfo {year}
  {2010})}\BibitemShut {NoStop}%
\bibitem [{\citenamefont {Jariwala}\ \emph
  {et~al.}(2014{\natexlab{a}})\citenamefont {Jariwala}, \citenamefont
  {Sangwan}, \citenamefont {Lauhon}, \citenamefont {Marks},\ and\ \citenamefont
  {Hersam}}]{RN3020}%
  \BibitemOpen
  \bibfield  {author} {\bibinfo {author} {\bibfnamefont {D.}~\bibnamefont
  {Jariwala}}, \bibinfo {author} {\bibfnamefont {V.~K.}\ \bibnamefont
  {Sangwan}}, \bibinfo {author} {\bibfnamefont {L.~J.}\ \bibnamefont {Lauhon}},
  \bibinfo {author} {\bibfnamefont {T.~J.}\ \bibnamefont {Marks}},\ and\
  \bibinfo {author} {\bibfnamefont {M.~C.}\ \bibnamefont {Hersam}},\ }\href
  {https://doi.org/10.1021/nn500064s} {\bibfield  {journal} {\bibinfo
  {journal} {Acs Nano}\ }\textbf {\bibinfo {volume} {8}},\ \bibinfo {pages}
  {1102} (\bibinfo {year} {2014}{\natexlab{a}})}\BibitemShut {NoStop}%
\bibitem [{\citenamefont {Mueller}\ \emph {et~al.}(2010)\citenamefont
  {Mueller}, \citenamefont {Xia},\ and\ \citenamefont {Avouris}}]{RN3019}%
  \BibitemOpen
  \bibfield  {author} {\bibinfo {author} {\bibfnamefont {T.}~\bibnamefont
  {Mueller}}, \bibinfo {author} {\bibfnamefont {F.~N.~A.}\ \bibnamefont
  {Xia}},\ and\ \bibinfo {author} {\bibfnamefont {P.}~\bibnamefont {Avouris}},\
  }\href {https://doi.org/10.1038/nphoton.2010.40} {\bibfield  {journal}
  {\bibinfo  {journal} {Nature Photonics}\ }\textbf {\bibinfo {volume} {4}},\
  \bibinfo {pages} {297} (\bibinfo {year} {2010})}\BibitemShut {NoStop}%
\bibitem [{\citenamefont {Qiao}\ \emph {et~al.}(2014)\citenamefont {Qiao},
  \citenamefont {Kong}, \citenamefont {Hu}, \citenamefont {Yang},\ and\
  \citenamefont {Ji}}]{RN3018}%
  \BibitemOpen
  \bibfield  {author} {\bibinfo {author} {\bibfnamefont {J.~S.}\ \bibnamefont
  {Qiao}}, \bibinfo {author} {\bibfnamefont {X.~H.}\ \bibnamefont {Kong}},
  \bibinfo {author} {\bibfnamefont {Z.~X.}\ \bibnamefont {Hu}}, \bibinfo
  {author} {\bibfnamefont {F.}~\bibnamefont {Yang}},\ and\ \bibinfo {author}
  {\bibfnamefont {W.}~\bibnamefont {Ji}},\ }\href
  {https://doi.org/10.1038/ncomms5475} {\bibfield  {journal} {\bibinfo
  {journal} {Nature Communications}\ }\textbf {\bibinfo {volume} {5}},\
  \bibinfo {pages} {7} (\bibinfo {year} {2014})}\BibitemShut {NoStop}%
\bibitem [{\citenamefont {Radisavljevic}\ \emph
  {et~al.}(2011{\natexlab{b}})\citenamefont {Radisavljevic}, \citenamefont
  {Whitwick},\ and\ \citenamefont {Kis}}]{RN3021}%
  \BibitemOpen
  \bibfield  {author} {\bibinfo {author} {\bibfnamefont {B.}~\bibnamefont
  {Radisavljevic}}, \bibinfo {author} {\bibfnamefont {M.~B.}\ \bibnamefont
  {Whitwick}},\ and\ \bibinfo {author} {\bibfnamefont {A.}~\bibnamefont
  {Kis}},\ }\href {https://doi.org/10.1021/nn203715c} {\bibfield  {journal}
  {\bibinfo  {journal} {Acs Nano}\ }\textbf {\bibinfo {volume} {5}},\ \bibinfo
  {pages} {9934} (\bibinfo {year} {2011}{\natexlab{b}})}\BibitemShut {NoStop}%
\bibitem [{\citenamefont {Xia}\ \emph {et~al.}(2010)\citenamefont {Xia},
  \citenamefont {Farmer}, \citenamefont {Lin},\ and\ \citenamefont
  {Avouris}}]{RN3025}%
  \BibitemOpen
  \bibfield  {author} {\bibinfo {author} {\bibfnamefont {F.~N.}\ \bibnamefont
  {Xia}}, \bibinfo {author} {\bibfnamefont {D.~B.}\ \bibnamefont {Farmer}},
  \bibinfo {author} {\bibfnamefont {Y.~M.}\ \bibnamefont {Lin}},\ and\ \bibinfo
  {author} {\bibfnamefont {P.}~\bibnamefont {Avouris}},\ }\href
  {https://doi.org/10.1021/nl9039636} {\bibfield  {journal} {\bibinfo
  {journal} {Nano Letters}\ }\textbf {\bibinfo {volume} {10}},\ \bibinfo
  {pages} {715} (\bibinfo {year} {2010})}\BibitemShut {NoStop}%
\bibitem [{\citenamefont {Shichman}\ and\ \citenamefont
  {Hodges}(1968)}]{RN3023}%
  \BibitemOpen
  \bibfield  {author} {\bibinfo {author} {\bibfnamefont {H.}~\bibnamefont
  {Shichman}}\ and\ \bibinfo {author} {\bibfnamefont {D.~A.}\ \bibnamefont
  {Hodges}},\ }\href {https://doi.org/10.1109/jssc.1968.1049902} {\bibfield
  {journal} {\bibinfo  {journal} {Ieee Journal of Solid-State Circuits}\
  }\textbf {\bibinfo {volume} {SC 3}},\ \bibinfo {pages} {285} (\bibinfo {year}
  {1968})}\BibitemShut {NoStop}%
\bibitem [{\citenamefont {Geurst}(1966)}]{RN3028}%
  \BibitemOpen
  \bibfield  {author} {\bibinfo {author} {\bibfnamefont {J.~A.}\ \bibnamefont
  {Geurst}},\ }\href {https://doi.org/10.1016/0038-1101(66)90084-0} {\bibfield
  {journal} {\bibinfo  {journal} {Solid-State Electronics}\ }\textbf {\bibinfo
  {volume} {9}},\ \bibinfo {pages} {129} (\bibinfo {year} {1966})}\BibitemShut
  {NoStop}%
\bibitem [{\citenamefont {Ben-Sasson}\ and\ \citenamefont
  {Tessler}(2011)}]{RN3026}%
  \BibitemOpen
  \bibfield  {author} {\bibinfo {author} {\bibfnamefont {A.~J.}\ \bibnamefont
  {Ben-Sasson}}\ and\ \bibinfo {author} {\bibfnamefont {N.}~\bibnamefont
  {Tessler}},\ }\href {https://doi.org/10.1063/1.3622291} {\bibfield  {journal}
  {\bibinfo  {journal} {Journal of Applied Physics}\ }\textbf {\bibinfo
  {volume} {110}},\ \bibinfo {pages} {12} (\bibinfo {year} {2011})}\BibitemShut
  {NoStop}%
\bibitem [{\citenamefont {Koswatta}\ \emph {et~al.}(2005)\citenamefont
  {Koswatta}, \citenamefont {Lundstrom}, \citenamefont {Anantram},\ and\
  \citenamefont {Nikonov}}]{RN3027}%
  \BibitemOpen
  \bibfield  {author} {\bibinfo {author} {\bibfnamefont {S.~O.}\ \bibnamefont
  {Koswatta}}, \bibinfo {author} {\bibfnamefont {M.~S.}\ \bibnamefont
  {Lundstrom}}, \bibinfo {author} {\bibfnamefont {M.~P.}\ \bibnamefont
  {Anantram}},\ and\ \bibinfo {author} {\bibfnamefont {D.~E.}\ \bibnamefont
  {Nikonov}},\ }\href {https://doi.org/10.1063/1.2146065} {\bibfield  {journal}
  {\bibinfo  {journal} {Applied Physics Letters}\ }\textbf {\bibinfo {volume}
  {87}},\ \bibinfo {pages} {3} (\bibinfo {year} {2005})}\BibitemShut {NoStop}%
\bibitem [{\citenamefont {Wang}\ and\ \citenamefont
  {Cheng}(2015{\natexlab{a}})}]{RN2199}%
  \BibitemOpen
  \bibfield  {author} {\bibinfo {author} {\bibfnamefont {Y.~P.}\ \bibnamefont
  {Wang}}\ and\ \bibinfo {author} {\bibfnamefont {H.~P.}\ \bibnamefont
  {Cheng}},\ }\href {https://doi.org/10.1103/PhysRevB.91.245307} {\bibfield
  {journal} {\bibinfo  {journal} {Physical Review B}\ }\textbf {\bibinfo
  {volume} {91}},\ \bibinfo {pages} {245307} (\bibinfo {year}
  {2015}{\natexlab{a}})}\BibitemShut {NoStop}%
\bibitem [{\citenamefont {Wang}\ \emph
  {et~al.}(2016{\natexlab{a}})\citenamefont {Wang}, \citenamefont {Li},
  \citenamefont {Fry},\ and\ \citenamefont {Cheng}}]{RN2201}%
  \BibitemOpen
  \bibfield  {author} {\bibinfo {author} {\bibfnamefont {Y.~P.}\ \bibnamefont
  {Wang}}, \bibinfo {author} {\bibfnamefont {X.~G.}\ \bibnamefont {Li}},
  \bibinfo {author} {\bibfnamefont {J.~N.}\ \bibnamefont {Fry}},\ and\ \bibinfo
  {author} {\bibfnamefont {H.~P.}\ \bibnamefont {Cheng}},\ }\href
  {https://doi.org/10.1103/PhysRevB.94.165428} {\bibfield  {journal} {\bibinfo
  {journal} {Physical Review B}\ }\textbf {\bibinfo {volume} {94}},\ \bibinfo
  {pages} {165428} (\bibinfo {year} {2016}{\natexlab{a}})}\BibitemShut
  {NoStop}%
\bibitem [{\citenamefont {Otani}\ and\ \citenamefont {Sugino}(2006)}]{RN92}%
  \BibitemOpen
  \bibfield  {author} {\bibinfo {author} {\bibfnamefont {M.}~\bibnamefont
  {Otani}}\ and\ \bibinfo {author} {\bibfnamefont {O.}~\bibnamefont {Sugino}},\
  }\href {https://doi.org/10.1103/PhysRevB.73.115407} {\bibfield  {journal}
  {\bibinfo  {journal} {Physical Review B}\ }\textbf {\bibinfo {volume} {73}},\
  \bibinfo {pages} {11} (\bibinfo {year} {2006})}\BibitemShut {NoStop}%
\bibitem [{\citenamefont {Chen}\ \emph {et~al.}(2017)\citenamefont {Chen},
  \citenamefont {Li}, \citenamefont {Wang}, \citenamefont {Fry},\ and\
  \citenamefont {Cheng}}]{RN2193}%
  \BibitemOpen
  \bibfield  {author} {\bibinfo {author} {\bibfnamefont {G.~X.}\ \bibnamefont
  {Chen}}, \bibinfo {author} {\bibfnamefont {X.~G.}\ \bibnamefont {Li}},
  \bibinfo {author} {\bibfnamefont {Y.~P.}\ \bibnamefont {Wang}}, \bibinfo
  {author} {\bibfnamefont {J.~N.}\ \bibnamefont {Fry}},\ and\ \bibinfo {author}
  {\bibfnamefont {H.~P.}\ \bibnamefont {Cheng}},\ }\href
  {https://doi.org/10.1103/PhysRevB.95.045302} {\bibfield  {journal} {\bibinfo
  {journal} {Physical Review B}\ }\textbf {\bibinfo {volume} {95}},\ \bibinfo
  {pages} {045302} (\bibinfo {year} {2017})}\BibitemShut {NoStop}%
\bibitem [{\citenamefont {Li}\ \emph {et~al.}(2019{\natexlab{a}})\citenamefont
  {Li}, \citenamefont {Wang}, \citenamefont {Fry}, \citenamefont {Zhang},\ and\
  \citenamefont {Cheng}}]{RN3009}%
  \BibitemOpen
  \bibfield  {author} {\bibinfo {author} {\bibfnamefont {X.~G.}\ \bibnamefont
  {Li}}, \bibinfo {author} {\bibfnamefont {Y.~P.}\ \bibnamefont {Wang}},
  \bibinfo {author} {\bibfnamefont {J.~N.}\ \bibnamefont {Fry}}, \bibinfo
  {author} {\bibfnamefont {X.~G.}\ \bibnamefont {Zhang}},\ and\ \bibinfo
  {author} {\bibfnamefont {H.~P.}\ \bibnamefont {Cheng}},\ }\href
  {https://doi.org/10.1016/j.jpcs.2017.12.005} {\bibfield  {journal} {\bibinfo
  {journal} {Journal of Physics and Chemistry of Solids}\ }\textbf {\bibinfo
  {volume} {128}},\ \bibinfo {pages} {343} (\bibinfo {year}
  {2019}{\natexlab{a}})}\BibitemShut {NoStop}%
\bibitem [{\citenamefont {Liu}\ \emph {et~al.}(2019{\natexlab{a}})\citenamefont
  {Liu}, \citenamefont {Wang}, \citenamefont {Fry},\ and\ \citenamefont
  {Cheng}}]{RN3010}%
  \BibitemOpen
  \bibfield  {author} {\bibinfo {author} {\bibfnamefont {S.~L.}\ \bibnamefont
  {Liu}}, \bibinfo {author} {\bibfnamefont {Y.~P.}\ \bibnamefont {Wang}},
  \bibinfo {author} {\bibfnamefont {J.~N.}\ \bibnamefont {Fry}},\ and\ \bibinfo
  {author} {\bibfnamefont {H.~P.}\ \bibnamefont {Cheng}},\ }\href
  {https://doi.org/10.1016/j.carbon.2018.12.035} {\bibfield  {journal}
  {\bibinfo  {journal} {Carbon}\ }\textbf {\bibinfo {volume} {144}},\ \bibinfo
  {pages} {362} (\bibinfo {year} {2019}{\natexlab{a}})}\BibitemShut {NoStop}%
\bibitem [{\citenamefont {Wang}\ \emph {et~al.}(2017)\citenamefont {Wang},
  \citenamefont {Fry},\ and\ \citenamefont {Cheng}}]{RN2888}%
  \BibitemOpen
  \bibfield  {author} {\bibinfo {author} {\bibfnamefont {Y.~P.}\ \bibnamefont
  {Wang}}, \bibinfo {author} {\bibfnamefont {J.~N.}\ \bibnamefont {Fry}},\ and\
  \bibinfo {author} {\bibfnamefont {H.~P.}\ \bibnamefont {Cheng}},\ }\href
  {https://doi.org/10.1021/acsomega.7b00856} {\bibfield  {journal} {\bibinfo
  {journal} {Acs Omega}\ }\textbf {\bibinfo {volume} {2}},\ \bibinfo {pages}
  {5824} (\bibinfo {year} {2017})}\BibitemShut {NoStop}%
\bibitem [{\citenamefont {Wang}\ \emph
  {et~al.}(2018{\natexlab{a}})\citenamefont {Wang}, \citenamefont {Li},
  \citenamefont {Liu}, \citenamefont {Fry},\ and\ \citenamefont
  {Cheng}}]{RN3060}%
  \BibitemOpen
  \bibfield  {author} {\bibinfo {author} {\bibfnamefont {Y.~P.}\ \bibnamefont
  {Wang}}, \bibinfo {author} {\bibfnamefont {X.~G.}\ \bibnamefont {Li}},
  \bibinfo {author} {\bibfnamefont {S.~L.}\ \bibnamefont {Liu}}, \bibinfo
  {author} {\bibfnamefont {J.~N.}\ \bibnamefont {Fry}},\ and\ \bibinfo {author}
  {\bibfnamefont {H.~P.}\ \bibnamefont {Cheng}},\ }\href
  {https://doi.org/10.1103/PhysRevB.97.115419} {\bibfield  {journal} {\bibinfo
  {journal} {Physical Review B}\ }\textbf {\bibinfo {volume} {97}},\ \bibinfo
  {pages} {115419} (\bibinfo {year} {2018}{\natexlab{a}})}\BibitemShut
  {NoStop}%
\bibitem [{\citenamefont {Liu}\ \emph {et~al.}(2020)\citenamefont {Liu},
  \citenamefont {Xu}, \citenamefont {Wang}, \citenamefont {P.}, \citenamefont
  {Fry},\ and\ \citenamefont {Cheng}}]{RN3033}%
  \BibitemOpen
  \bibfield  {author} {\bibinfo {author} {\bibfnamefont {S.-L.}\ \bibnamefont
  {Liu}}, \bibinfo {author} {\bibfnamefont {Y.}~\bibnamefont {Xu}}, \bibinfo
  {author} {\bibfnamefont {Y.-P.}\ \bibnamefont {Wang}}, \bibinfo {author}
  {\bibfnamefont {C.~Y.}\ \bibnamefont {P.}}, \bibinfo {author} {\bibfnamefont
  {J.~N.}\ \bibnamefont {Fry}},\ and\ \bibinfo {author} {\bibfnamefont {H.-P.}\
  \bibnamefont {Cheng}},\ }\href
  {https://doi.org/https://doi.org/10.1063/1.5127065} {\bibfield  {journal}
  {\bibinfo  {journal} {Applied Physics Letters}\ }\textbf {\bibinfo {volume}
  {116}},\ \bibinfo {pages} {031601} (\bibinfo {year} {2020})}\BibitemShut
  {NoStop}%
\bibitem [{\citenamefont {Stokbro}\ \emph {et~al.}(2003)\citenamefont
  {Stokbro}, \citenamefont {Taylor}, \citenamefont {Brandbyge},\ and\
  \citenamefont {Ordejon}}]{RN1155}%
  \BibitemOpen
  \bibfield  {author} {\bibinfo {author} {\bibfnamefont {K.}~\bibnamefont
  {Stokbro}}, \bibinfo {author} {\bibfnamefont {J.}~\bibnamefont {Taylor}},
  \bibinfo {author} {\bibfnamefont {M.}~\bibnamefont {Brandbyge}},\ and\
  \bibinfo {author} {\bibfnamefont {P.}~\bibnamefont {Ordejon}},\ }\href
  {https://doi.org/10.1196/annals.1292.014} {\bibfield  {journal} {\bibinfo
  {journal} {Molecular Electronics Iii}\ }\textbf {\bibinfo {volume} {1006}},\
  \bibinfo {pages} {212} (\bibinfo {year} {2003})}\BibitemShut {NoStop}%
\bibitem [{\citenamefont {Giannozzi}\ \emph {et~al.}(2009)\citenamefont
  {Giannozzi}, \citenamefont {Baroni}, \citenamefont {Bonini}, \citenamefont
  {Calandra}, \citenamefont {Car}, \citenamefont {Cavazzoni}, \citenamefont
  {Ceresoli}, \citenamefont {Chiarotti}, \citenamefont {Cococcioni},
  \citenamefont {Dabo}, \citenamefont {Dal~Corso}, \citenamefont
  {de~Gironcoli}, \citenamefont {Fabris}, \citenamefont {Fratesi},
  \citenamefont {Gebauer}, \citenamefont {Gerstmann}, \citenamefont
  {Gougoussis}, \citenamefont {Kokalj}, \citenamefont {Lazzeri}, \citenamefont
  {Martin-Samos}, \citenamefont {Marzari}, \citenamefont {Mauri}, \citenamefont
  {Mazzarello}, \citenamefont {Paolini}, \citenamefont {Pasquarello},
  \citenamefont {Paulatto}, \citenamefont {Sbraccia}, \citenamefont {Scandolo},
  \citenamefont {Sclauzero}, \citenamefont {Seitsonen}, \citenamefont
  {Smogunov}, \citenamefont {Umari},\ and\ \citenamefont
  {Wentzcovitch}}]{RN1694}%
  \BibitemOpen
  \bibfield  {author} {\bibinfo {author} {\bibfnamefont {P.}~\bibnamefont
  {Giannozzi}}, \bibinfo {author} {\bibfnamefont {S.}~\bibnamefont {Baroni}},
  \bibinfo {author} {\bibfnamefont {N.}~\bibnamefont {Bonini}}, \bibinfo
  {author} {\bibfnamefont {M.}~\bibnamefont {Calandra}}, \bibinfo {author}
  {\bibfnamefont {R.}~\bibnamefont {Car}}, \bibinfo {author} {\bibfnamefont
  {C.}~\bibnamefont {Cavazzoni}}, \bibinfo {author} {\bibfnamefont
  {D.}~\bibnamefont {Ceresoli}}, \bibinfo {author} {\bibfnamefont {G.~L.}\
  \bibnamefont {Chiarotti}}, \bibinfo {author} {\bibfnamefont {M.}~\bibnamefont
  {Cococcioni}}, \bibinfo {author} {\bibfnamefont {I.}~\bibnamefont {Dabo}},
  \bibinfo {author} {\bibfnamefont {A.}~\bibnamefont {Dal~Corso}}, \bibinfo
  {author} {\bibfnamefont {S.}~\bibnamefont {de~Gironcoli}}, \bibinfo {author}
  {\bibfnamefont {S.}~\bibnamefont {Fabris}}, \bibinfo {author} {\bibfnamefont
  {G.}~\bibnamefont {Fratesi}}, \bibinfo {author} {\bibfnamefont
  {R.}~\bibnamefont {Gebauer}}, \bibinfo {author} {\bibfnamefont
  {U.}~\bibnamefont {Gerstmann}}, \bibinfo {author} {\bibfnamefont
  {C.}~\bibnamefont {Gougoussis}}, \bibinfo {author} {\bibfnamefont
  {A.}~\bibnamefont {Kokalj}}, \bibinfo {author} {\bibfnamefont
  {M.}~\bibnamefont {Lazzeri}}, \bibinfo {author} {\bibfnamefont
  {L.}~\bibnamefont {Martin-Samos}}, \bibinfo {author} {\bibfnamefont
  {N.}~\bibnamefont {Marzari}}, \bibinfo {author} {\bibfnamefont
  {F.}~\bibnamefont {Mauri}}, \bibinfo {author} {\bibfnamefont
  {R.}~\bibnamefont {Mazzarello}}, \bibinfo {author} {\bibfnamefont
  {S.}~\bibnamefont {Paolini}}, \bibinfo {author} {\bibfnamefont
  {A.}~\bibnamefont {Pasquarello}}, \bibinfo {author} {\bibfnamefont
  {L.}~\bibnamefont {Paulatto}}, \bibinfo {author} {\bibfnamefont
  {C.}~\bibnamefont {Sbraccia}}, \bibinfo {author} {\bibfnamefont
  {S.}~\bibnamefont {Scandolo}}, \bibinfo {author} {\bibfnamefont
  {G.}~\bibnamefont {Sclauzero}}, \bibinfo {author} {\bibfnamefont {A.~P.}\
  \bibnamefont {Seitsonen}}, \bibinfo {author} {\bibfnamefont {A.}~\bibnamefont
  {Smogunov}}, \bibinfo {author} {\bibfnamefont {P.}~\bibnamefont {Umari}},\
  and\ \bibinfo {author} {\bibfnamefont {R.~M.}\ \bibnamefont {Wentzcovitch}},\
  }\href {https://doi.org/Artn 395502 Doi 10.1088/0953-8984/21/39/395502}
  {\bibfield  {journal} {\bibinfo  {journal} {Journal of Physics-Condensed
  Matter}\ }\textbf {\bibinfo {volume} {21}},\ \bibinfo {pages} {395502 (19
  pages)} (\bibinfo {year} {2009})}\BibitemShut {NoStop}%
\bibitem [{\citenamefont {Smidstrup}\ \emph {et~al.}(2020)\citenamefont
  {Smidstrup}, \citenamefont {Markussen}, \citenamefont {Vancraeyveld},
  \citenamefont {Wellendorff}, \citenamefont {Schneider}, \citenamefont
  {GunstL}, \citenamefont {Verstichel}, \citenamefont {Stradi}, \citenamefont
  {Khomyakov}, \citenamefont {Vej-Hansen}, \citenamefont {Lee}, \citenamefont
  {Chill}, \citenamefont {Rasmussen}, \citenamefont {Penazzi}, \citenamefont
  {Corsetti}, \citenamefont {Ojanpera}, \citenamefont {Jensen}, \citenamefont
  {Palsgaard}, \citenamefont {Martinez}, \citenamefont {Blom}, \citenamefont
  {Brandbyge},\ and\ \citenamefont {Stokbro}}]{RN3047}%
  \BibitemOpen
  \bibfield  {author} {\bibinfo {author} {\bibfnamefont {S.}~\bibnamefont
  {Smidstrup}}, \bibinfo {author} {\bibfnamefont {T.}~\bibnamefont
  {Markussen}}, \bibinfo {author} {\bibfnamefont {P.}~\bibnamefont
  {Vancraeyveld}}, \bibinfo {author} {\bibfnamefont {J.}~\bibnamefont
  {Wellendorff}}, \bibinfo {author} {\bibfnamefont {J.}~\bibnamefont
  {Schneider}}, \bibinfo {author} {\bibfnamefont {T.}~\bibnamefont {GunstL}},
  \bibinfo {author} {\bibfnamefont {B.}~\bibnamefont {Verstichel}}, \bibinfo
  {author} {\bibfnamefont {D.}~\bibnamefont {Stradi}}, \bibinfo {author}
  {\bibfnamefont {P.~A.}\ \bibnamefont {Khomyakov}}, \bibinfo {author}
  {\bibfnamefont {U.~G.}\ \bibnamefont {Vej-Hansen}}, \bibinfo {author}
  {\bibfnamefont {M.~E.}\ \bibnamefont {Lee}}, \bibinfo {author} {\bibfnamefont
  {S.~T.}\ \bibnamefont {Chill}}, \bibinfo {author} {\bibfnamefont
  {F.}~\bibnamefont {Rasmussen}}, \bibinfo {author} {\bibfnamefont
  {G.}~\bibnamefont {Penazzi}}, \bibinfo {author} {\bibfnamefont
  {F.}~\bibnamefont {Corsetti}}, \bibinfo {author} {\bibfnamefont
  {A.}~\bibnamefont {Ojanpera}}, \bibinfo {author} {\bibfnamefont
  {K.}~\bibnamefont {Jensen}}, \bibinfo {author} {\bibfnamefont {M.~L.~N.}\
  \bibnamefont {Palsgaard}}, \bibinfo {author} {\bibfnamefont {U.}~\bibnamefont
  {Martinez}}, \bibinfo {author} {\bibfnamefont {A.}~\bibnamefont {Blom}},
  \bibinfo {author} {\bibfnamefont {M.}~\bibnamefont {Brandbyge}},\ and\
  \bibinfo {author} {\bibfnamefont {K.}~\bibnamefont {Stokbro}},\ }\bibfield
  {journal} {\bibinfo  {journal} {Journal of Physics-Condensed Matter}\
  }\textbf {\bibinfo {volume} {32}},\ \href {https://doi.org/ARTN 015901
  10.1088/1361-648X/ab4007} {ARTN 015901 10.1088/1361-648X/ab4007} (\bibinfo
  {year} {2020})\BibitemShut {NoStop}%
\bibitem [{\citenamefont {Bani-Hashemian}\ \emph {et~al.}(2016)\citenamefont
  {Bani-Hashemian}, \citenamefont {Bruck}, \citenamefont {Luisier},\ and\
  \citenamefont {VandeVondele}}]{RN158}%
  \BibitemOpen
  \bibfield  {author} {\bibinfo {author} {\bibfnamefont {M.~H.}\ \bibnamefont
  {Bani-Hashemian}}, \bibinfo {author} {\bibfnamefont {S.}~\bibnamefont
  {Bruck}}, \bibinfo {author} {\bibfnamefont {M.}~\bibnamefont {Luisier}},\
  and\ \bibinfo {author} {\bibfnamefont {J.}~\bibnamefont {VandeVondele}},\
  }\href {https://doi.org/10.1063/1.4940796} {\bibfield  {journal} {\bibinfo
  {journal} {Journal of Chemical Physics}\ }\textbf {\bibinfo {volume} {144}},\
  \bibinfo {pages} {12} (\bibinfo {year} {2016})}\BibitemShut {NoStop}%
\bibitem [{\citenamefont {Sohier}\ \emph {et~al.}(2017)\citenamefont {Sohier},
  \citenamefont {Calandra},\ and\ \citenamefont {Mauri}}]{RN3032}%
  \BibitemOpen
  \bibfield  {author} {\bibinfo {author} {\bibfnamefont {T.}~\bibnamefont
  {Sohier}}, \bibinfo {author} {\bibfnamefont {M.}~\bibnamefont {Calandra}},\
  and\ \bibinfo {author} {\bibfnamefont {F.}~\bibnamefont {Mauri}},\ }\bibfield
   {journal} {\bibinfo  {journal} {Physical Review B}\ }\textbf {\bibinfo
  {volume} {96}},\ \href {https://doi.org/ARTN 075448
  10.1103/PhysRevB.96.075448} {ARTN 075448 10.1103/PhysRevB.96.075448}
  (\bibinfo {year} {2017})\BibitemShut {NoStop}%
\bibitem [{\citenamefont {Brumme}\ \emph {et~al.}(2014)\citenamefont {Brumme},
  \citenamefont {Calandra},\ and\ \citenamefont {Mauri}}]{RN162}%
  \BibitemOpen
  \bibfield  {author} {\bibinfo {author} {\bibfnamefont {T.}~\bibnamefont
  {Brumme}}, \bibinfo {author} {\bibfnamefont {M.}~\bibnamefont {Calandra}},\
  and\ \bibinfo {author} {\bibfnamefont {F.}~\bibnamefont {Mauri}},\ }\href
  {https://doi.org/10.1103/PhysRevB.89.245406} {\bibfield  {journal} {\bibinfo
  {journal} {Physical Review B}\ }\textbf {\bibinfo {volume} {89}},\ \bibinfo
  {pages} {11} (\bibinfo {year} {2014})}\BibitemShut {NoStop}%
\bibitem [{\citenamefont {Lazic}\ \emph {et~al.}(2016)\citenamefont {Lazic},
  \citenamefont {Belashchenko},\ and\ \citenamefont {Zutic}}]{RN3051}%
  \BibitemOpen
  \bibfield  {author} {\bibinfo {author} {\bibfnamefont {P.}~\bibnamefont
  {Lazic}}, \bibinfo {author} {\bibfnamefont {K.~D.}\ \bibnamefont
  {Belashchenko}},\ and\ \bibinfo {author} {\bibfnamefont {I.}~\bibnamefont
  {Zutic}},\ }\bibfield  {journal} {\bibinfo  {journal} {Physical Review B}\
  }\textbf {\bibinfo {volume} {93}},\ \href {https://doi.org/ARTN 241401
  10.1103/PhysRevB.93.241401} {ARTN 241401 10.1103/PhysRevB.93.241401}
  (\bibinfo {year} {2016})\BibitemShut {NoStop}%
\bibitem [{\citenamefont {Bokdam}\ \emph {et~al.}(2011)\citenamefont {Bokdam},
  \citenamefont {Khomyakov}, \citenamefont {Brocks}, \citenamefont {Zhong},\
  and\ \citenamefont {Kelly}}]{RN3050}%
  \BibitemOpen
  \bibfield  {author} {\bibinfo {author} {\bibfnamefont {M.}~\bibnamefont
  {Bokdam}}, \bibinfo {author} {\bibfnamefont {P.~A.}\ \bibnamefont
  {Khomyakov}}, \bibinfo {author} {\bibfnamefont {G.}~\bibnamefont {Brocks}},
  \bibinfo {author} {\bibfnamefont {Z.~C.}\ \bibnamefont {Zhong}},\ and\
  \bibinfo {author} {\bibfnamefont {P.~J.}\ \bibnamefont {Kelly}},\ }\href
  {https://doi.org/10.1021/nl202131q} {\bibfield  {journal} {\bibinfo
  {journal} {Nano Letters}\ }\textbf {\bibinfo {volume} {11}},\ \bibinfo
  {pages} {4631} (\bibinfo {year} {2011})}\BibitemShut {NoStop}%
\bibitem [{\citenamefont {Bokdam}\ \emph {et~al.}(2013)\citenamefont {Bokdam},
  \citenamefont {Khomyakov}, \citenamefont {Brocks},\ and\ \citenamefont
  {Kelly}}]{RN3049}%
  \BibitemOpen
  \bibfield  {author} {\bibinfo {author} {\bibfnamefont {M.}~\bibnamefont
  {Bokdam}}, \bibinfo {author} {\bibfnamefont {P.~A.}\ \bibnamefont
  {Khomyakov}}, \bibinfo {author} {\bibfnamefont {G.}~\bibnamefont {Brocks}},\
  and\ \bibinfo {author} {\bibfnamefont {P.~J.}\ \bibnamefont {Kelly}},\
  }\bibfield  {journal} {\bibinfo  {journal} {Physical Review B}\ }\textbf
  {\bibinfo {volume} {87}},\ \href {https://doi.org/ARTN 075414
  10.1103/PhysRevB.87.075414} {ARTN 075414 10.1103/PhysRevB.87.075414}
  (\bibinfo {year} {2013})\BibitemShut {NoStop}%
\bibitem [{\citenamefont {Wang}\ and\ \citenamefont
  {Cheng}(2015{\natexlab{b}})}]{RN20}%
  \BibitemOpen
  \bibfield  {author} {\bibinfo {author} {\bibfnamefont {Y.~P.}\ \bibnamefont
  {Wang}}\ and\ \bibinfo {author} {\bibfnamefont {H.~P.}\ \bibnamefont
  {Cheng}},\ }\href {https://doi.org/10.1103/PhysRevB.91.245307} {\bibfield
  {journal} {\bibinfo  {journal} {Physical Review B}\ }\textbf {\bibinfo
  {volume} {91}},\ \bibinfo {pages} {6} (\bibinfo {year}
  {2015}{\natexlab{b}})}\BibitemShut {NoStop}%
\bibitem [{\citenamefont {Otani}\ \emph {et~al.}(2010)\citenamefont {Otani},
  \citenamefont {Takagi}, \citenamefont {Koshino},\ and\ \citenamefont
  {Okada}}]{RN265}%
  \BibitemOpen
  \bibfield  {author} {\bibinfo {author} {\bibfnamefont {M.}~\bibnamefont
  {Otani}}, \bibinfo {author} {\bibfnamefont {Y.}~\bibnamefont {Takagi}},
  \bibinfo {author} {\bibfnamefont {M.}~\bibnamefont {Koshino}},\ and\ \bibinfo
  {author} {\bibfnamefont {S.}~\bibnamefont {Okada}},\ }\href
  {https://doi.org/10.1063/1.3455069} {\bibfield  {journal} {\bibinfo
  {journal} {Applied Physics Letters}\ }\textbf {\bibinfo {volume} {96}},\
  \bibinfo {pages} {3} (\bibinfo {year} {2010})}\BibitemShut {NoStop}%
\bibitem [{\citenamefont {Wang}\ \emph
  {et~al.}(2016{\natexlab{b}})\citenamefont {Wang}, \citenamefont {Li},
  \citenamefont {Fry},\ and\ \citenamefont {Cheng}}]{RN14}%
  \BibitemOpen
  \bibfield  {author} {\bibinfo {author} {\bibfnamefont {Y.}~\bibnamefont
  {Wang}}, \bibinfo {author} {\bibfnamefont {X.}~\bibnamefont {Li}}, \bibinfo
  {author} {\bibfnamefont {J.}~\bibnamefont {Fry}},\ and\ \bibinfo {author}
  {\bibfnamefont {H.}~\bibnamefont {Cheng}},\ }\bibfield  {journal} {\bibinfo
  {journal} {Physical Review B}\ }\textbf {\bibinfo {volume} {94}},\ \href
  {https://doi.org/10.1103/PhysRevB.94.165428} {10.1103/PhysRevB.94.165428}
  (\bibinfo {year} {2016}{\natexlab{b}})\BibitemShut {NoStop}%
\bibitem [{\citenamefont {Liu}\ \emph {et~al.}(2019{\natexlab{b}})\citenamefont
  {Liu}, \citenamefont {Wang}, \citenamefont {Fry},\ and\ \citenamefont
  {Cheng}}]{RN103}%
  \BibitemOpen
  \bibfield  {author} {\bibinfo {author} {\bibfnamefont {S.~L.}\ \bibnamefont
  {Liu}}, \bibinfo {author} {\bibfnamefont {Y.~P.}\ \bibnamefont {Wang}},
  \bibinfo {author} {\bibfnamefont {J.~N.}\ \bibnamefont {Fry}},\ and\ \bibinfo
  {author} {\bibfnamefont {H.~P.}\ \bibnamefont {Cheng}},\ }\href
  {https://doi.org/10.1016/j.carbon.2018.12.035} {\bibfield  {journal}
  {\bibinfo  {journal} {Carbon}\ }\textbf {\bibinfo {volume} {144}},\ \bibinfo
  {pages} {362} (\bibinfo {year} {2019}{\natexlab{b}})}\BibitemShut {NoStop}%
\bibitem [{\citenamefont {Sohier}\ \emph {et~al.}(2018)\citenamefont {Sohier},
  \citenamefont {Campi}, \citenamefont {Marzari},\ and\ \citenamefont
  {Gibertini}}]{RN154}%
  \BibitemOpen
  \bibfield  {author} {\bibinfo {author} {\bibfnamefont {T.}~\bibnamefont
  {Sohier}}, \bibinfo {author} {\bibfnamefont {D.}~\bibnamefont {Campi}},
  \bibinfo {author} {\bibfnamefont {N.}~\bibnamefont {Marzari}},\ and\ \bibinfo
  {author} {\bibfnamefont {M.}~\bibnamefont {Gibertini}},\ }\href
  {https://doi.org/10.1103/PhysRevMaterials.2.114010} {\bibfield  {journal}
  {\bibinfo  {journal} {Physical Review Materials}\ }\textbf {\bibinfo {volume}
  {2}},\ \bibinfo {pages} {21} (\bibinfo {year} {2018})}\BibitemShut {NoStop}%
\bibitem [{\citenamefont {Otani}\ and\ \citenamefont {Okada}(2011)}]{RN233}%
  \BibitemOpen
  \bibfield  {author} {\bibinfo {author} {\bibfnamefont {M.}~\bibnamefont
  {Otani}}\ and\ \bibinfo {author} {\bibfnamefont {S.}~\bibnamefont {Okada}},\
  }\href {https://doi.org/10.1103/PhysRevB.83.073405} {\bibfield  {journal}
  {\bibinfo  {journal} {Physical Review B}\ }\textbf {\bibinfo {volume} {83}},\
  \bibinfo {pages} {4} (\bibinfo {year} {2011})}\BibitemShut {NoStop}%
\bibitem [{\citenamefont {Brumme}\ \emph {et~al.}(2015)\citenamefont {Brumme},
  \citenamefont {Calandra},\ and\ \citenamefont {Mauri}}]{RN157}%
  \BibitemOpen
  \bibfield  {author} {\bibinfo {author} {\bibfnamefont {T.}~\bibnamefont
  {Brumme}}, \bibinfo {author} {\bibfnamefont {M.}~\bibnamefont {Calandra}},\
  and\ \bibinfo {author} {\bibfnamefont {F.}~\bibnamefont {Mauri}},\ }\href
  {https://doi.org/10.1103/PhysRevB.91.155436} {\bibfield  {journal} {\bibinfo
  {journal} {Physical Review B}\ }\textbf {\bibinfo {volume} {91}},\ \bibinfo
  {pages} {19} (\bibinfo {year} {2015})}\BibitemShut {NoStop}%
\bibitem [{\citenamefont {Novko}(2020)}]{RN160}%
  \BibitemOpen
  \bibfield  {author} {\bibinfo {author} {\bibfnamefont {D.}~\bibnamefont
  {Novko}},\ }\bibfield  {journal} {\bibinfo  {journal} {Communications
  Physics}\ }\textbf {\bibinfo {volume} {3}},\ \href
  {https://doi.org/10.1038/s42005-020-0299-1} {10.1038/s42005-020-0299-1}
  (\bibinfo {year} {2020})\BibitemShut {NoStop}%
\bibitem [{\citenamefont {Li}\ \emph {et~al.}(2019{\natexlab{b}})\citenamefont
  {Li}, \citenamefont {Wang}, \citenamefont {Fry}, \citenamefont {Zhang},\ and\
  \citenamefont {Cheng}}]{RN100}%
  \BibitemOpen
  \bibfield  {author} {\bibinfo {author} {\bibfnamefont {X.~G.}\ \bibnamefont
  {Li}}, \bibinfo {author} {\bibfnamefont {Y.~P.}\ \bibnamefont {Wang}},
  \bibinfo {author} {\bibfnamefont {J.~N.}\ \bibnamefont {Fry}}, \bibinfo
  {author} {\bibfnamefont {X.~G.}\ \bibnamefont {Zhang}},\ and\ \bibinfo
  {author} {\bibfnamefont {H.~P.}\ \bibnamefont {Cheng}},\ }\href
  {https://doi.org/10.1016/j.jpcs.2017.12.005} {\bibfield  {journal} {\bibinfo
  {journal} {Journal of Physics and Chemistry of Solids}\ }\textbf {\bibinfo
  {volume} {128}},\ \bibinfo {pages} {343} (\bibinfo {year}
  {2019}{\natexlab{b}})}\BibitemShut {NoStop}%
\bibitem [{\citenamefont {Piatti}\ \emph {et~al.}(2019)\citenamefont {Piatti},
  \citenamefont {Romanin},\ and\ \citenamefont {Gonneili}}]{RN155}%
  \BibitemOpen
  \bibfield  {author} {\bibinfo {author} {\bibfnamefont {E.}~\bibnamefont
  {Piatti}}, \bibinfo {author} {\bibfnamefont {D.}~\bibnamefont {Romanin}},\
  and\ \bibinfo {author} {\bibfnamefont {R.~S.}\ \bibnamefont {Gonneili}},\
  }\href {https://doi.org/10.1088/1361-648X/aaf981} {\bibfield  {journal}
  {\bibinfo  {journal} {Journal of Physics-Condensed Matter}\ }\textbf
  {\bibinfo {volume} {31}},\ \bibinfo {pages} {8} (\bibinfo {year}
  {2019})}\BibitemShut {NoStop}%
\bibitem [{\citenamefont {Liu}\ \emph {et~al.}(2018{\natexlab{a}})\citenamefont
  {Liu}, \citenamefont {Wang}, \citenamefont {Li}, \citenamefont {Fry},\ and\
  \citenamefont {Cheng}}]{RN101}%
  \BibitemOpen
  \bibfield  {author} {\bibinfo {author} {\bibfnamefont {S.~L.}\ \bibnamefont
  {Liu}}, \bibinfo {author} {\bibfnamefont {Y.~P.}\ \bibnamefont {Wang}},
  \bibinfo {author} {\bibfnamefont {X.~G.}\ \bibnamefont {Li}}, \bibinfo
  {author} {\bibfnamefont {J.~N.}\ \bibnamefont {Fry}},\ and\ \bibinfo {author}
  {\bibfnamefont {H.~P.}\ \bibnamefont {Cheng}},\ }\href
  {https://doi.org/10.1103/PhysRevB.97.035409} {\bibfield  {journal} {\bibinfo
  {journal} {Physical Review B}\ }\textbf {\bibinfo {volume} {97}},\ \bibinfo
  {pages} {8} (\bibinfo {year} {2018}{\natexlab{a}})}\BibitemShut {NoStop}%
\bibitem [{\citenamefont {Wang}\ \emph
  {et~al.}(2018{\natexlab{b}})\citenamefont {Wang}, \citenamefont {Li},
  \citenamefont {Liu}, \citenamefont {Fry},\ and\ \citenamefont
  {Cheng}}]{RN102}%
  \BibitemOpen
  \bibfield  {author} {\bibinfo {author} {\bibfnamefont {Y.~P.}\ \bibnamefont
  {Wang}}, \bibinfo {author} {\bibfnamefont {X.~G.}\ \bibnamefont {Li}},
  \bibinfo {author} {\bibfnamefont {S.~L.}\ \bibnamefont {Liu}}, \bibinfo
  {author} {\bibfnamefont {J.~N.}\ \bibnamefont {Fry}},\ and\ \bibinfo {author}
  {\bibfnamefont {H.~P.}\ \bibnamefont {Cheng}},\ }\href
  {https://doi.org/10.1103/PhysRevB.97.115419} {\bibfield  {journal} {\bibinfo
  {journal} {Physical Review B}\ }\textbf {\bibinfo {volume} {97}},\ \bibinfo
  {pages} {7} (\bibinfo {year} {2018}{\natexlab{b}})}\BibitemShut {NoStop}%
\bibitem [{\citenamefont {Kohn}\ and\ \citenamefont {Sham}(1965)}]{RN925}%
  \BibitemOpen
  \bibfield  {author} {\bibinfo {author} {\bibfnamefont {W.}~\bibnamefont
  {Kohn}}\ and\ \bibinfo {author} {\bibfnamefont {L.~J.}\ \bibnamefont
  {Sham}},\ }\href {<Go to ISI>://A19657000000015} {\bibfield  {journal}
  {\bibinfo  {journal} {Physical Review}\ }\textbf {\bibinfo {volume} {140}},\
  \bibinfo {pages} {1133} (\bibinfo {year} {1965})}\BibitemShut {NoStop}%
\bibitem [{\citenamefont {Perdew}\ \emph {et~al.}(1996)\citenamefont {Perdew},
  \citenamefont {Burke},\ and\ \citenamefont {Ernzerhof}}]{RN8}%
  \BibitemOpen
  \bibfield  {author} {\bibinfo {author} {\bibfnamefont {J.~P.}\ \bibnamefont
  {Perdew}}, \bibinfo {author} {\bibfnamefont {K.}~\bibnamefont {Burke}},\ and\
  \bibinfo {author} {\bibfnamefont {M.}~\bibnamefont {Ernzerhof}},\ }\href
  {https://doi.org/10.1103/PhysRevLett.77.3865} {\bibfield  {journal} {\bibinfo
   {journal} {Physical Review Letters}\ }\textbf {\bibinfo {volume} {77}},\
  \bibinfo {pages} {3865} (\bibinfo {year} {1996})}\BibitemShut {NoStop}%
\bibitem [{\citenamefont {Mostofi}\ \emph {et~al.}(2008)\citenamefont
  {Mostofi}, \citenamefont {Yates}, \citenamefont {Lee}, \citenamefont {Souza},
  \citenamefont {Vanderbilt},\ and\ \citenamefont {Marzari}}]{RN278}%
  \BibitemOpen
  \bibfield  {author} {\bibinfo {author} {\bibfnamefont {A.~A.}\ \bibnamefont
  {Mostofi}}, \bibinfo {author} {\bibfnamefont {J.~R.}\ \bibnamefont {Yates}},
  \bibinfo {author} {\bibfnamefont {Y.~S.}\ \bibnamefont {Lee}}, \bibinfo
  {author} {\bibfnamefont {I.}~\bibnamefont {Souza}}, \bibinfo {author}
  {\bibfnamefont {D.}~\bibnamefont {Vanderbilt}},\ and\ \bibinfo {author}
  {\bibfnamefont {N.}~\bibnamefont {Marzari}},\ }\href
  {https://doi.org/10.1016/j.cpc.2007.11.016} {\bibfield  {journal} {\bibinfo
  {journal} {Computer Physics Communications}\ }\textbf {\bibinfo {volume}
  {178}},\ \bibinfo {pages} {685} (\bibinfo {year} {2008})}\BibitemShut
  {NoStop}%
\bibitem [{\citenamefont {Stengel}\ and\ \citenamefont
  {Spaldin}(2007)}]{RN3053}%
  \BibitemOpen
  \bibfield  {author} {\bibinfo {author} {\bibfnamefont {M.}~\bibnamefont
  {Stengel}}\ and\ \bibinfo {author} {\bibfnamefont {N.~A.}\ \bibnamefont
  {Spaldin}},\ }\bibfield  {journal} {\bibinfo  {journal} {Physical Review B}\
  }\textbf {\bibinfo {volume} {75}},\ \href {https://doi.org/ARTN 205121
  10.1103/PhysRevB.75.205121} {ARTN 205121 10.1103/PhysRevB.75.205121}
  (\bibinfo {year} {2007})\BibitemShut {NoStop}%
\bibitem [{\citenamefont {Wang}\ and\ \citenamefont
  {Cheng}(2015{\natexlab{c}})}]{RN2887}%
  \BibitemOpen
  \bibfield  {author} {\bibinfo {author} {\bibfnamefont {Y.~P.}\ \bibnamefont
  {Wang}}\ and\ \bibinfo {author} {\bibfnamefont {H.~P.}\ \bibnamefont
  {Cheng}},\ }\href {https://doi.org/10.1103/PhysRevB.91.245307} {\bibfield
  {journal} {\bibinfo  {journal} {Physical Review B}\ }\textbf {\bibinfo
  {volume} {91}},\ \bibinfo {pages} {245307} (\bibinfo {year}
  {2015}{\natexlab{c}})}\BibitemShut {NoStop}%
\bibitem [{\citenamefont {Aryasetiawan}\ \emph {et~al.}(2004)\citenamefont
  {Aryasetiawan}, \citenamefont {Imada}, \citenamefont {Georges}, \citenamefont
  {Kotliar}, \citenamefont {Biermann},\ and\ \citenamefont
  {Lichtenstein}}]{PhysRevB_frequ_depend_U}%
  \BibitemOpen
  \bibfield  {author} {\bibinfo {author} {\bibfnamefont {F.}~\bibnamefont
  {Aryasetiawan}}, \bibinfo {author} {\bibfnamefont {M.}~\bibnamefont {Imada}},
  \bibinfo {author} {\bibfnamefont {A.}~\bibnamefont {Georges}}, \bibinfo
  {author} {\bibfnamefont {G.}~\bibnamefont {Kotliar}}, \bibinfo {author}
  {\bibfnamefont {S.}~\bibnamefont {Biermann}},\ and\ \bibinfo {author}
  {\bibfnamefont {A.~I.}\ \bibnamefont {Lichtenstein}},\ }\href
  {https://doi.org/10.1103/PhysRevB.70.195104} {\bibfield  {journal} {\bibinfo
  {journal} {Phys. Rev. B}\ }\textbf {\bibinfo {volume} {70}},\ \bibinfo
  {pages} {195104} (\bibinfo {year} {2004})}\BibitemShut {NoStop}%
\bibitem [{\citenamefont {Aryasetiawan}\ \emph {et~al.}(2006)\citenamefont
  {Aryasetiawan}, \citenamefont {Karlsson}, \citenamefont {Jepsen},\ and\
  \citenamefont {Sch\"onberger}}]{U_from_cRPA}%
  \BibitemOpen
  \bibfield  {author} {\bibinfo {author} {\bibfnamefont {F.}~\bibnamefont
  {Aryasetiawan}}, \bibinfo {author} {\bibfnamefont {K.}~\bibnamefont
  {Karlsson}}, \bibinfo {author} {\bibfnamefont {O.}~\bibnamefont {Jepsen}},\
  and\ \bibinfo {author} {\bibfnamefont {U.}~\bibnamefont {Sch\"onberger}},\
  }\href {https://doi.org/10.1103/PhysRevB.74.125106} {\bibfield  {journal}
  {\bibinfo  {journal} {Phys. Rev. B}\ }\textbf {\bibinfo {volume} {74}},\
  \bibinfo {pages} {125106} (\bibinfo {year} {2006})}\BibitemShut {NoStop}%
\bibitem [{\citenamefont {Petersilka}\ \emph {et~al.}(1996)\citenamefont
  {Petersilka}, \citenamefont {Gossmann},\ and\ \citenamefont
  {Gross}}]{PhysRevLett.76.1212}%
  \BibitemOpen
  \bibfield  {author} {\bibinfo {author} {\bibfnamefont {M.}~\bibnamefont
  {Petersilka}}, \bibinfo {author} {\bibfnamefont {U.~J.}\ \bibnamefont
  {Gossmann}},\ and\ \bibinfo {author} {\bibfnamefont {E.~K.~U.}\ \bibnamefont
  {Gross}},\ }\href {https://doi.org/10.1103/PhysRevLett.76.1212} {\bibfield
  {journal} {\bibinfo  {journal} {Phys. Rev. Lett.}\ }\textbf {\bibinfo
  {volume} {76}},\ \bibinfo {pages} {1212} (\bibinfo {year}
  {1996})}\BibitemShut {NoStop}%
\bibitem [{\citenamefont {Craciun}\ \emph {et~al.}(2009)\citenamefont
  {Craciun}, \citenamefont {Russo}, \citenamefont {Yamamoto}, \citenamefont
  {Oostinga}, \citenamefont {Morpurgo},\ and\ \citenamefont {Tarucha}}]{RN312}%
  \BibitemOpen
  \bibfield  {author} {\bibinfo {author} {\bibfnamefont {M.~F.}\ \bibnamefont
  {Craciun}}, \bibinfo {author} {\bibfnamefont {S.}~\bibnamefont {Russo}},
  \bibinfo {author} {\bibfnamefont {M.}~\bibnamefont {Yamamoto}}, \bibinfo
  {author} {\bibfnamefont {J.~B.}\ \bibnamefont {Oostinga}}, \bibinfo {author}
  {\bibfnamefont {A.~F.}\ \bibnamefont {Morpurgo}},\ and\ \bibinfo {author}
  {\bibfnamefont {S.}~\bibnamefont {Tarucha}},\ }\href
  {https://doi.org/10.1038/nnano.2009.89} {\bibfield  {journal} {\bibinfo
  {journal} {Nat Nanotechnol}\ }\textbf {\bibinfo {volume} {4}},\ \bibinfo
  {pages} {383} (\bibinfo {year} {2009})}\BibitemShut {NoStop}%
\bibitem [{\citenamefont {Yankowitz}\ \emph {et~al.}(2013)\citenamefont
  {Yankowitz}, \citenamefont {Wang}, \citenamefont {Lau},\ and\ \citenamefont
  {LeRoy}}]{RN2610}%
  \BibitemOpen
  \bibfield  {author} {\bibinfo {author} {\bibfnamefont {M.}~\bibnamefont
  {Yankowitz}}, \bibinfo {author} {\bibfnamefont {F.}~\bibnamefont {Wang}},
  \bibinfo {author} {\bibfnamefont {C.~N.}\ \bibnamefont {Lau}},\ and\ \bibinfo
  {author} {\bibfnamefont {B.~J.}\ \bibnamefont {LeRoy}},\ }\href
  {https://doi.org/10.1103/PhysRevB.87.165102} {\bibfield  {journal} {\bibinfo
  {journal} {Physical Review B}\ }\textbf {\bibinfo {volume} {87}},\ \bibinfo
  {pages} {165102} (\bibinfo {year} {2013})}\BibitemShut {NoStop}%
\bibitem [{\citenamefont {Seo}\ \emph {et~al.}(2013)\citenamefont {Seo},
  \citenamefont {Min}, \citenamefont {Lee},\ and\ \citenamefont
  {Lee}}]{RN3056}%
  \BibitemOpen
  \bibfield  {author} {\bibinfo {author} {\bibfnamefont {S.}~\bibnamefont
  {Seo}}, \bibinfo {author} {\bibfnamefont {M.}~\bibnamefont {Min}}, \bibinfo
  {author} {\bibfnamefont {S.~M.}\ \bibnamefont {Lee}},\ and\ \bibinfo {author}
  {\bibfnamefont {H.}~\bibnamefont {Lee}},\ }\href
  {https://doi.org/10.1038/ncomms2937} {\bibfield  {journal} {\bibinfo
  {journal} {Nature Communications}\ }\textbf {\bibinfo {volume} {4}},\
  \bibinfo {pages} {7} (\bibinfo {year} {2013})}\BibitemShut {NoStop}%
\bibitem [{\citenamefont {Feng}\ \emph {et~al.}(2001)\citenamefont {Feng},
  \citenamefont {Zhang}, \citenamefont {Jin}, \citenamefont {Song},
  \citenamefont {Xie}, \citenamefont {Qu}, \citenamefont {Jiang},\ and\
  \citenamefont {Zhu}}]{RN2364}%
  \BibitemOpen
  \bibfield  {author} {\bibinfo {author} {\bibfnamefont {C.~L.}\ \bibnamefont
  {Feng}}, \bibinfo {author} {\bibfnamefont {Y.~J.}\ \bibnamefont {Zhang}},
  \bibinfo {author} {\bibfnamefont {J.}~\bibnamefont {Jin}}, \bibinfo {author}
  {\bibfnamefont {Y.~L.}\ \bibnamefont {Song}}, \bibinfo {author}
  {\bibfnamefont {L.~Y.}\ \bibnamefont {Xie}}, \bibinfo {author} {\bibfnamefont
  {G.~R.}\ \bibnamefont {Qu}}, \bibinfo {author} {\bibfnamefont
  {L.}~\bibnamefont {Jiang}},\ and\ \bibinfo {author} {\bibfnamefont {D.~B.}\
  \bibnamefont {Zhu}},\ }\href {https://doi.org/10.1021/la010071r} {\bibfield
  {journal} {\bibinfo  {journal} {Langmuir}\ }\textbf {\bibinfo {volume}
  {17}},\ \bibinfo {pages} {4593} (\bibinfo {year} {2001})}\BibitemShut
  {NoStop}%
\bibitem [{\citenamefont {Hugel}\ \emph {et~al.}(2002)\citenamefont {Hugel},
  \citenamefont {Holland}, \citenamefont {Cattani}, \citenamefont {Moroder},
  \citenamefont {Seitz},\ and\ \citenamefont {Gaub}}]{RN1616}%
  \BibitemOpen
  \bibfield  {author} {\bibinfo {author} {\bibfnamefont {T.}~\bibnamefont
  {Hugel}}, \bibinfo {author} {\bibfnamefont {N.~B.}\ \bibnamefont {Holland}},
  \bibinfo {author} {\bibfnamefont {A.}~\bibnamefont {Cattani}}, \bibinfo
  {author} {\bibfnamefont {L.}~\bibnamefont {Moroder}}, \bibinfo {author}
  {\bibfnamefont {M.}~\bibnamefont {Seitz}},\ and\ \bibinfo {author}
  {\bibfnamefont {H.~E.}\ \bibnamefont {Gaub}},\ }\href
  {https://doi.org/10.1126/science.1069856} {\bibfield  {journal} {\bibinfo
  {journal} {Science}\ }\textbf {\bibinfo {volume} {296}},\ \bibinfo {pages}
  {1103} (\bibinfo {year} {2002})}\BibitemShut {NoStop}%
\bibitem [{\citenamefont {Zhang}\ \emph {et~al.}(2004)\citenamefont {Zhang},
  \citenamefont {Du}, \citenamefont {Cheng}, \citenamefont {Zhang},
  \citenamefont {Roitberg},\ and\ \citenamefont {Krause}}]{RN484}%
  \BibitemOpen
  \bibfield  {author} {\bibinfo {author} {\bibfnamefont {C.}~\bibnamefont
  {Zhang}}, \bibinfo {author} {\bibfnamefont {M.~H.}\ \bibnamefont {Du}},
  \bibinfo {author} {\bibfnamefont {H.~P.}\ \bibnamefont {Cheng}}, \bibinfo
  {author} {\bibfnamefont {X.~G.}\ \bibnamefont {Zhang}}, \bibinfo {author}
  {\bibfnamefont {A.~E.}\ \bibnamefont {Roitberg}},\ and\ \bibinfo {author}
  {\bibfnamefont {J.~L.}\ \bibnamefont {Krause}},\ }\bibfield  {journal}
  {\bibinfo  {journal} {Physical Review Letters}\ }\textbf {\bibinfo {volume}
  {92}},\ \href {https://doi.org/158301 10.1103/PhysRevLett.92.158301} {158301
  10.1103/PhysRevLett.92.158301} (\bibinfo {year} {2004})\BibitemShut {NoStop}%
\bibitem [{\citenamefont {Zhang}\ \emph {et~al.}(2006)\citenamefont {Zhang},
  \citenamefont {He}, \citenamefont {Cheng}, \citenamefont {Xue}, \citenamefont
  {Ratner}, \citenamefont {Zhang},\ and\ \citenamefont {Krstic}}]{RN1807}%
  \BibitemOpen
  \bibfield  {author} {\bibinfo {author} {\bibfnamefont {C.}~\bibnamefont
  {Zhang}}, \bibinfo {author} {\bibfnamefont {Y.}~\bibnamefont {He}}, \bibinfo
  {author} {\bibfnamefont {H.~P.}\ \bibnamefont {Cheng}}, \bibinfo {author}
  {\bibfnamefont {Y.~Q.}\ \bibnamefont {Xue}}, \bibinfo {author} {\bibfnamefont
  {M.~A.}\ \bibnamefont {Ratner}}, \bibinfo {author} {\bibfnamefont {X.~G.}\
  \bibnamefont {Zhang}},\ and\ \bibinfo {author} {\bibfnamefont
  {P.}~\bibnamefont {Krstic}},\ }\href {https://doi.org/125445
  10.1103/PhysRevB.73.125445} {\bibfield  {journal} {\bibinfo  {journal}
  {Physical Review B}\ }\textbf {\bibinfo {volume} {73}},\ \bibinfo {pages}
  {125445} (\bibinfo {year} {2006})}\BibitemShut {NoStop}%
\bibitem [{\citenamefont {Wang}\ and\ \citenamefont {Cheng}(2012)}]{RN2606}%
  \BibitemOpen
  \bibfield  {author} {\bibinfo {author} {\bibfnamefont {Y.}~\bibnamefont
  {Wang}}\ and\ \bibinfo {author} {\bibfnamefont {H.-P.}\ \bibnamefont
  {Cheng}},\ }\href {https://doi.org/10.1103/PhysRevB.86.035444} {\bibfield
  {journal} {\bibinfo  {journal} {Physical Review B}\ }\textbf {\bibinfo
  {volume} {86}},\ \bibinfo {pages} {035444} (\bibinfo {year}
  {2012})}\BibitemShut {NoStop}%
\bibitem [{\citenamefont {Kim}\ \emph {et~al.}(2012)\citenamefont {Kim},
  \citenamefont {Garcia-Lekue}, \citenamefont {Sysoiev}, \citenamefont
  {Frederiksen}, \citenamefont {Groth},\ and\ \citenamefont {Scheer}}]{RN3057}%
  \BibitemOpen
  \bibfield  {author} {\bibinfo {author} {\bibfnamefont {Y.}~\bibnamefont
  {Kim}}, \bibinfo {author} {\bibfnamefont {A.}~\bibnamefont {Garcia-Lekue}},
  \bibinfo {author} {\bibfnamefont {D.}~\bibnamefont {Sysoiev}}, \bibinfo
  {author} {\bibfnamefont {T.}~\bibnamefont {Frederiksen}}, \bibinfo {author}
  {\bibfnamefont {U.}~\bibnamefont {Groth}},\ and\ \bibinfo {author}
  {\bibfnamefont {E.}~\bibnamefont {Scheer}},\ }\href {https://doi.org/ARTN
  226801 10.1103/PhysRevLett.109.226801} {\bibfield  {journal} {\bibinfo
  {journal} {Physical Review Letters}\ }\textbf {\bibinfo {volume} {109}},\
  \bibinfo {pages} {226801} (\bibinfo {year} {2012})}\BibitemShut {NoStop}%
\bibitem [{\citenamefont {Chu}\ \emph {et~al.}(2014)\citenamefont {Chu},
  \citenamefont {Trinastic}, \citenamefont {Wang},\ and\ \citenamefont
  {Cheng}}]{RN2194}%
  \BibitemOpen
  \bibfield  {author} {\bibinfo {author} {\bibfnamefont {I.~H.}\ \bibnamefont
  {Chu}}, \bibinfo {author} {\bibfnamefont {J.}~\bibnamefont {Trinastic}},
  \bibinfo {author} {\bibfnamefont {L.~W.}\ \bibnamefont {Wang}},\ and\
  \bibinfo {author} {\bibfnamefont {H.~P.}\ \bibnamefont {Cheng}},\ }\href
  {https://doi.org/10.1103/PhysRevB.89.115415} {\bibfield  {journal} {\bibinfo
  {journal} {Physical Review B}\ }\textbf {\bibinfo {volume} {89}},\ \bibinfo
  {pages} {115415} (\bibinfo {year} {2014})}\BibitemShut {NoStop}%
\bibitem [{\citenamefont {Trinastic}\ and\ \citenamefont
  {Cheng}(2014)}]{RN1966}%
  \BibitemOpen
  \bibfield  {author} {\bibinfo {author} {\bibfnamefont {J.~P.}\ \bibnamefont
  {Trinastic}}\ and\ \bibinfo {author} {\bibfnamefont {H.-P.}\ \bibnamefont
  {Cheng}},\ }\bibfield  {journal} {\bibinfo  {journal} {Physical Review B}\
  }\textbf {\bibinfo {volume} {89}},\ \href
  {https://doi.org/10.1103/PhysRevB.89.245447} {10.1103/PhysRevB.89.245447}
  (\bibinfo {year} {2014})\BibitemShut {NoStop}%
\bibitem [{\citenamefont {Bhimanapati}\ \emph {et~al.}(2015)\citenamefont
  {Bhimanapati}, \citenamefont {Lin}, \citenamefont {Meunier}, \citenamefont
  {Jung}, \citenamefont {Cha}, \citenamefont {Das}, \citenamefont {Xiao},
  \citenamefont {Son}, \citenamefont {Strano}, \citenamefont {Cooper},
  \citenamefont {Liang}, \citenamefont {Louie}, \citenamefont {Ringe},
  \citenamefont {Zhou}, \citenamefont {Kim}, \citenamefont {Naik},
  \citenamefont {Sumpter}, \citenamefont {Terrones}, \citenamefont {Xia},
  \citenamefont {Wang}, \citenamefont {Zhu}, \citenamefont {Akinwande},
  \citenamefont {Alem}, \citenamefont {Schuller}, \citenamefont {Schaak},
  \citenamefont {Terrones},\ and\ \citenamefont {Robinson}}]{RN3063}%
  \BibitemOpen
  \bibfield  {author} {\bibinfo {author} {\bibfnamefont {G.~R.}\ \bibnamefont
  {Bhimanapati}}, \bibinfo {author} {\bibfnamefont {Z.}~\bibnamefont {Lin}},
  \bibinfo {author} {\bibfnamefont {V.}~\bibnamefont {Meunier}}, \bibinfo
  {author} {\bibfnamefont {Y.}~\bibnamefont {Jung}}, \bibinfo {author}
  {\bibfnamefont {J.}~\bibnamefont {Cha}}, \bibinfo {author} {\bibfnamefont
  {S.}~\bibnamefont {Das}}, \bibinfo {author} {\bibfnamefont {D.}~\bibnamefont
  {Xiao}}, \bibinfo {author} {\bibfnamefont {Y.}~\bibnamefont {Son}}, \bibinfo
  {author} {\bibfnamefont {M.~S.}\ \bibnamefont {Strano}}, \bibinfo {author}
  {\bibfnamefont {V.~R.}\ \bibnamefont {Cooper}}, \bibinfo {author}
  {\bibfnamefont {L.~B.}\ \bibnamefont {Liang}}, \bibinfo {author}
  {\bibfnamefont {S.~G.}\ \bibnamefont {Louie}}, \bibinfo {author}
  {\bibfnamefont {E.}~\bibnamefont {Ringe}}, \bibinfo {author} {\bibfnamefont
  {W.}~\bibnamefont {Zhou}}, \bibinfo {author} {\bibfnamefont {S.~S.}\
  \bibnamefont {Kim}}, \bibinfo {author} {\bibfnamefont {R.~R.}\ \bibnamefont
  {Naik}}, \bibinfo {author} {\bibfnamefont {B.~G.}\ \bibnamefont {Sumpter}},
  \bibinfo {author} {\bibfnamefont {H.}~\bibnamefont {Terrones}}, \bibinfo
  {author} {\bibfnamefont {F.~N.}\ \bibnamefont {Xia}}, \bibinfo {author}
  {\bibfnamefont {Y.~L.}\ \bibnamefont {Wang}}, \bibinfo {author}
  {\bibfnamefont {J.}~\bibnamefont {Zhu}}, \bibinfo {author} {\bibfnamefont
  {D.}~\bibnamefont {Akinwande}}, \bibinfo {author} {\bibfnamefont
  {N.}~\bibnamefont {Alem}}, \bibinfo {author} {\bibfnamefont {J.~A.}\
  \bibnamefont {Schuller}}, \bibinfo {author} {\bibfnamefont {R.~E.}\
  \bibnamefont {Schaak}}, \bibinfo {author} {\bibfnamefont {M.}~\bibnamefont
  {Terrones}},\ and\ \bibinfo {author} {\bibfnamefont {J.~A.}\ \bibnamefont
  {Robinson}},\ }\href {https://doi.org/10.1021/acsnano.5b05556} {\bibfield
  {journal} {\bibinfo  {journal} {Acs Nano}\ }\textbf {\bibinfo {volume} {9}},\
  \bibinfo {pages} {11509} (\bibinfo {year} {2015})}\BibitemShut {NoStop}%
\bibitem [{\citenamefont {Chhowalla}\ \emph {et~al.}(2013)\citenamefont
  {Chhowalla}, \citenamefont {Shin}, \citenamefont {Eda}, \citenamefont {Li},
  \citenamefont {Loh},\ and\ \citenamefont {Zhang}}]{RN3064}%
  \BibitemOpen
  \bibfield  {author} {\bibinfo {author} {\bibfnamefont {M.}~\bibnamefont
  {Chhowalla}}, \bibinfo {author} {\bibfnamefont {H.~S.}\ \bibnamefont {Shin}},
  \bibinfo {author} {\bibfnamefont {G.}~\bibnamefont {Eda}}, \bibinfo {author}
  {\bibfnamefont {L.~J.}\ \bibnamefont {Li}}, \bibinfo {author} {\bibfnamefont
  {K.~P.}\ \bibnamefont {Loh}},\ and\ \bibinfo {author} {\bibfnamefont
  {H.}~\bibnamefont {Zhang}},\ }\href {https://doi.org/10.1038/Nchem.1589}
  {\bibfield  {journal} {\bibinfo  {journal} {Nature Chemistry}\ }\textbf
  {\bibinfo {volume} {5}},\ \bibinfo {pages} {263} (\bibinfo {year}
  {2013})}\BibitemShut {NoStop}%
\bibitem [{\citenamefont {Jariwala}\ \emph
  {et~al.}(2014{\natexlab{b}})\citenamefont {Jariwala}, \citenamefont
  {Sangwan}, \citenamefont {Lauhon}, \citenamefont {Marks},\ and\ \citenamefont
  {Hersam}}]{RN3065}%
  \BibitemOpen
  \bibfield  {author} {\bibinfo {author} {\bibfnamefont {D.}~\bibnamefont
  {Jariwala}}, \bibinfo {author} {\bibfnamefont {V.~K.}\ \bibnamefont
  {Sangwan}}, \bibinfo {author} {\bibfnamefont {L.~J.}\ \bibnamefont {Lauhon}},
  \bibinfo {author} {\bibfnamefont {T.~J.}\ \bibnamefont {Marks}},\ and\
  \bibinfo {author} {\bibfnamefont {M.~C.}\ \bibnamefont {Hersam}},\ }\href
  {https://doi.org/10.1021/nn500064s} {\bibfield  {journal} {\bibinfo
  {journal} {Acs Nano}\ }\textbf {\bibinfo {volume} {8}},\ \bibinfo {pages}
  {1102} (\bibinfo {year} {2014}{\natexlab{b}})}\BibitemShut {NoStop}%
\bibitem [{\citenamefont {Splendiani}\ \emph {et~al.}(2010)\citenamefont
  {Splendiani}, \citenamefont {Sun}, \citenamefont {Zhang}, \citenamefont {Li},
  \citenamefont {Kim}, \citenamefont {Chim}, \citenamefont {Galli},\ and\
  \citenamefont {Wang}}]{RN3066}%
  \BibitemOpen
  \bibfield  {author} {\bibinfo {author} {\bibfnamefont {A.}~\bibnamefont
  {Splendiani}}, \bibinfo {author} {\bibfnamefont {L.}~\bibnamefont {Sun}},
  \bibinfo {author} {\bibfnamefont {Y.~B.}\ \bibnamefont {Zhang}}, \bibinfo
  {author} {\bibfnamefont {T.~S.}\ \bibnamefont {Li}}, \bibinfo {author}
  {\bibfnamefont {J.}~\bibnamefont {Kim}}, \bibinfo {author} {\bibfnamefont
  {C.~Y.}\ \bibnamefont {Chim}}, \bibinfo {author} {\bibfnamefont
  {G.}~\bibnamefont {Galli}},\ and\ \bibinfo {author} {\bibfnamefont
  {F.}~\bibnamefont {Wang}},\ }\href {https://doi.org/10.1021/nl903868w}
  {\bibfield  {journal} {\bibinfo  {journal} {Nano Letters}\ }\textbf {\bibinfo
  {volume} {10}},\ \bibinfo {pages} {1271} (\bibinfo {year}
  {2010})}\BibitemShut {NoStop}%
\bibitem [{\citenamefont {Georgiou}\ \emph {et~al.}(2013)\citenamefont
  {Georgiou}, \citenamefont {Jalil}, \citenamefont {Belle}, \citenamefont
  {Britnell}, \citenamefont {Gorbachev}, \citenamefont {Morozov}, \citenamefont
  {Kim}, \citenamefont {Gholinia}, \citenamefont {Haigh}, \citenamefont
  {Makarovsky}, \citenamefont {Eaves}, \citenamefont {Ponomarenko},
  \citenamefont {Geim}, \citenamefont {Novoselov},\ and\ \citenamefont
  {Mishchenko}}]{RN2452}%
  \BibitemOpen
  \bibfield  {author} {\bibinfo {author} {\bibfnamefont {T.}~\bibnamefont
  {Georgiou}}, \bibinfo {author} {\bibfnamefont {R.}~\bibnamefont {Jalil}},
  \bibinfo {author} {\bibfnamefont {B.~D.}\ \bibnamefont {Belle}}, \bibinfo
  {author} {\bibfnamefont {L.}~\bibnamefont {Britnell}}, \bibinfo {author}
  {\bibfnamefont {R.~V.}\ \bibnamefont {Gorbachev}}, \bibinfo {author}
  {\bibfnamefont {S.~V.}\ \bibnamefont {Morozov}}, \bibinfo {author}
  {\bibfnamefont {Y.-J.}\ \bibnamefont {Kim}}, \bibinfo {author} {\bibfnamefont
  {A.}~\bibnamefont {Gholinia}}, \bibinfo {author} {\bibfnamefont {S.~J.}\
  \bibnamefont {Haigh}}, \bibinfo {author} {\bibfnamefont {O.}~\bibnamefont
  {Makarovsky}}, \bibinfo {author} {\bibfnamefont {L.}~\bibnamefont {Eaves}},
  \bibinfo {author} {\bibfnamefont {L.~A.}\ \bibnamefont {Ponomarenko}},
  \bibinfo {author} {\bibfnamefont {A.~K.}\ \bibnamefont {Geim}}, \bibinfo
  {author} {\bibfnamefont {K.~S.}\ \bibnamefont {Novoselov}},\ and\ \bibinfo
  {author} {\bibfnamefont {A.}~\bibnamefont {Mishchenko}},\ }\href
  {https://doi.org/10.1038/nnano.2012.224} {\bibfield  {journal} {\bibinfo
  {journal} {Nature Nanotechnology}\ }\textbf {\bibinfo {volume} {8}},\
  \bibinfo {pages} {100} (\bibinfo {year} {2013})}\BibitemShut {NoStop}%
\bibitem [{\citenamefont {Kuc}\ \emph {et~al.}(2011)\citenamefont {Kuc},
  \citenamefont {Zibouche},\ and\ \citenamefont {Heine}}]{RN3067}%
  \BibitemOpen
  \bibfield  {author} {\bibinfo {author} {\bibfnamefont {A.}~\bibnamefont
  {Kuc}}, \bibinfo {author} {\bibfnamefont {N.}~\bibnamefont {Zibouche}},\ and\
  \bibinfo {author} {\bibfnamefont {T.}~\bibnamefont {Heine}},\ }\bibfield
  {journal} {\bibinfo  {journal} {Physical Review B}\ }\textbf {\bibinfo
  {volume} {83}},\ \href {https://doi.org/ARTN 245213
  10.1103/PhysRevB.83.245213} {ARTN 245213 10.1103/PhysRevB.83.245213}
  (\bibinfo {year} {2011})\BibitemShut {NoStop}%
\bibitem [{\citenamefont {Abel}\ \emph {et~al.}(2011)\citenamefont {Abel},
  \citenamefont {Clair}, \citenamefont {Ourdjini}, \citenamefont {Mossoyan},\
  and\ \citenamefont {Porte}}]{RN2124}%
  \BibitemOpen
  \bibfield  {author} {\bibinfo {author} {\bibfnamefont {M.}~\bibnamefont
  {Abel}}, \bibinfo {author} {\bibfnamefont {S.}~\bibnamefont {Clair}},
  \bibinfo {author} {\bibfnamefont {O.}~\bibnamefont {Ourdjini}}, \bibinfo
  {author} {\bibfnamefont {M.}~\bibnamefont {Mossoyan}},\ and\ \bibinfo
  {author} {\bibfnamefont {L.}~\bibnamefont {Porte}},\ }\href
  {https://doi.org/10.1021/ja108628r} {\bibfield  {journal} {\bibinfo
  {journal} {Journal of the American Chemical Society}\ }\textbf {\bibinfo
  {volume} {133}},\ \bibinfo {pages} {1203} (\bibinfo {year}
  {2011})}\BibitemShut {NoStop}%
\bibitem [{\citenamefont {Koudia}\ and\ \citenamefont {Abel}(2014)}]{RN2117}%
  \BibitemOpen
  \bibfield  {author} {\bibinfo {author} {\bibfnamefont {M.}~\bibnamefont
  {Koudia}}\ and\ \bibinfo {author} {\bibfnamefont {M.}~\bibnamefont {Abel}},\
  }\href {https://doi.org/10.1039/c4cc02792b} {\bibfield  {journal} {\bibinfo
  {journal} {Chemical Communications}\ }\textbf {\bibinfo {volume} {50}},\
  \bibinfo {pages} {8565} (\bibinfo {year} {2014})}\BibitemShut {NoStop}%
\bibitem [{\citenamefont {Koudia}\ \emph {et~al.}(2017)\citenamefont {Koudia},
  \citenamefont {Nardi}, \citenamefont {Siri},\ and\ \citenamefont
  {Abel}}]{RN3059}%
  \BibitemOpen
  \bibfield  {author} {\bibinfo {author} {\bibfnamefont {M.}~\bibnamefont
  {Koudia}}, \bibinfo {author} {\bibfnamefont {E.}~\bibnamefont {Nardi}},
  \bibinfo {author} {\bibfnamefont {O.}~\bibnamefont {Siri}},\ and\ \bibinfo
  {author} {\bibfnamefont {M.}~\bibnamefont {Abel}},\ }\href
  {https://doi.org/10.1007/s12274-016-1352-y} {\bibfield  {journal} {\bibinfo
  {journal} {Nano Research}\ }\textbf {\bibinfo {volume} {10}},\ \bibinfo
  {pages} {933} (\bibinfo {year} {2017})}\BibitemShut {NoStop}%
\bibitem [{\citenamefont {Liu}\ \emph {et~al.}(2018{\natexlab{b}})\citenamefont
  {Liu}, \citenamefont {Wang}, \citenamefont {Li}, \citenamefont {Fry},\ and\
  \citenamefont {Cheng}}]{RN3062}%
  \BibitemOpen
  \bibfield  {author} {\bibinfo {author} {\bibfnamefont {S.~L.}\ \bibnamefont
  {Liu}}, \bibinfo {author} {\bibfnamefont {Y.~P.}\ \bibnamefont {Wang}},
  \bibinfo {author} {\bibfnamefont {X.~G.}\ \bibnamefont {Li}}, \bibinfo
  {author} {\bibfnamefont {J.~N.}\ \bibnamefont {Fry}},\ and\ \bibinfo {author}
  {\bibfnamefont {H.~P.}\ \bibnamefont {Cheng}},\ }\href {https://doi.org/ARTN
  035409 10.1103/PhysRevB.97.035409} {\bibfield  {journal} {\bibinfo  {journal}
  {Physical Review B}\ }\textbf {\bibinfo {volume} {97}},\ \bibinfo {pages}
  {035409} (\bibinfo {year} {2018}{\natexlab{b}})}\BibitemShut {NoStop}%
\bibitem [{\citenamefont {Iijima}\ and\ \citenamefont
  {Ichihashi}(1993)}]{RN1232}%
  \BibitemOpen
  \bibfield  {author} {\bibinfo {author} {\bibfnamefont {S.}~\bibnamefont
  {Iijima}}\ and\ \bibinfo {author} {\bibfnamefont {T.}~\bibnamefont
  {Ichihashi}},\ }\href {https://doi.org/10.1038/363603a0} {\bibfield
  {journal} {\bibinfo  {journal} {Nature}\ }\textbf {\bibinfo {volume} {363}},\
  \bibinfo {pages} {603} (\bibinfo {year} {1993})}\BibitemShut {NoStop}%
\bibitem [{\citenamefont {Geim}\ and\ \citenamefont
  {Novoselov}(2007)}]{RN1336}%
  \BibitemOpen
  \bibfield  {author} {\bibinfo {author} {\bibfnamefont {A.~K.}\ \bibnamefont
  {Geim}}\ and\ \bibinfo {author} {\bibfnamefont {K.~S.}\ \bibnamefont
  {Novoselov}},\ }\href {https://doi.org/Doi 10.1038/Nmat1849} {\bibfield
  {journal} {\bibinfo  {journal} {Nature Materials}\ }\textbf {\bibinfo
  {volume} {6}},\ \bibinfo {pages} {183} (\bibinfo {year} {2007})}\BibitemShut
  {NoStop}%
\bibitem [{\citenamefont {Lee}\ \emph {et~al.}(2016)\citenamefont {Lee},
  \citenamefont {Lee}, \citenamefont {Ryoo}, \citenamefont {Kang},
  \citenamefont {Kim}, \citenamefont {Kim}, \citenamefont {Park}, \citenamefont
  {Park},\ and\ \citenamefont {Cheong}}]{RN250}%
  \BibitemOpen
  \bibfield  {author} {\bibinfo {author} {\bibfnamefont {J.~U.}\ \bibnamefont
  {Lee}}, \bibinfo {author} {\bibfnamefont {S.}~\bibnamefont {Lee}}, \bibinfo
  {author} {\bibfnamefont {J.~H.}\ \bibnamefont {Ryoo}}, \bibinfo {author}
  {\bibfnamefont {S.}~\bibnamefont {Kang}}, \bibinfo {author} {\bibfnamefont
  {T.~Y.}\ \bibnamefont {Kim}}, \bibinfo {author} {\bibfnamefont
  {P.}~\bibnamefont {Kim}}, \bibinfo {author} {\bibfnamefont {C.~H.}\
  \bibnamefont {Park}}, \bibinfo {author} {\bibfnamefont {J.~G.}\ \bibnamefont
  {Park}},\ and\ \bibinfo {author} {\bibfnamefont {H.}~\bibnamefont {Cheong}},\
  }\href {https://doi.org/10.1021/acs.nanolett.6b03052} {\bibfield  {journal}
  {\bibinfo  {journal} {Nano Letters}\ }\textbf {\bibinfo {volume} {16}},\
  \bibinfo {pages} {7433} (\bibinfo {year} {2016})}\BibitemShut {NoStop}%
\bibitem [{\citenamefont {Gong}\ \emph {et~al.}(2017)\citenamefont {Gong},
  \citenamefont {Li}, \citenamefont {Li}, \citenamefont {Ji}, \citenamefont
  {Stern}, \citenamefont {Xia}, \citenamefont {Cao}, \citenamefont {Bao},
  \citenamefont {Wang}, \citenamefont {Wang}, \citenamefont {Qiu},
  \citenamefont {Cava}, \citenamefont {Louie}, \citenamefont {Xia},\ and\
  \citenamefont {Zhang}}]{RN246}%
  \BibitemOpen
  \bibfield  {author} {\bibinfo {author} {\bibfnamefont {C.}~\bibnamefont
  {Gong}}, \bibinfo {author} {\bibfnamefont {L.}~\bibnamefont {Li}}, \bibinfo
  {author} {\bibfnamefont {Z.~L.}\ \bibnamefont {Li}}, \bibinfo {author}
  {\bibfnamefont {H.~W.}\ \bibnamefont {Ji}}, \bibinfo {author} {\bibfnamefont
  {A.}~\bibnamefont {Stern}}, \bibinfo {author} {\bibfnamefont
  {Y.}~\bibnamefont {Xia}}, \bibinfo {author} {\bibfnamefont {T.}~\bibnamefont
  {Cao}}, \bibinfo {author} {\bibfnamefont {W.}~\bibnamefont {Bao}}, \bibinfo
  {author} {\bibfnamefont {C.~Z.}\ \bibnamefont {Wang}}, \bibinfo {author}
  {\bibfnamefont {Y.~A.}\ \bibnamefont {Wang}}, \bibinfo {author}
  {\bibfnamefont {Z.~Q.}\ \bibnamefont {Qiu}}, \bibinfo {author} {\bibfnamefont
  {R.~J.}\ \bibnamefont {Cava}}, \bibinfo {author} {\bibfnamefont {S.~G.}\
  \bibnamefont {Louie}}, \bibinfo {author} {\bibfnamefont {J.}~\bibnamefont
  {Xia}},\ and\ \bibinfo {author} {\bibfnamefont {X.}~\bibnamefont {Zhang}},\
  }\href {https://doi.org/10.1038/nature22060} {\bibfield  {journal} {\bibinfo
  {journal} {Nature}\ }\textbf {\bibinfo {volume} {546}},\ \bibinfo {pages}
  {265} (\bibinfo {year} {2017})}\BibitemShut {NoStop}%
\bibitem [{\citenamefont {O'Hara}\ \emph {et~al.}(2018)\citenamefont {O'Hara},
  \citenamefont {Zhu}, \citenamefont {Trout}, \citenamefont {Ahmed},
  \citenamefont {Luo}, \citenamefont {Lee}, \citenamefont {Brenner},
  \citenamefont {Rajan}, \citenamefont {Gupta}, \citenamefont {McComb},\ and\
  \citenamefont {Kawakami}}]{RN248}%
  \BibitemOpen
  \bibfield  {author} {\bibinfo {author} {\bibfnamefont {D.~J.}\ \bibnamefont
  {O'Hara}}, \bibinfo {author} {\bibfnamefont {T.~C.}\ \bibnamefont {Zhu}},
  \bibinfo {author} {\bibfnamefont {A.~H.}\ \bibnamefont {Trout}}, \bibinfo
  {author} {\bibfnamefont {A.~S.}\ \bibnamefont {Ahmed}}, \bibinfo {author}
  {\bibfnamefont {Y.~K.}\ \bibnamefont {Luo}}, \bibinfo {author} {\bibfnamefont
  {C.~H.}\ \bibnamefont {Lee}}, \bibinfo {author} {\bibfnamefont {M.~R.}\
  \bibnamefont {Brenner}}, \bibinfo {author} {\bibfnamefont {S.}~\bibnamefont
  {Rajan}}, \bibinfo {author} {\bibfnamefont {J.~A.}\ \bibnamefont {Gupta}},
  \bibinfo {author} {\bibfnamefont {D.~W.}\ \bibnamefont {McComb}},\ and\
  \bibinfo {author} {\bibfnamefont {R.~K.}\ \bibnamefont {Kawakami}},\ }\href
  {https://doi.org/10.1021/acs.nanolett.8b00683} {\bibfield  {journal}
  {\bibinfo  {journal} {Nano Letters}\ }\textbf {\bibinfo {volume} {18}},\
  \bibinfo {pages} {3125} (\bibinfo {year} {2018})}\BibitemShut {NoStop}%
\bibitem [{\citenamefont {Huang}\ \emph {et~al.}(2017)\citenamefont {Huang},
  \citenamefont {Clark}, \citenamefont {Navarro-Moratalla}, \citenamefont
  {Klein}, \citenamefont {Cheng}, \citenamefont {Seyler}, \citenamefont
  {Zhong}, \citenamefont {Schmidgall}, \citenamefont {McGuire}, \citenamefont
  {Cobden}, \citenamefont {Yao}, \citenamefont {Xiao}, \citenamefont
  {Jarillo-Herrero},\ and\ \citenamefont {Xu}}]{RN247}%
  \BibitemOpen
  \bibfield  {author} {\bibinfo {author} {\bibfnamefont {B.}~\bibnamefont
  {Huang}}, \bibinfo {author} {\bibfnamefont {G.}~\bibnamefont {Clark}},
  \bibinfo {author} {\bibfnamefont {E.}~\bibnamefont {Navarro-Moratalla}},
  \bibinfo {author} {\bibfnamefont {D.~R.}\ \bibnamefont {Klein}}, \bibinfo
  {author} {\bibfnamefont {R.}~\bibnamefont {Cheng}}, \bibinfo {author}
  {\bibfnamefont {K.~L.}\ \bibnamefont {Seyler}}, \bibinfo {author}
  {\bibfnamefont {D.}~\bibnamefont {Zhong}}, \bibinfo {author} {\bibfnamefont
  {E.}~\bibnamefont {Schmidgall}}, \bibinfo {author} {\bibfnamefont {M.~A.}\
  \bibnamefont {McGuire}}, \bibinfo {author} {\bibfnamefont {D.~H.}\
  \bibnamefont {Cobden}}, \bibinfo {author} {\bibfnamefont {W.}~\bibnamefont
  {Yao}}, \bibinfo {author} {\bibfnamefont {D.}~\bibnamefont {Xiao}}, \bibinfo
  {author} {\bibfnamefont {P.}~\bibnamefont {Jarillo-Herrero}},\ and\ \bibinfo
  {author} {\bibfnamefont {X.~D.}\ \bibnamefont {Xu}},\ }\href
  {https://doi.org/10.1038/nature22391} {\bibfield  {journal} {\bibinfo
  {journal} {Nature}\ }\textbf {\bibinfo {volume} {546}},\ \bibinfo {pages}
  {270} (\bibinfo {year} {2017})}\BibitemShut {NoStop}%
\bibitem [{\citenamefont {McGuire}\ \emph {et~al.}(2015)\citenamefont
  {McGuire}, \citenamefont {Dixit}, \citenamefont {Cooper},\ and\ \citenamefont
  {Sales}}]{RN255}%
  \BibitemOpen
  \bibfield  {author} {\bibinfo {author} {\bibfnamefont {M.~A.}\ \bibnamefont
  {McGuire}}, \bibinfo {author} {\bibfnamefont {H.}~\bibnamefont {Dixit}},
  \bibinfo {author} {\bibfnamefont {V.~R.}\ \bibnamefont {Cooper}},\ and\
  \bibinfo {author} {\bibfnamefont {B.~C.}\ \bibnamefont {Sales}},\ }\href
  {https://doi.org/10.1021/cm504242t} {\bibfield  {journal} {\bibinfo
  {journal} {Chemistry of Materials}\ }\textbf {\bibinfo {volume} {27}},\
  \bibinfo {pages} {612} (\bibinfo {year} {2015})}\BibitemShut {NoStop}%
\bibitem [{\citenamefont {Jiang}\ \emph
  {et~al.}(2018{\natexlab{a}})\citenamefont {Jiang}, \citenamefont {Shan},\
  and\ \citenamefont {Mak}}]{RN251}%
  \BibitemOpen
  \bibfield  {author} {\bibinfo {author} {\bibfnamefont {S.~W.}\ \bibnamefont
  {Jiang}}, \bibinfo {author} {\bibfnamefont {J.}~\bibnamefont {Shan}},\ and\
  \bibinfo {author} {\bibfnamefont {K.~F.}\ \bibnamefont {Mak}},\ }\href
  {https://doi.org/10.1038/s41563-018-0040-6} {\bibfield  {journal} {\bibinfo
  {journal} {Nature Materials}\ }\textbf {\bibinfo {volume} {17}},\ \bibinfo
  {pages} {406} (\bibinfo {year} {2018}{\natexlab{a}})}\BibitemShut {NoStop}%
\bibitem [{\citenamefont {Jiang}\ \emph
  {et~al.}(2018{\natexlab{b}})\citenamefont {Jiang}, \citenamefont {Li},
  \citenamefont {Wang}, \citenamefont {Mak},\ and\ \citenamefont
  {Shan}}]{RN254}%
  \BibitemOpen
  \bibfield  {author} {\bibinfo {author} {\bibfnamefont {S.~W.}\ \bibnamefont
  {Jiang}}, \bibinfo {author} {\bibfnamefont {L.~Z.}\ \bibnamefont {Li}},
  \bibinfo {author} {\bibfnamefont {Z.~F.}\ \bibnamefont {Wang}}, \bibinfo
  {author} {\bibfnamefont {K.~F.}\ \bibnamefont {Mak}},\ and\ \bibinfo {author}
  {\bibfnamefont {J.}~\bibnamefont {Shan}},\ }\href
  {https://doi.org/10.1038/s41565-018-0135-x} {\bibfield  {journal} {\bibinfo
  {journal} {Nature Nanotechnology}\ }\textbf {\bibinfo {volume} {13}},\
  \bibinfo {pages} {549} (\bibinfo {year} {2018}{\natexlab{b}})}\BibitemShut
  {NoStop}%
\bibitem [{\citenamefont {Song}\ \emph {et~al.}(2019)\citenamefont {Song},
  \citenamefont {Fei}, \citenamefont {Yankowitz}, \citenamefont {Lin},
  \citenamefont {Jiang}, \citenamefont {Hwangbo}, \citenamefont {Zhang},
  \citenamefont {Sun}, \citenamefont {Taniguchi}, \citenamefont {Watanabe},
  \citenamefont {McGuire}, \citenamefont {Graf}, \citenamefont {Cao},
  \citenamefont {Chu}, \citenamefont {Cobden}, \citenamefont {Dean},
  \citenamefont {Xiao},\ and\ \citenamefont {Xu}}]{RN252}%
  \BibitemOpen
  \bibfield  {author} {\bibinfo {author} {\bibfnamefont {T.~C.}\ \bibnamefont
  {Song}}, \bibinfo {author} {\bibfnamefont {Z.~Y.}\ \bibnamefont {Fei}},
  \bibinfo {author} {\bibfnamefont {M.}~\bibnamefont {Yankowitz}}, \bibinfo
  {author} {\bibfnamefont {Z.}~\bibnamefont {Lin}}, \bibinfo {author}
  {\bibfnamefont {Q.~N.}\ \bibnamefont {Jiang}}, \bibinfo {author}
  {\bibfnamefont {K.}~\bibnamefont {Hwangbo}}, \bibinfo {author} {\bibfnamefont
  {Q.}~\bibnamefont {Zhang}}, \bibinfo {author} {\bibfnamefont {B.~S.}\
  \bibnamefont {Sun}}, \bibinfo {author} {\bibfnamefont {T.}~\bibnamefont
  {Taniguchi}}, \bibinfo {author} {\bibfnamefont {K.}~\bibnamefont {Watanabe}},
  \bibinfo {author} {\bibfnamefont {M.~A.}\ \bibnamefont {McGuire}}, \bibinfo
  {author} {\bibfnamefont {D.}~\bibnamefont {Graf}}, \bibinfo {author}
  {\bibfnamefont {T.}~\bibnamefont {Cao}}, \bibinfo {author} {\bibfnamefont
  {J.~H.}\ \bibnamefont {Chu}}, \bibinfo {author} {\bibfnamefont {D.~H.}\
  \bibnamefont {Cobden}}, \bibinfo {author} {\bibfnamefont {C.~R.}\
  \bibnamefont {Dean}}, \bibinfo {author} {\bibfnamefont {D.}~\bibnamefont
  {Xiao}},\ and\ \bibinfo {author} {\bibfnamefont {X.~D.}\ \bibnamefont {Xu}},\
  }\href {https://doi.org/10.1038/s41563-019-0505-2} {\bibfield  {journal}
  {\bibinfo  {journal} {Nature Materials}\ }\textbf {\bibinfo {volume} {18}},\
  \bibinfo {pages} {1298} (\bibinfo {year} {2019})}\BibitemShut {NoStop}%
\bibitem [{\citenamefont {Li}\ \emph {et~al.}(2019{\natexlab{c}})\citenamefont
  {Li}, \citenamefont {Jiang}, \citenamefont {Sivadas}, \citenamefont {Wang},
  \citenamefont {Xu}, \citenamefont {Weber}, \citenamefont {Goldberger},
  \citenamefont {Watanabe}, \citenamefont {Taniguchi}, \citenamefont {Fennie},
  \citenamefont {Mak},\ and\ \citenamefont {Shan}}]{ISI:000497968400013}%
  \BibitemOpen
  \bibfield  {author} {\bibinfo {author} {\bibfnamefont {T.}~\bibnamefont
  {Li}}, \bibinfo {author} {\bibfnamefont {S.}~\bibnamefont {Jiang}}, \bibinfo
  {author} {\bibfnamefont {N.}~\bibnamefont {Sivadas}}, \bibinfo {author}
  {\bibfnamefont {Z.}~\bibnamefont {Wang}}, \bibinfo {author} {\bibfnamefont
  {Y.}~\bibnamefont {Xu}}, \bibinfo {author} {\bibfnamefont {D.}~\bibnamefont
  {Weber}}, \bibinfo {author} {\bibfnamefont {J.}~\bibnamefont {Goldberger}},
  \bibinfo {author} {\bibfnamefont {K.}~\bibnamefont {Watanabe}}, \bibinfo
  {author} {\bibfnamefont {T.}~\bibnamefont {Taniguchi}}, \bibinfo {author}
  {\bibfnamefont {C.}~\bibnamefont {Fennie}}, \bibinfo {author} {\bibfnamefont
  {K.}~\bibnamefont {Mak}},\ and\ \bibinfo {author} {\bibfnamefont
  {J.}~\bibnamefont {Shan}},\ }\href@noop {} {\bibfield  {journal} {\bibinfo
  {journal} {Nature Materials}\ }\textbf {\bibinfo {volume} {18}},\ \bibinfo
  {pages} {1303} (\bibinfo {year} {2019}{\natexlab{c}})}\BibitemShut {NoStop}%
\bibitem [{\citenamefont {Sivadas}\ \emph {et~al.}(2018)\citenamefont
  {Sivadas}, \citenamefont {Okamoto}, \citenamefont {Xu}, \citenamefont
  {Fennie},\ and\ \citenamefont {Xiao}}]{RN257}%
  \BibitemOpen
  \bibfield  {author} {\bibinfo {author} {\bibfnamefont {N.}~\bibnamefont
  {Sivadas}}, \bibinfo {author} {\bibfnamefont {S.}~\bibnamefont {Okamoto}},
  \bibinfo {author} {\bibfnamefont {X.~D.}\ \bibnamefont {Xu}}, \bibinfo
  {author} {\bibfnamefont {C.~J.}\ \bibnamefont {Fennie}},\ and\ \bibinfo
  {author} {\bibfnamefont {D.}~\bibnamefont {Xiao}},\ }\href
  {https://doi.org/10.1021/acs.nanolett.8b03321} {\bibfield  {journal}
  {\bibinfo  {journal} {Nano Letters}\ }\textbf {\bibinfo {volume} {18}},\
  \bibinfo {pages} {7658} (\bibinfo {year} {2018})}\BibitemShut {NoStop}%
\bibitem [{\citenamefont {Jang}\ \emph {et~al.}(2019)\citenamefont {Jang},
  \citenamefont {Jeong}, \citenamefont {Yoon}, \citenamefont {Ryee},\ and\
  \citenamefont {Han}}]{RN258}%
  \BibitemOpen
  \bibfield  {author} {\bibinfo {author} {\bibfnamefont {S.~W.}\ \bibnamefont
  {Jang}}, \bibinfo {author} {\bibfnamefont {M.~Y.}\ \bibnamefont {Jeong}},
  \bibinfo {author} {\bibfnamefont {H.}~\bibnamefont {Yoon}}, \bibinfo {author}
  {\bibfnamefont {S.}~\bibnamefont {Ryee}},\ and\ \bibinfo {author}
  {\bibfnamefont {M.~J.}\ \bibnamefont {Han}},\ }\href
  {https://doi.org/10.1103/PhysRevMaterials.3.031001} {\bibfield  {journal}
  {\bibinfo  {journal} {Physical Review Materials}\ }\textbf {\bibinfo {volume}
  {3}},\ \bibinfo {pages} {6} (\bibinfo {year} {2019})}\BibitemShut {NoStop}%
\bibitem [{\citenamefont {Song}\ \emph {et~al.}(2018)\citenamefont {Song},
  \citenamefont {Cai}, \citenamefont {Tu}, \citenamefont {Zhang}, \citenamefont
  {Huang}, \citenamefont {Wilson}, \citenamefont {Seyler}, \citenamefont {Zhu},
  \citenamefont {Taniguchi}, \citenamefont {Watanabe}, \citenamefont {McGuire},
  \citenamefont {Cobden}, \citenamefont {Xiao}, \citenamefont {Yao},\ and\
  \citenamefont {Xu}}]{RN253}%
  \BibitemOpen
  \bibfield  {author} {\bibinfo {author} {\bibfnamefont {T.~C.}\ \bibnamefont
  {Song}}, \bibinfo {author} {\bibfnamefont {X.~H.}\ \bibnamefont {Cai}},
  \bibinfo {author} {\bibfnamefont {M.~W.~Y.}\ \bibnamefont {Tu}}, \bibinfo
  {author} {\bibfnamefont {X.~O.}\ \bibnamefont {Zhang}}, \bibinfo {author}
  {\bibfnamefont {B.~V.}\ \bibnamefont {Huang}}, \bibinfo {author}
  {\bibfnamefont {N.~P.}\ \bibnamefont {Wilson}}, \bibinfo {author}
  {\bibfnamefont {K.~L.}\ \bibnamefont {Seyler}}, \bibinfo {author}
  {\bibfnamefont {L.}~\bibnamefont {Zhu}}, \bibinfo {author} {\bibfnamefont
  {T.}~\bibnamefont {Taniguchi}}, \bibinfo {author} {\bibfnamefont
  {K.}~\bibnamefont {Watanabe}}, \bibinfo {author} {\bibfnamefont {M.~A.}\
  \bibnamefont {McGuire}}, \bibinfo {author} {\bibfnamefont {D.~H.}\
  \bibnamefont {Cobden}}, \bibinfo {author} {\bibfnamefont {D.}~\bibnamefont
  {Xiao}}, \bibinfo {author} {\bibfnamefont {W.}~\bibnamefont {Yao}},\ and\
  \bibinfo {author} {\bibfnamefont {X.~D.}\ \bibnamefont {Xu}},\ }\href
  {https://doi.org/10.1126/science.aar4851} {\bibfield  {journal} {\bibinfo
  {journal} {Science}\ }\textbf {\bibinfo {volume} {360}},\ \bibinfo {pages}
  {1214} (\bibinfo {year} {2018})}\BibitemShut {NoStop}%
\bibitem [{\citenamefont {Kresse}\ and\ \citenamefont
  {Furthmuller}(1996)}]{RN2539}%
  \BibitemOpen
  \bibfield  {author} {\bibinfo {author} {\bibfnamefont {G.}~\bibnamefont
  {Kresse}}\ and\ \bibinfo {author} {\bibfnamefont {J.}~\bibnamefont
  {Furthmuller}},\ }\href {https://doi.org/10.1103/PhysRevB.54.11169}
  {\bibfield  {journal} {\bibinfo  {journal} {Physical Review B}\ }\textbf
  {\bibinfo {volume} {54}},\ \bibinfo {pages} {11169} (\bibinfo {year}
  {1996})}\BibitemShut {NoStop}%
\bibitem [{Note1()}]{Note1}%
  \BibitemOpen
  \bibinfo {note} {We set an energy cutoff of $450 \protect \tmspace
  +\thinmuskip {.1667em}\protect \textrm {eV}$ for plane waves and adopted the
  Perdew-Burke-Ernzerhof exchange correlation energy functional \cite {RN8}
  together with PAW pseudopotentials. \cite {RN2495} A $9\times 9 \times 1$
  Monkhorst-Pack mesh for sampling the first Brillouin zone was applied. The
  van der Waals interaction was taken into account via the PBE-D3 method. An
  energy tolerance of $1\times 10^{-6} \protect \tmspace +\thinmuskip
  {.1667em}\protect \textrm {eV}$ and a force tolerance of $0.001 \protect
  \tmspace +\thinmuskip {.1667em} \protect \textrm {eV}/\protect \textrm {\r
  A}$ were used for self-consistent and ionic relaxations, respectively. A
  vacuum region separates periodic images of the 2D system along the
  out-of-plane direction by at least $12\protect \tmspace +\thinmuskip
  {.1667em}\protect \textrm {\r A}$ to eliminate any interaction.}\BibitemShut
  {Stop}%
\bibitem [{\citenamefont {Dudarev}\ \emph {et~al.}(1998)\citenamefont
  {Dudarev}, \citenamefont {Botton}, \citenamefont {Savrasov}, \citenamefont
  {Humphreys},\ and\ \citenamefont {Sutton}}]{RN262}%
  \BibitemOpen
  \bibfield  {author} {\bibinfo {author} {\bibfnamefont {S.~L.}\ \bibnamefont
  {Dudarev}}, \bibinfo {author} {\bibfnamefont {G.~A.}\ \bibnamefont {Botton}},
  \bibinfo {author} {\bibfnamefont {S.~Y.}\ \bibnamefont {Savrasov}}, \bibinfo
  {author} {\bibfnamefont {C.~J.}\ \bibnamefont {Humphreys}},\ and\ \bibinfo
  {author} {\bibfnamefont {A.~P.}\ \bibnamefont {Sutton}},\ }\href
  {https://doi.org/10.1103/PhysRevB.57.1505} {\bibfield  {journal} {\bibinfo
  {journal} {Physical Review B}\ }\textbf {\bibinfo {volume} {57}},\ \bibinfo
  {pages} {1505} (\bibinfo {year} {1998})}\BibitemShut {NoStop}%
\bibitem [{\citenamefont {Amadon}(2012)}]{Amadon_2012}%
  \BibitemOpen
  \bibfield  {author} {\bibinfo {author} {\bibfnamefont {B.}~\bibnamefont
  {Amadon}},\ }\href {https://doi.org/10.1088/0953-8984/24/7/075604} {\bibfield
   {journal} {\bibinfo  {journal} {Journal of Physics: Condensed Matter}\
  }\textbf {\bibinfo {volume} {24}},\ \bibinfo {pages} {075604} (\bibinfo
  {year} {2012})}\BibitemShut {NoStop}%
\bibitem [{\citenamefont {Park}\ \emph {et~al.}(2014)\citenamefont {Park},
  \citenamefont {Millis},\ and\ \citenamefont
  {Marianetti}}]{PhysRevB.90.235103}%
  \BibitemOpen
  \bibfield  {author} {\bibinfo {author} {\bibfnamefont {H.}~\bibnamefont
  {Park}}, \bibinfo {author} {\bibfnamefont {A.~J.}\ \bibnamefont {Millis}},\
  and\ \bibinfo {author} {\bibfnamefont {C.~A.}\ \bibnamefont {Marianetti}},\
  }\href {https://doi.org/10.1103/PhysRevB.90.235103} {\bibfield  {journal}
  {\bibinfo  {journal} {Phys. Rev. B}\ }\textbf {\bibinfo {volume} {90}},\
  \bibinfo {pages} {235103} (\bibinfo {year} {2014})}\BibitemShut {NoStop}%
\bibitem [{\citenamefont {Bhandary}\ \emph {et~al.}(2016)\citenamefont
  {Bhandary}, \citenamefont {Assmann}, \citenamefont {Aichhorn},\ and\
  \citenamefont {Held}}]{PhysRevB.94.155131}%
  \BibitemOpen
  \bibfield  {author} {\bibinfo {author} {\bibfnamefont {S.}~\bibnamefont
  {Bhandary}}, \bibinfo {author} {\bibfnamefont {E.}~\bibnamefont {Assmann}},
  \bibinfo {author} {\bibfnamefont {M.}~\bibnamefont {Aichhorn}},\ and\
  \bibinfo {author} {\bibfnamefont {K.}~\bibnamefont {Held}},\ }\href
  {https://doi.org/10.1103/PhysRevB.94.155131} {\bibfield  {journal} {\bibinfo
  {journal} {Phys. Rev. B}\ }\textbf {\bibinfo {volume} {94}},\ \bibinfo
  {pages} {155131} (\bibinfo {year} {2016})}\BibitemShut {NoStop}%
\bibitem [{ELK(2018)}]{ELK_code}%
  \BibitemOpen
  \href@noop {} {\bibinfo {title} {{ELK:} an all-electron full-potential
  linearised augmented-plane wave code}},\ \bibinfo {howpublished}
  {\url{http://elk.sourceforge.net/}} (\bibinfo {year} {2018})\BibitemShut
  {NoStop}%
\bibitem [{\citenamefont {Kozhevnikov}\ \emph {et~al.}(2010)\citenamefont
  {Kozhevnikov}, \citenamefont {Eguiluz},\ and\ \citenamefont
  {Schulthess}}]{Anton_IEEE_paper_2010}%
  \BibitemOpen
  \bibfield  {author} {\bibinfo {author} {\bibfnamefont {A.}~\bibnamefont
  {Kozhevnikov}}, \bibinfo {author} {\bibfnamefont {A.~G.}\ \bibnamefont
  {Eguiluz}},\ and\ \bibinfo {author} {\bibfnamefont {T.~C.}\ \bibnamefont
  {Schulthess}},\ }\href {https://doi.org/10.1109/SC.2010.55} {\bibfield
  {journal} {\bibinfo  {journal} {2010 ACM/IEEE International Conference for
  High Performance Computing, Networking, Storage and Analysis}\ ,\ \bibinfo
  {pages} {1}} (\bibinfo {year} {2010})}\BibitemShut {NoStop}%
\bibitem [{\citenamefont {Zhang}\ \emph {et~al.}(2019)\citenamefont {Zhang},
  \citenamefont {Staar}, \citenamefont {Kozhevnikov}, \citenamefont {Wang},
  \citenamefont {Trinastic}, \citenamefont {Schulthess},\ and\ \citenamefont
  {Cheng}}]{PhysRevB.100.035104}%
  \BibitemOpen
  \bibfield  {author} {\bibinfo {author} {\bibfnamefont {L.}~\bibnamefont
  {Zhang}}, \bibinfo {author} {\bibfnamefont {P.}~\bibnamefont {Staar}},
  \bibinfo {author} {\bibfnamefont {A.}~\bibnamefont {Kozhevnikov}}, \bibinfo
  {author} {\bibfnamefont {Y.-P.}\ \bibnamefont {Wang}}, \bibinfo {author}
  {\bibfnamefont {J.}~\bibnamefont {Trinastic}}, \bibinfo {author}
  {\bibfnamefont {T.}~\bibnamefont {Schulthess}},\ and\ \bibinfo {author}
  {\bibfnamefont {H.-P.}\ \bibnamefont {Cheng}},\ }\href
  {https://doi.org/10.1103/PhysRevB.100.035104} {\bibfield  {journal} {\bibinfo
   {journal} {Phys. Rev. B}\ }\textbf {\bibinfo {volume} {100}},\ \bibinfo
  {pages} {035104} (\bibinfo {year} {2019})}\BibitemShut {NoStop}%
\bibitem [{\citenamefont {Sakuma}\ and\ \citenamefont
  {Aryasetiawan}(2013)}]{U_from_MLWF_cRPA_NiOCoOFeOMnO_2013}%
  \BibitemOpen
  \bibfield  {author} {\bibinfo {author} {\bibfnamefont {R.}~\bibnamefont
  {Sakuma}}\ and\ \bibinfo {author} {\bibfnamefont {F.}~\bibnamefont
  {Aryasetiawan}},\ }\href {https://doi.org/10.1103/PhysRevB.87.165118}
  {\bibfield  {journal} {\bibinfo  {journal} {Phys. Rev. B}\ }\textbf {\bibinfo
  {volume} {87}},\ \bibinfo {pages} {165118} (\bibinfo {year}
  {2013})}\BibitemShut {NoStop}%
\bibitem [{\citenamefont {Soler}\ \emph {et~al.}(2002)\citenamefont {Soler},
  \citenamefont {Artacho}, \citenamefont {Gale}, \citenamefont {Garc\'{i}a},
  \citenamefont {Junquera1}, \citenamefont {Ordejón},\ and\ \citenamefont
  {Sánchez-Portal}}]{RN304}%
  \BibitemOpen
  \bibfield  {author} {\bibinfo {author} {\bibfnamefont {J.~M.}\ \bibnamefont
  {Soler}}, \bibinfo {author} {\bibfnamefont {E.}~\bibnamefont {Artacho}},
  \bibinfo {author} {\bibfnamefont {J.~D.}\ \bibnamefont {Gale}}, \bibinfo
  {author} {\bibfnamefont {A.}~\bibnamefont {Garc\'{i}a}}, \bibinfo {author}
  {\bibfnamefont {J.}~\bibnamefont {Junquera1}}, \bibinfo {author}
  {\bibfnamefont {P.}~\bibnamefont {Ordejón}},\ and\ \bibinfo {author}
  {\bibfnamefont {D.}~\bibnamefont {Sánchez-Portal}},\ }\href
  {https://doi.org/https://doi.org/10.1088/0953-8984/14/11/302} {\bibfield
  {journal} {\bibinfo  {journal} {Journal of Physics: Condensed Matter}\
  }\textbf {\bibinfo {volume} {14}},\ \bibinfo {pages} {2745} (\bibinfo {year}
  {2002})}\BibitemShut {NoStop}%
\bibitem [{Note2()}]{Note2}%
  \BibitemOpen
  \bibinfo {note} {We used a double-$\zeta $ basis set for \ce {Cr} $3d$
  orbitals, a single-$\zeta $ polarized basis set for \ce {Cr} $4s$ orbitals,
  and a single-$\zeta $ basis set for \ce {I} $5s$ and $5p$ orbitals. We
  applied the Perdew-Burke-Ernzerhof exchange correlation energy functional and
  norm-conserving pseudo-potentials. A $51 \times 51 \times 1$ Monkhorst-Pack
  $k$-mesh was used to sample the reciprocal space. Such a $k$-mesh was tested
  to be dense enough to capture the interlayer charge transfer between graphene
  and \ce {CrI_3}{}. A MeshCutoff of $150 \protect \tmspace +\thinmuskip
  {.1667em}\protect \textrm {Ry}{}$ was applied for the real space sampling.
  The Hubbard $U$ parameter in the DFT$+U$ method was set to $4\protect
  \tmspace +\thinmuskip {.1667em}\protect \textrm {eV}$. For insulating or
  semiconducting systems, we adopted a Fermi-Dirac function with $T=10\protect
  \tmspace +\thinmuskip {.1667em} \protect \textrm {K}$ to determine the
  occupation of Kohn-Sham orbitals. For metallic systems, we adopted the $4$th
  order Methfessel-Paxton smearing method with $T=200 \protect \tmspace
  +\thinmuskip {.1667em} \protect \textrm {K}$ to calculate the electron
  distribution accurately.}\BibitemShut {Stop}%
\bibitem [{Note3()}]{Note3}%
  \BibitemOpen
  \bibinfo {note} {The energy difference $E_{\protect \textrm {AFM}} -
  E_{\protect \textrm {FM}}$ in Figure~\ref {fig:CrI3a}b differs from that in
  \ref {fig:CrI3a}d because the former is calculated using a localized basis
  set (SIESTA package) but the latter using plane waves (VASP package). Plane
  wave results are considered to be more accurate.}\BibitemShut {Stop}%
\bibitem [{\citenamefont {Morell}\ \emph {et~al.}(2019)\citenamefont {Morell},
  \citenamefont {Leon}, \citenamefont {Miwa},\ and\ \citenamefont
  {Vargas}}]{RN256}%
  \BibitemOpen
  \bibfield  {author} {\bibinfo {author} {\bibfnamefont {E.~S.}\ \bibnamefont
  {Morell}}, \bibinfo {author} {\bibfnamefont {A.}~\bibnamefont {Leon}},
  \bibinfo {author} {\bibfnamefont {R.~H.}\ \bibnamefont {Miwa}},\ and\
  \bibinfo {author} {\bibfnamefont {P.}~\bibnamefont {Vargas}},\ }\href
  {https://doi.org/10.1088/2053-1583/ab04fb} {\bibfield  {journal} {\bibinfo
  {journal} {2d Materials}\ }\textbf {\bibinfo {volume} {6}},\ \bibinfo {pages}
  {6} (\bibinfo {year} {2019})}\BibitemShut {NoStop}%
\bibitem [{\citenamefont {Zyazin}\ \emph {et~al.}(2010)\citenamefont {Zyazin},
  \citenamefont {van~den Berg}, \citenamefont {Osorio}, \citenamefont {van~der
  Zant}, \citenamefont {Konstantinidis}, \citenamefont {Leijnse}, \citenamefont
  {Wegewijs}, \citenamefont {May}, \citenamefont {Hofstetter}, \citenamefont
  {Danieli},\ and\ \citenamefont {Cornia}}]{RN104}%
  \BibitemOpen
  \bibfield  {author} {\bibinfo {author} {\bibfnamefont {A.~S.}\ \bibnamefont
  {Zyazin}}, \bibinfo {author} {\bibfnamefont {J.~W.~G.}\ \bibnamefont {van~den
  Berg}}, \bibinfo {author} {\bibfnamefont {E.~A.}\ \bibnamefont {Osorio}},
  \bibinfo {author} {\bibfnamefont {H.~S.~J.}\ \bibnamefont {van~der Zant}},
  \bibinfo {author} {\bibfnamefont {N.~P.}\ \bibnamefont {Konstantinidis}},
  \bibinfo {author} {\bibfnamefont {M.}~\bibnamefont {Leijnse}}, \bibinfo
  {author} {\bibfnamefont {M.~R.}\ \bibnamefont {Wegewijs}}, \bibinfo {author}
  {\bibfnamefont {F.}~\bibnamefont {May}}, \bibinfo {author} {\bibfnamefont
  {W.}~\bibnamefont {Hofstetter}}, \bibinfo {author} {\bibfnamefont
  {C.}~\bibnamefont {Danieli}},\ and\ \bibinfo {author} {\bibfnamefont
  {A.}~\bibnamefont {Cornia}},\ }\href {https://doi.org/10.1021/nl1009603}
  {\bibfield  {journal} {\bibinfo  {journal} {Nano Letters}\ }\textbf {\bibinfo
  {volume} {10}},\ \bibinfo {pages} {3307} (\bibinfo {year}
  {2010})}\BibitemShut {NoStop}%
\bibitem [{\citenamefont {Li}\ \emph {et~al.}(2014)\citenamefont {Li},
  \citenamefont {Fry},\ and\ \citenamefont {Cheng}}]{RN259}%
  \BibitemOpen
  \bibfield  {author} {\bibinfo {author} {\bibfnamefont {X.~G.}\ \bibnamefont
  {Li}}, \bibinfo {author} {\bibfnamefont {J.~N.}\ \bibnamefont {Fry}},\ and\
  \bibinfo {author} {\bibfnamefont {H.~P.}\ \bibnamefont {Cheng}},\ }\href
  {https://doi.org/10.1103/PhysRevB.90.125447} {\bibfield  {journal} {\bibinfo
  {journal} {Physical Review B}\ }\textbf {\bibinfo {volume} {90}},\ \bibinfo
  {pages} {7} (\bibinfo {year} {2014})}\BibitemShut {NoStop}%
\bibitem [{\citenamefont {Zhang}\ \emph {et~al.}(2009)\citenamefont {Zhang},
  \citenamefont {Ma}, \citenamefont {Li}, \citenamefont {Qian}, \citenamefont
  {Shen}, \citenamefont {Zhao}, \citenamefont {Hou},\ and\ \citenamefont
  {Sanvito}}]{RN263}%
  \BibitemOpen
  \bibfield  {author} {\bibinfo {author} {\bibfnamefont {R.~X.}\ \bibnamefont
  {Zhang}}, \bibinfo {author} {\bibfnamefont {G.~H.}\ \bibnamefont {Ma}},
  \bibinfo {author} {\bibfnamefont {R.}~\bibnamefont {Li}}, \bibinfo {author}
  {\bibfnamefont {Z.~K.}\ \bibnamefont {Qian}}, \bibinfo {author}
  {\bibfnamefont {Z.~Y.}\ \bibnamefont {Shen}}, \bibinfo {author}
  {\bibfnamefont {X.~Y.}\ \bibnamefont {Zhao}}, \bibinfo {author}
  {\bibfnamefont {S.~M.}\ \bibnamefont {Hou}},\ and\ \bibinfo {author}
  {\bibfnamefont {S.}~\bibnamefont {Sanvito}},\ }\href
  {https://doi.org/10.1088/0953-8984/21/33/335301} {\bibfield  {journal}
  {\bibinfo  {journal} {Journal of Physics-Condensed Matter}\ }\textbf
  {\bibinfo {volume} {21}},\ \bibinfo {pages} {9} (\bibinfo {year}
  {2009})}\BibitemShut {NoStop}%
\bibitem [{\citenamefont {Yuan}\ \emph {et~al.}(2016)\citenamefont {Yuan},
  \citenamefont {Yan}, \citenamefont {Xiao}, \citenamefont {Guo},\ and\
  \citenamefont {Dai}}]{RN264}%
  \BibitemOpen
  \bibfield  {author} {\bibinfo {author} {\bibfnamefont {J.~R.}\ \bibnamefont
  {Yuan}}, \bibinfo {author} {\bibfnamefont {X.~H.}\ \bibnamefont {Yan}},
  \bibinfo {author} {\bibfnamefont {Y.}~\bibnamefont {Xiao}}, \bibinfo {author}
  {\bibfnamefont {Y.~D.}\ \bibnamefont {Guo}},\ and\ \bibinfo {author}
  {\bibfnamefont {C.~J.}\ \bibnamefont {Dai}},\ }\href
  {https://doi.org/10.1088/0957-4484/27/47/475202} {\bibfield  {journal}
  {\bibinfo  {journal} {Nanotechnology}\ }\textbf {\bibinfo {volume} {27}},\
  \bibinfo {pages} {7} (\bibinfo {year} {2016})}\BibitemShut {NoStop}%
\bibitem [{\citenamefont {Zhang}\ \emph {et~al.}(2017)\citenamefont {Zhang},
  \citenamefont {Jia}, \citenamefont {Kholmanov}, \citenamefont {Dong},
  \citenamefont {Er}, \citenamefont {Chen}, \citenamefont {Guo}, \citenamefont
  {Jin}, \citenamefont {Shenoy}, \citenamefont {Shi},\ and\ \citenamefont
  {Lou}}]{RN260}%
  \BibitemOpen
  \bibfield  {author} {\bibinfo {author} {\bibfnamefont {J.}~\bibnamefont
  {Zhang}}, \bibinfo {author} {\bibfnamefont {S.}~\bibnamefont {Jia}}, \bibinfo
  {author} {\bibfnamefont {I.}~\bibnamefont {Kholmanov}}, \bibinfo {author}
  {\bibfnamefont {L.}~\bibnamefont {Dong}}, \bibinfo {author} {\bibfnamefont
  {D.~Q.}\ \bibnamefont {Er}}, \bibinfo {author} {\bibfnamefont {W.~B.}\
  \bibnamefont {Chen}}, \bibinfo {author} {\bibfnamefont {H.}~\bibnamefont
  {Guo}}, \bibinfo {author} {\bibfnamefont {Z.~H.}\ \bibnamefont {Jin}},
  \bibinfo {author} {\bibfnamefont {V.~B.}\ \bibnamefont {Shenoy}}, \bibinfo
  {author} {\bibfnamefont {L.}~\bibnamefont {Shi}},\ and\ \bibinfo {author}
  {\bibfnamefont {J.}~\bibnamefont {Lou}},\ }\href
  {https://doi.org/10.1021/acsnano.7b03186} {\bibfield  {journal} {\bibinfo
  {journal} {Acs Nano}\ }\textbf {\bibinfo {volume} {11}},\ \bibinfo {pages}
  {8192} (\bibinfo {year} {2017})}\BibitemShut {NoStop}%
\bibitem [{\citenamefont {Lu}\ \emph {et~al.}(2017)\citenamefont {Lu},
  \citenamefont {Zhu}, \citenamefont {Xiao}, \citenamefont {Chuu},
  \citenamefont {Han}, \citenamefont {Chiu}, \citenamefont {Cheng},
  \citenamefont {Yang}, \citenamefont {Wei}, \citenamefont {Yang},
  \citenamefont {Wang}, \citenamefont {Sokaras}, \citenamefont {Nordlund},
  \citenamefont {Yang}, \citenamefont {Muller}, \citenamefont {Chou},
  \citenamefont {Zhang},\ and\ \citenamefont {Li}}]{RN261}%
  \BibitemOpen
  \bibfield  {author} {\bibinfo {author} {\bibfnamefont {A.~Y.}\ \bibnamefont
  {Lu}}, \bibinfo {author} {\bibfnamefont {H.~Y.}\ \bibnamefont {Zhu}},
  \bibinfo {author} {\bibfnamefont {J.}~\bibnamefont {Xiao}}, \bibinfo {author}
  {\bibfnamefont {C.~P.}\ \bibnamefont {Chuu}}, \bibinfo {author}
  {\bibfnamefont {Y.~M.}\ \bibnamefont {Han}}, \bibinfo {author} {\bibfnamefont
  {M.~H.}\ \bibnamefont {Chiu}}, \bibinfo {author} {\bibfnamefont {C.~C.}\
  \bibnamefont {Cheng}}, \bibinfo {author} {\bibfnamefont {C.~W.}\ \bibnamefont
  {Yang}}, \bibinfo {author} {\bibfnamefont {K.~H.}\ \bibnamefont {Wei}},
  \bibinfo {author} {\bibfnamefont {Y.~M.}\ \bibnamefont {Yang}}, \bibinfo
  {author} {\bibfnamefont {Y.}~\bibnamefont {Wang}}, \bibinfo {author}
  {\bibfnamefont {D.}~\bibnamefont {Sokaras}}, \bibinfo {author} {\bibfnamefont
  {D.}~\bibnamefont {Nordlund}}, \bibinfo {author} {\bibfnamefont {P.~D.}\
  \bibnamefont {Yang}}, \bibinfo {author} {\bibfnamefont {D.~A.}\ \bibnamefont
  {Muller}}, \bibinfo {author} {\bibfnamefont {M.~Y.}\ \bibnamefont {Chou}},
  \bibinfo {author} {\bibfnamefont {X.}~\bibnamefont {Zhang}},\ and\ \bibinfo
  {author} {\bibfnamefont {L.~J.}\ \bibnamefont {Li}},\ }\href
  {https://doi.org/10.1038/nnano.2017.100} {\bibfield  {journal} {\bibinfo
  {journal} {Nature Nanotechnology}\ }\textbf {\bibinfo {volume} {12}},\
  \bibinfo {pages} {744} (\bibinfo {year} {2017})}\BibitemShut {NoStop}%
\bibitem [{\citenamefont {Blochl}(1994)}]{RN2495}%
  \BibitemOpen
  \bibfield  {author} {\bibinfo {author} {\bibfnamefont {P.~E.}\ \bibnamefont
  {Blochl}},\ }\href {https://doi.org/10.1103/PhysRevB.50.17953} {\bibfield
  {journal} {\bibinfo  {journal} {Physical Review B}\ }\textbf {\bibinfo
  {volume} {50}},\ \bibinfo {pages} {17953} (\bibinfo {year}
  {1994})}\BibitemShut {NoStop}%
\end{thebibliography}%

\end{document}